\documentclass[AER, reqno]{AEA}

\usepackage{multicol}
\usepackage{mathtools} 
\usepackage{rotating}
\usepackage{pdflscape} 
\usepackage{amssymb}
\usepackage{newtxtext,newtxmath} 
\providecommand{\medium}{\normalsize} 
\providecommand{\sym}[1]{\ifmmode^{#1}\else\(^{#1}\)\fi} 
\usepackage{multirow}
\usepackage{colortbl}
\draftSpacing{1.5}
\usepackage{xr-hyper} 
\usepackage[dvipsnames]{xcolor}
\definecolor{burgundy}{rgb}{0.5, 0.0, 0.13}
\definecolor{darkblue}{rgb}{0.0, 0.0, 0.55}
\RequirePackage[colorlinks,citecolor=burgundy,linkcolor=darkblue,urlcolor=darkblue,pagebackref]{hyperref}
\usepackage[autostyle]{csquotes}
\usepackage[english]{babel}
\usepackage{longtable}
\usepackage{ragged2e}
\usepackage{array}
\usepackage{rotating}
\usepackage{graphicx}
\usepackage{pdfpages}
\usepackage{lscape}
\usepackage{afterpage}
\usepackage{fancyhdr}
\usepackage{geometry}
\usepackage{wrapfig}
\usepackage{subcaption}
\usepackage{booktabs}
\usepackage{tabularx}
\usepackage{threeparttable}
\usepackage{dcolumn}
\usepackage{caption}
\usepackage{float}
\usepackage[export]{adjustbox} 
\usepackage{url}
\usepackage{breakurl}
\usepackage[round]{natbib}
\usepackage{setspace}

\AtBeginDocument{%
\setstretch{2} 
\let\oldfootnotesize\footnotesize
\renewcommand{\footnotesize}{\oldfootnotesize\setstretch{1}}
}

\usepackage{indentfirst}
\usepackage{enumitem}
\setlist{nosep}
\setlist[itemize]{noitemsep, topsep=0pt}
\setlist[enumerate]{noitemsep, topsep=0pt}

\renewcommand{\thefigure}{\Roman{figure}}
\renewenvironment{figurenotes}{
  \vspace{1ex}
  \small
  \raggedright
}{
  \par
}

\makeatletter
\renewcommand{\fnum@figure}{\textup{FIGURE \thefigure}}
\makeatother

\makeatletter
\ifx\figurenotes\undefined
\newenvironment{figurenotes}
{
  \small
  \setlength{\parindent}{0pt}
  \setlength{\parskip}{0pt}
  \par
}{
  \par
}
\fi
\makeatother

\usepackage{titlesec}
\usepackage[splitrule]{footmisc}

\titleformat{\subsubsection}[runin]{\normalfont\normalsize\itshape}{\thesubsubsection}{1em}{}
\renewcommand{\thesection}{\Roman{section}}
\renewcommand{\thesubsection}{\Alph{subsection}. }
\renewcommand{\thesubsubsection}{\thesubsection\Roman{subsubsection}. }
\titleformat{\section}{\bfseries\large\centering}{\thesection}{1em}{\MakeUppercase}
\titlespacing*{\section}{0pt}{0ex plus 0ex minus 0ex}{0pt}
\titleformat{\subsection}{\centering\bfseries\normalsize}{\thesubsection}{1em}{}
\titlespacing{\subsection}{0pt}{1em plus 0ex minus 0ex}{0pt}
\titleformat{\subsubsection}{\normalfont\normalsize}{\thesubsubsection}{1em}{\itshape}
\titlespacing{\subsubsection}{0pt}{0ex plus 0ex minus 0ex}{0pt}
\titlespacing*{\section}{0pt}{0ex plus 0.1ex minus .2ex}{0ex plus .2ex}
\titlespacing{\subsubsection}{0pt}{0ex plus 0.1ex minus .5ex}{0.1ex plus .1ex}

\usepackage{etoolbox}
\makeatletter
\patchcmd{\section}{@afterheading}{}{}{}
\makeatother




\newcommand*{\Rom}[1]{\uppercase\expandafter{\romannumeral #1\relax}}
\renewcommand{\thetable}{\Roman{table}}

\fancypagestyle{plain}{
  \fancyhf{}
  \fancyhead[LE,RO]{\thepage}

}

\newcounter{prop}
\renewcommand{\theprop}{\arabic{prop}}
\newcommand{\proposition}[1]{%
  \refstepcounter{prop}%
  \textbf{Proposition \theprop}\label{#1}%
}

\usepackage[switch*, pagewise]{lineno}

\modulolinenumbers[1]

\makeatletter
\let\old@footnotetext@footnotetext
\makeatother

\renewcommand{\thefigure}{\Roman{figure}}
\renewcommand{\thetable}{\Roman{table}}
\draftSpacing{1.5}
\makeatletter
\renewcommand\footnoterule{%
  \kern-3\p@
  \hrule\@width 0.4\textwidth
  \kern2.6\p@}
\makeatother

\makeatletter
\newcommand{\bibliographytitle}{\markboth{}{}\begin{center}\normalfont\large\bfseries REFERENCES\end{center}}

\makeatother

\begin{document}
\title{
  \normalfont\normalsize
  \bfseries How Do Regulations and Technology Affect Service Allocation and Market Structure?
}
\shortTitle{
  \normalfont\normalsize Effects of Price Regulations and Technology on Healthcare Service Allocation: Evidence from Parity Laws
}

\thispagestyle{plain}
\author{\normalfont Piyush Akimitsu\thanks{Akimitsu: SUNY at Albany, pgade@albany.edu. I thank João Santos Silva, Jonathan Roth, Ulrich Hounyo, John List, Zach Brown, Adrian Masters, Chun-Yu Ho, Pinka Chatterji, Chris Taber, Kate Ho, Casey Mulligan, Katja Seim, Jeffrey Wooldridge, Brent Hickman, Kim Ruhl and Amanda Starc for their helpful comments. I also thank the seminar participants at\textemdash{}University of Chicago, Voltage Lab; 2024 Asia Meeting of the Econometric Society (AMES), Ho Chi Minh City, Vietnam; and Eastern Economic Association (EEA), CSWEP session, Boston, MA.}}

\date{\normalsize \today}

\date{\today}
\pubMonth{Month}
\pubYear{Year}
\pubVolume{Vol}
\pubIssue{Issue}

\begin{abstract}

The paper estimates the effects of Price Controls and Cost Controls on healthcare service quantity and their role in spatial restructuring of physician market. Exploiting the quasi-experimental variation in regulatory-technological environment generated by telehealth parity laws and broadband internet, I find that for aggregate sample and metro areas, Price Floor has a conducive effect on service quantity and physician density, amplified progressively with broadband, and Price Ceiling, unfavorable at mean broadband, converges to the same conducive effect as broadband rises, while Cost Parity has an unfavorable effect, with a negative broadband gradient. For non-metro areas, Price Floor and Price Ceiling have conducive effect on service quantity and unfavorable effect on physician density, indicating efficiency gains. 

\vspace{2em}
\end{abstract}
\vspace{2em}

\maketitle
\skip\footins=\bigskipamount
\pagenumbering{gobble}

\clearpage
\pagenumbering{arabic}
\setcounter{page}{1} 
\section{Introduction}\label{s1}
\interfootnotelinepenalty=10000



\hspace{.5cm} Price Controls are widely used but controversial tools. Conventional economic theory posits that Price Ceilings, commonly discussed in the context of rent-control, result in shortages as they cause suppliers to reduce quantity supplied. Price Floors, discussed mostly in the context of minimum-wage, can lead to surpluses as they encourage higher than equilibrium quantity supplied. Price Controls, such as Price Ceilings or Price Floors, can lead to market inefficiencies and misallocation of goods and services. Consumers often pay an amount approximately equal to the posted price or more. However, in healthcare, if the reimbursement obtained by the physician were to be considered the posted price of a service, the insured consumers don't pay the posted price. These consumers are responsible for ``deductibles, copay and coinsurance" (out-of-pocket expenses) and insurance premium.\footnote{Throughout the paper, the words ``Cost" or ``Cost Control" pertain to the ``out-of-pocket" costs incurred by the consumer for telehealth in the form of ``deductibles, co-pays \& coinsurance". ``Price" or ``Price Control" pertains to the posted price of the service, the reimbursement received by the physician. The ``full monetary price" paid by the consumers includes total out-of-pocket costs and insurance premium, both for telehealth and in-person. These are different from the input costs incurred by the provider or the claim and administrative costs incurred by the insurer.} There is a tripartite relationship, involving a third-party insurer, instead of the conventional biparty ``provider-consumer" relationship. Thus, Price and Cost become differentiated.\footnote{Such a disentanglement due to a ``wedge" or mismatch between price quoted and price actually paid, occurs in sectors or markets other than healthcare. For instance, government programs such as food stamps or Supplemental Nutrition Assistance Program (SNAP), section 8 housing vouchers, energy assistance programs, Affordable Connectivity Program, etc., can help reduce the effective monetary costs incurred by consumers. Other notable examples are discounts, cashback programs, drug pricing for the insured such as Medicare Part D, and government schemes such as Minimum Support Price (MSP) for farmers in India, among others.}

Telehealth Parity Laws (henceforth TPL) introduced a unique quasi-experimental variation in the regulatory environment and have distinct regulatory specifications for the already differentiated ``Price" and ``Cost" in healthcare. These specifications differed for each state owing to the distinct ``framing" of regulation by each state. With the onset of the global pandemic and social distancing norms, healthcare rapidly shifted towards telehealth. The surge in telehealth usage led to the swift adoption of parity laws across various states.\footnote{Telehealth visits increased by 154\% from March 2019 to March 2020 \citep{koonin2020}. Telehealth usage grew by 60\% from 2012 to 2013, with 40-50\% rise in institutional adoption by 2016. Usage varied by demographics, socioeconomic status, and geography \citep{lucas2022}.} This shift was facilitated by the already increasing use of the internet for health-related purposes. Adopted in a staggered manner by various states, TPL sought to establish parity between \enquote{reimbursements} received by physicians for telehealth services with those for in-person services, and the \enquote{out-of-pocket costs} (deductibles, co-pays, and coinsurance) incurred by consumers for telehealth services with those for in-person services. Distinct framing of laws resulted in varied combinations of a type of Price Control—such as Price Ceiling, Price Floor, or Price Parity—with a type of Cost Control—such as Cost Ceiling or Cost Parity—across states. For instance, a state could specify just a Price Ceiling (by specifying that ``physician reimbursement" for telehealth ``cannot exceed" that for in-person), while another state might specify a combination of Price Floor (by specifying that ``physician reimbursement" for telehealth be ``at least as much as" that for in-person) with Cost Ceiling (by specifying that ``out-of-pocket costs" for telehealth cannot exceed that for in-person). The Parity Laws, thus, introduce quasi-experimental state-level variation in the regulatory environment owing to distinct framing of the laws by each state, which can be leveraged to address confounding factors and draw causal inferences. This setting provides a unique opportunity to study Cost Controls along with Price Controls, while highlighting the non-price factors at play. Since broadband internet is crucial for telehealth, its presence as a technological mediator affects the input supply elasticities and opportunity cost of time, substantially altering the regulation policy outcomes. This setting provides a rare opportunity to study regulation-technology (``reg-tech'') interactions\textemdash{}an understudied but increasingly significant area of economic inquiry as the proliferation of artificial intelligence and other transformative technologies intensifies the need to understand how technology and regulation jointly shape equilibrium outcomes across markets.

This setting merits investigation because telehealth is a rapidly expanding sector addressing access barriers in underserved areas and during crises, with the global market valued at USD 196.81 billion in 2025 and projected to reach USD 1,211.14 billion by 2034.\footnote{See: \url{https://www.precedenceresearch.com/telehealth-market}} In the U.S., it is estimated at USD 41.54 billion in 2025, growing to USD 160.45 billion by 2034.\footnote{See: \url{https://www.novaoneadvisor.com/report/us-telemedicine-market}} Despite its scale, pre-COVID effects of TPL remain understudied, as do interactions with technology and geography.

I estimate Average Treatment Effects on the Treated ($ATT$), at specific levels of broadband, for each treatment type consisting of types of Price Control and Cost Control, for metro and non-metro areas separately. First, the heterogeneous treatment effects on the data-driven ``Composite Healthcare Service Provision Index" (CHSPI) provide a composite measure of shifts in overall service provision for each policy type, which I interpret as a proxy for the equilibrium quantity shifts predicted by the theory. Second, the heterogeneous effects on outpatient visits show the demand side effects for each policy type. Third, the heterogeneous effects on physician counts suggest spatial physician market restructuring, at the extensive margin.\footnote{These are partial equilibrium causal estimates since the analysis is restricted to physician market \citep{MWG1995}.} The comparison between effects on service quantity and physician supply shows whether there is efficiency gain or loss in service provision. In addition, I estimate the Average Causal Response on Treated ($ACRT$) which shows how the $ATT$ for each policy type varies with the spatio-temporal variation in broadband post-treatment. Thus, the $ACRT$ shows how the conducive or unfavorable effect of a policy type varies with broadband.

The results show that a binding Price Floor has a conducive impact on service quantity (CHSPI) and physician counts that is amplified with broadband, and a non-binding Price Ceiling, unfavorable at mean broadband, converges to the same conducive impact as broadband rises, nationwide and for metro areas. The binding Cost Parity has an unfavorable impact on service quantity (CHSPI) and physician counts that carries a negative broadband gradient, nationwide and for metro areas. However, for non-metro areas, Price Floor and Price Ceiling show positive effect on the service quantity (CHSPI) but negative effect on physician counts. This suggests increased workload on existing physicians or efficiency gain, since the effect of Price Floor and Price Ceiling on outpatient visits is negligible. The impact of combination of Cost Parity with a Price Control is unfavorable, since Cost Parity's unfavorable impact dominates. At lower levels of broadband, these causal responses become diluted. Furthermore, the estimates for binding regulations such as Price Floor or Cost Parity, are more pronounced for physicians with specialties that involve heavy telehealth use and interaction with patients (e.g., Radiology). However, for specialty which uses telehealth lightly, and for that which uses telehealth heavily but doesn't involve patient interaction, the causal response estimates for binding regulations are insignificant. 

I build upon the supply chain approach \citep{Mulligan2024}, where production runs through an upstream and a downstream input, a vertical structure long established in economics \citep{Spengler1950, Tirole1988}, to study the role of non-price or quality competition in this regulated market, where the Price Controls distort the input mix away from the cost-minimizing level, which causes production inefficiency and rotate the supply curve. I also build upon the theory of non-monetary factors of demand to show how the Cost Controls change the consumption mix away from the utility-maximizing level, which causes consumption inefficiency and rotate the demand curve. The rotations in the supply and demand curves determine the ``regulated" equilibrium healthcare service quantity. For each type of regulation, the respective difference between post-regulation and pre-regulation equilibrium quantities denotes the equilibrium quantity shift.

The reasons for studying the impact of parity regulations on physician market are convincing. First, TPL specify regulatory controls on ``physician reimbursement" (Price Controls), in addition to consumer out-of-pocket costs (Cost Controls). Second, causal estimates demonstrate changes in physician-provided healthcare service quantity and physician counts. Third, there is a high correlation between physician density and service availability. The changes in physician counts capture a substantial part of changes in equilibrium healthcare service quantities.\footnote{While the idea of physician location being influenced by regulatory conduciveness or friction and demand shifts might seem extreme, it is a realistic outcome, especially for marginal physicians---those for whom the net benefit of staying in a particular area is minimal. Similar to how negative shocks in the labor market disproportionately affect low-wage workers, leading to separations \citep{CP1985}, regulatory friction can push marginal physicians to exit the market or relocate to areas with regulatory conduciveness. This process is neither immediate nor widespread, as they involve search and matching costs. However, empirical data substantiates the prevalence of physician turnover and relocation \citep{turnover2023}. Between 2010 and 2018, the annual physician turnover rate rose from 5.3\% to 7.6\%. Physicians in rural areas, encompassing non-metro regions, demonstrate higher rates of movement (5.1\% vs 3.9\% in urban areas) and practice exit (3.3\% vs 2.7\%). Additionally, professional services such as \textit{MD Match}, which specialize in facilitating physician relocation, highlight the practicality and accessibility of these transitions. Given that these constitute common occurrences, regulatory environment created by TPL quite plausibly account for a considerable variation in such transitions.} Lastly, any restructuring in the physician market can significantly impact the overall economy and healthcare service delivery.\footnote{Physician services constitute about 3.6\% of the U.S. GDP \citep{GHT2015}. Financial incentives for physicians can increase the provision of healthcare. For example, \cite{CG2014} show that areas with higher payment shocks experience increases in overall health care provision. While \cite{EFM2017} study the effect of financial incentives to providers, I study the effect of financial incentives and disincentives, for both providers and consumers.}

The metro and non-metro areas differ in terms of development, technological infrastructure, opportunity cost of time, and baseline demand and supply, which would lead to distinct changes in physician and consumer behavior under regulation. The physician responses to Price Controls depend on the type of Price Control and on broadband, which determines the supply elasticity of telehealth input. For example, a higher physician reimbursement rate for telehealth (owing to Price Floor) may encourage telehealth investment, while a restrictive Cost Control (such as Cost Parity) can deter telehealth utilization. Physicians in non-metro areas or newly entering physicians can relocate to metro areas to take advantage of the financial surplus created by a Price Floor, where competition would further drive up investment in telehealth and raise overall supply and quality of healthcare services. Alternately, physicians in non-metro areas can substitute more towards in-person services provided there is enough demand for in-person services. Thus, Cost Controls assume significance. The changes in out-of-pocket costs play a crucial role in metro areas where opportunity cost of time is higher owing to higher wages, and in non-metro areas where the dis-utility of accessing healthcare in-person due to distance, exacerbated by physician shortage, is higher.

\textbf{\textit{Related literature and contribution:}} This paper makes several novel contributions. To the best of my knowledge, this is the first paper to study both theoretically and empirically, the impact of ``Price Controls" along with consumer ``Cost Controls". This is also the first study to discuss the implications of state-level TPL framing. Thus, the paper fills an empirical gap in the literature on price regulations in healthcare. Additionally, this is the first study to model the interaction of technology, in this case broadband, with Price and Cost Controls. Thus, this paper fills theoretical and conceptual gaps since it addresses the discrepancy between conventional models of Price Controls and empirical results. Studies examining these effects before the COVID-19 pandemic period are sparse. This paper addresses this temporal gap by examining the pre-COVID-19 implications of TPL. By analyzing the effects of TPL, separately for metro and non-metro areas, the paper fills the gap related to lack of spatial analysis in the economics of regulation.

TPL introduce unique scenarios where the Market Equilibrium Reimbursement for in-person ($MERR-I$) can function as either a Price Ceiling or a Price Floor, while the Market Equilibrium Cost Rate for in-person ($MECR-I$) could function as a Cost Ceiling or Cost Parity, based on each state's specification. Prior to TPL, the physician reimbursement for telehealth ($MERR-T$) was generally below that for in-person ($MERR-I$), making the Price Ceiling non-binding in a conventional sense.\footnote{In the pre-pandemic era in places without TPL, telehealth reimbursement rates were established beneath $MERR-I$, with averages for telehealth consultations ranging between \$40 and \$50, contrasting with up to \$176 for in-person visits \citep{Yamamoto2014, mahar2018}.} This undervaluation of telehealth led to advocacy for TPL. Thus, when $MERR-I$ acts as the Price Ceiling on $MERR-T$, it is above $MERR-T$. The effect of such Price Ceiling can be similar to or opposite to that of Price Floor depending on whether the Price Control is binding or not \citep{Sweeney1977, Lee1980}.\footnote{In \cite{Lee1980}'s critique of \cite{Sweeney1977}, a binding price ceiling can increase current supply by lowering the cost of meeting current demand, potentially driving the market clearing price below the ceiling. However, such concerns do not arise in our case, since $MERR-I$ as Price Ceiling is above $MERR-T$, due to the framing of parity laws.} This is an important source of divergence of the results of this study from \cite{Mulligan2024}'s prediction that Price Ceilings and Price Floors inherently have effects opposite to each other. I show that Price Floor and the non-binding Price Ceiling have similar effects, which are amplified at higher levels of broadband. In addition, the supply chain framework of \cite{Mulligan2024} focuses on rotations of the supply curve under Price Controls as the sole determinant of the regulated equilibrium quantity along a static demand curve. This paper models the rotations in both the supply and demand curves. \textit{Figure \ref{fig:comparison}} in the Appendix shows this comparison.

The insurer-defined reimbursement rates or Physician Pay Schedules ($PPS$) and consumer monetary expenses, add layers of indirect Price Control. The theoretical part incorporates the influence of these factors with quality adjustments happening with distortion of input mix in response to regulations.\footnote{This approach is different from the framework in \cite{GHT2015}, where the quality levels are “given” in the first stage with unrestricted consumer choice in the absence of market constraints such as choice frictions \citep{HKMS2024}, restricted provider networks and plan offerings in less populated areas \citep{Freed2021}, regional monopolies \citep{Fulton2017}, and employer-sponsored insurance limitations \citep{KBL2023}.} The role of third-party insurers matters since the Price and Cost Controls can directly impact insurer's cost sharing structures and claim costs. The economic surplus held by insurers could be redistributed to providers through pay parity mandates. The main object of interest are the upper and lower limits of $PPS$. If $MERR-I$ is the Price Floor (or Ceiling), $PPS_{max}$ (or $PPS_{min}$) acts as the Price Ceiling (or Floor). Thus, the paper abstracts away from the mechanisms leading to settling of $PPS$, such as insurer networks, competition and bargaining.\footnote{The TPL would compel the upper and lower limit of $PPS$ to be same across the states due to need for uniformity in provider reimbursement across states \citep{CG2014} or bargaining by the hospitals/physicians for parity with other states for both public and private insurance. Thus, physician reimbursement might rise up to $MERR-I$ under a Price Ceiling if there is a Price Floor in another state. If $MERR-I$ itself varies, so would physician reimbursement with it to comply with regulation.} Additionally, the studies on insurance markets often overlook the out-of-pocket costs in the form of ``deductibles, copay, co-insurance". On the contrary, in this study, the Cost Controls directly regulate the out-of-pocket expenses.\footnote{Out-of-pocket costs are vital for understanding consumer behavior in insurance markets since they influence perceptions, utilization \citep{Wong2023, Cavalier2023}, choices \citep{CZ2000, Manning1987} and welfare \citep{HKMS2024, Davis2014, KFF2023, BJ2024}, help visualize the demand curve via the full price.}

Broadband rollout has been linked to economic benefits \citep{CANZIAN201987, CHEN2023102504, hsls2018} and health benefits \citep{vanparys2023, tomer2020}. However, the number of internet providers or broadband infrastructure have been used as proxies for broadband access or penetration. This study refines this metric using a novel county-level residential connection data. Telehealth, which depends on broadband, is crucial in reducing disease exposure during health crises, benefits patients with mobility issues or chronic conditions in Health Professional Shortage Areas (HPSAs), and can achieve outcomes similar to in-person care \citep{shaver2022}. Broadband increases telehealth supply elasticity, modifies opportunity costs of in-person access, and facilitates health-related information access, leading to increased provision and utilization of telehealth \citep{Amaral-Garcia2022, Okoye2021, Pandit2025}. These changes are not uniform as metropolitan areas often benefit more. Broadband is less accessible in rural areas and Indian reservations, adversely affecting telehealth deployment. Despite increased broadband penetration enhancing healthcare delivery, the digital divide remains.\footnote{\textit{Figure \ref{fig:bbdpen}} in the Appendix shows a noticeable increase in the county level broadband penetration from 2010 to 2019 in the U.S. \textit{Figure \ref{fig:bivariate}} in the Appendix reveals the compounded challenges faced by non-metro areas.} This study marks a departure from classical models of regulation as I account for the interaction of broadband with regulations. The shifts in equilibrium quantities and restructuring of physician market is indeed spatial reallocation of economic activity under regulation and technology constraints.

Contrary to previous policy studies that treat TPL as a uniform treatment \citep{restrepo2018, cornaggia2023}, I take into account the impact of diversity in the framing of these laws across states. The paper enriches the current discourse on healthcare consumption, physician response and the accessibility and financial implications of TPL \citep{bavafa2018, jama2021, phillips2023}. This study foregrounds the preeminence of Fee-for-Service (FFS) in the U.S. telehealth payment framework through which the financial implications of telehealth can be evaluated.\footnote{In FFS, the provider is reimbursed for each service, and the quantity of service determines the provider revenue. Value-Based Payment (VBP) rewards the value of care provided. However, the typical VBP compensated services are surgeries, which are not feasible to be carried out via telehealth. FFS is the main reimbursement model and channel for telehealth pay-parity policies.} In addition, this study also conducts a specialty-wise comparison, while explicitly accounting for licensure environment, further enriching the discourse.\footnote{\textit{Section \ref{sec:modalities} and \ref{sec:imlcdata}} in the Appendix provide additional details on telehealth modalities with results for specialty-wise estimates and licensure, respectively.}

The rest of the paper is organized as follows: Section \ref{sec:2} provides the background, Section \ref{sec:ef} outlines the empirical framework, Section \ref{sec:results} summarizes the results, Section \ref{sec:tf} lays out the theoretical framework, Section \ref{sec:rc} reviews the robustness checks, and Section \ref{sec:conclusion} concludes.

\vspace{1em}
\section{Institutional Background}\label{sec:2}

\subsection{The Framing of Telehealth Parity Laws Generated Various Types of Price and Cost Controls} 
\begin{figure}[h!]
    \centering
        \caption{State-wise Telehealth Parity Laws: Framing and Staggered Adoption}
    \label{fig:framingadoption}
    \begin{minipage}[t]{0.48\textwidth}
        \centering
        \includegraphics[width=\textwidth,height=0.23\textheight]{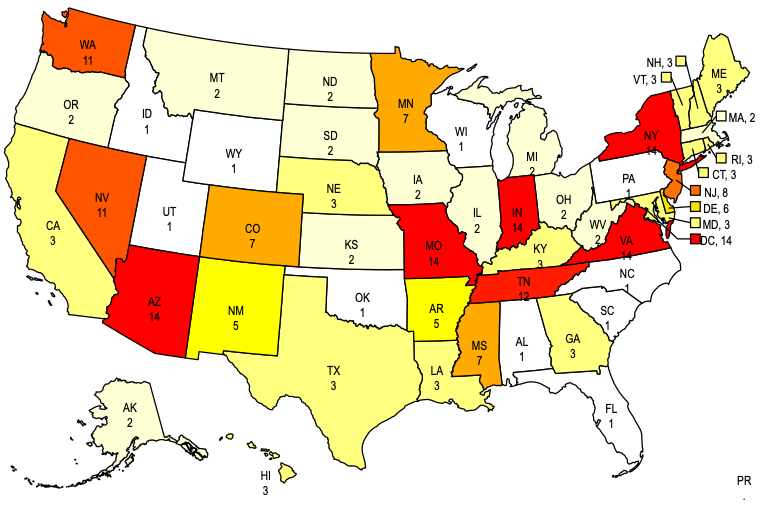}
        \vspace{0.1cm} 
        \includegraphics[width=0.6\textwidth,height=0.25\textheight]{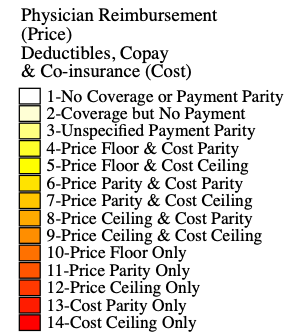}
        \subcaption[]{}
    \end{minipage}
    \hspace{0.01\textwidth} 
    \begin{minipage}[t]{0.48\textwidth}
        \centering
        \includegraphics[width=\textwidth,height=0.23\textheight]{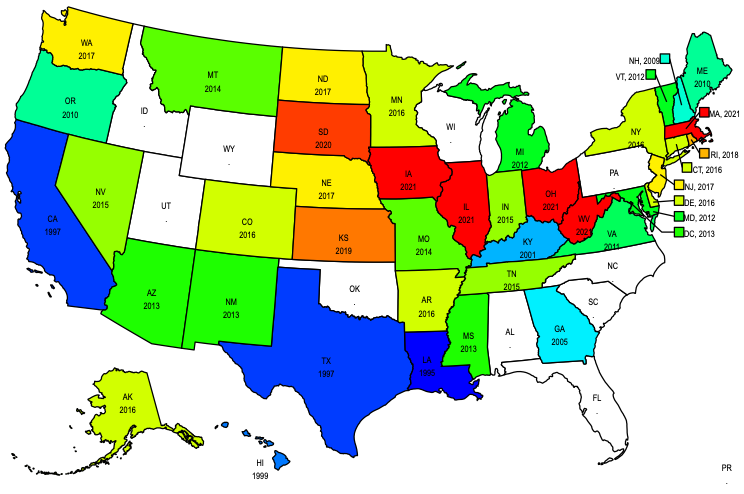}
        \vspace{0.1cm} 
        \includegraphics[width=0.4\textwidth,height=0.23\textheight]{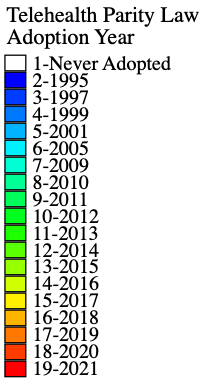}
        \vspace{1em}
        \subcaption[]{}
    \end{minipage}
    \noindent
    \begin{minipage}{\textwidth}
        \begin{spacing}{1.5}
        \footnotesize
        \justifying
        \textit{Note}: Panel (a) shows the state level framing for the states who adopted telehealth parity laws in the United States till 2021. Panel (b) shows the staggered adoption of Telehealth Parity Laws in the United States up to 2021. 
        \end{spacing}
    \end{minipage}
\end{figure}

Since 1995, 40 states and the District of Columbia have implemented telehealth Coverage Parity mandates for insurance plans (\textit{Panel (b), Figure \ref{fig:framingadoption}}). This means that if a service is covered when it is provided in-person, it is also covered when provided through telehealth. Many of the states who introduced Coverage Parity also introduced Payment Parity mandates, which ensure that telehealth has the same level of reimbursement for physicians and the same out-of-pocket costs for the consumer, as those for equivalent in-person services. These stipulations are not uniform across all states, largely due to differences in framing and communication of these laws (\textit{Panel (a), Figure \ref{fig:framingadoption}}). Neglecting the framing could lead to missing differential impacts. Framing effects cannot be overlooked even for the most experienced physicians \citep{tversky1986}. While behavioral economics emphasizes perceived value in decision-making in the face of non-statutory framing, in the case of TPL, the statutory framing could directly impact not only perceptions but also the actual current or expected financial incentives and cost structures faced by providers and patients. \textit{Table \ref{tab:framingtimeline}} in the Appendix lists each state's adoption year and framing category.

Distinct legal stipulations for Payment Parity within state laws can manifest as a type of Price Control (Price Floor, Price Ceiling, or Price Parity), or a type of Cost Control (Cost Ceiling or Cost Parity), or a combination of both, or neither.\footnote{This study predominantly centers on the regulatory effects of Price Ceiling and Price Floor. Hence, Price Parity becomes a secondary aspect rather than a direct subject of inquiry.} A state may stipulate that physician reimbursement for telehealth \enquote{may not exceed} that for in-person services, establishing a \enquote{Price Ceiling}; that it be \enquote{at least as much as} in-person services, establishing a \enquote{Price Floor}; or that it be the  \enquote{same rate} as in-person services, establishing exact  \enquote{Price Parity}. A state may stipulate that  \enquote{deductibles, co-pays, and coinsurance} for telehealth  \enquote{may not exceed} those for in-person services, establishing a \enquote{Cost Ceiling}, or be the  \enquote{same rate} as for in-person services, establishing  \enquote{Cost Parity}.\footnote{Once telehealth is covered under a plan, the premium is the same for the plan and doesn't differ by whether healthcare was accessed in-person or through telehealth. However, our measure of full price includes premium although TPL focus on out-of-pocket costs. This is because the Cost Controls can affect cost sharing, and hence premiums too, thus affecting the full price.} A state might not specify the details or framing. In that case, it would be deemed to just have a Payment Parity law. Each of the types within Price or Cost Control is mutually exclusive, meaning that a state can have only one: a Price Ceiling, Price Floor, or Price Parity. Similarly, a state might specify either a Cost Ceiling or Cost Parity, but not both. \textit{Figure \ref{fig:pcillustration}} illustrates the mechanism through which Price Controls are created by TPL. 

\subsection{Regulatory Compliance and Real Financial Effects}

Three independent lines of evidence establish that the statutory framings translated into actual changes in physician reimbursements and consumer cost-sharing. First, policy trackers confirm the exact language was enacted and maintained. The Center for Connected Health Policy reports that 23 states plus the District of Columbia retain explicit payment-parity mandates using the precise wording analyzed here (``at least as much as'' or ``same rate'' for floors and ``may not exceed'' for ceilings), with many states extending these requirements through 2026 or making them permanent.\footnote{Center for Connected Health Policy (CCHP), \textit{State Telehealth Laws and Reimbursement Policies Report, Fall 2025}, October 2025, \url{https://www.cchpca.org/resources/state-telehealth-laws-and-reimbursement-policies-report-fall-2025/}.} An independent Mercatus Center analysis confirms the economic interpretation, finding that ``not less than'' provisions function as price floors that raise rates above market levels, while ``not more than'' provisions function as non-binding ceilings \citep{dills2021}. Second, utilization responses demonstrate changed financial incentives. Using Truven MarketScan data from 2010--2015, \citet{harvey2019} find that parity states experienced 29.8 percent higher odds of a telehealth visit ($p < 0.0001$). More recent evidence from commercial claims reinforces this finding. States adopting payment-parity language saw significant increases in telehealth utilization with no offsetting decline in in-person volume \citep{zhangbundorf2025}. Cost-sharing provisions operated analogously, constraining out-of-pocket amounts. Third, enforcement is routine. State insurance departments conduct market-conduct examinations with authority to impose fines and corrective action plans against carriers that fail to comply with payment parity mandates. Although self-funded plans are generally exempt from state insurance mandates under ERISA preemption, \citet{zhangbundorf2025} find the parity effect was more pronounced among workers in such plans, indicating that parity norms extended beyond the legally mandated population. Taken together, the statutory language, behavioral responses, and enforcement record establish genuine variation in financial incentives during the 2010--2019 sample period.

\vspace{2em}
\section{Empirical Framework}\label{sec:ef}
\subsection{\textbf{Model Specification and Data Generating Process}}
The data-generating process (DGP) is specified as follows:
\vspace{-1.5ex}
\begin{align}
    Y_{j(i)t} = \exp[\alpha_0 + \alpha_1 C_{it} + \alpha_2 A_{j(i)t} - \alpha_3 U_{it} - \alpha_4 F_{j(i)t}] \varepsilon_{j(i)t} \hspace{.2cm}
\end{align}

Here, $\alpha_0$ represents the natural log of a constant. In this panel data framework, $i$ denotes the state, $j(i)$ the county within state $i$, and $t$ the year. The outcome variable $Y_{j(i)t}$ is either a composite index of healthcare service provision or aggregate or specialty-wise count of physicians in county $j(i)$ at time $t$, with the error term $\varepsilon_{j(i)t}$ having an expectation of 1. $\alpha_1$ signifies the degree of conducive nature of the state's policy environment, with positive sign indicating how favorable a policy type is to physician practice or location. For instance, a state specifying a Price Floor could be conducive to practice or location. $\alpha_2$ captures the degree of attractiveness of a county as signified by its characteristics, $A_{j(i)t}$, which includes population, standardized broadband, median household income, per capita risk-adjusted Medicare expenditure, the number of hospital admissions per population and the post Interstate Medical Licensure (IMLC) membership indicator.\footnote{The state licensure requirements and Interstate Medical Licensure Compact (IMLC) influence the physical location, remote practice capabilities of physicians and telehealth accessibility. Licensure requirements directly impact the number of physicians entering the profession, while the IMLC aims to make it easier for existing physicians to practice in multiple states.} In contrast, $\alpha_3$ measures the impact of state-level unfavorable policies $U_{it}$, such as Cost Parity, which would increase consumer cost, reduce demand and discourage practice or location decision. $\alpha_4$ measures the impact of county-level frictional factors $F_{j(i)t}$, such as the percentage of poverty or the percentage of people aged 65 and older without health insurance, which have a discouraging effect on physician practice or location. Unobserved time-invariant factors are captured in the error term $\varepsilon_{j(i)t}$.

\subsection{\textbf{Composite Healthcare Service Provision Index (CHSPI)}}
The theoretical framework predicts partial equilibrium quantity shifts in healthcare service provision following parity law adoption. Physician counts reflect extensive-margin supply responses but miss intensive-margin adjustments such as increased workload per provider, while outpatient visits capture demand-side utilization but not supply-side capacity. The CHSPI is constructed to serve as a direct empirical counterpart to this prediction, measuring the comprehensive level of realized healthcare service provision. Provision arises where physician capacity and demand culminate in patient contact. Physicians without patient contact do not constitute provision, and patient visits cannot occur without physicians. The index therefore combines capacity and utilization components, and the first principal component extracts their common factor, which is the object the index measures. The index is constructed through principal component analysis on standardized per-capita measures of healthcare infrastructure and service utilization.\footnote{The first principal component is extracted and rescaled to a 0--100 range for interpretability. Full loadings for all retained components are reported in Table~\ref{tab:pca_loadings}.} Auxiliary regressions using the disaggregated components as outcome variables then identify whether estimated shifts are driven by supply expansion or utilization gains, separating extensive from intensive margin responses. The index aggregates eight arcsinh-transformed, standardized per-capita measures: hospital admissions, hospitals, physician density, outpatient visits, inpatient days, emergency department visits per 1,000 beneficiaries, fee-for-service beneficiaries, and aged-and-disabled enrollment. CHSPI is the first principal component rescaled to 0--100, a dimensionless composite index of realized healthcare service provision rather than a count of services. The first component loads positively on the six provision measures and negatively on the two beneficiary-composition measures, so movements in CHSPI capture realized provision relative to enrollment composition.\footnote{PPML requires a nonnegative outcome with a correctly specified exponential conditional mean rather than count data \citep[e.g.,][]{silvatenreyro2006}. CHSPI is nonnegative by construction, and the RESET test (\textit{Table \ref{tab:reset}}) supports the exponential form.} Treatment effects on CHSPI therefore read as percentage changes in realized service provision. An ATT of 0.03 is a 3 percent increase in the index relative to the never-treated counterfactual.

\subsection{\textbf{Estimation and Identification}}

Physician count and outpatient visits, both from the Area Health Resource File (\textit{Section \ref{sec:ahrfdata}, Appendix}), are count variables, which frequently include zeros.\footnote{Estimating a log-linearized specification via ordinary least squares (OLS) can introduce biases, particularly when the underlying model is nonlinear in parameters, and may yield inconsistent estimates owing to heteroskedasticity. Logarithm of zero is undefined. Even if the counts are strictly greater than zero or transformed as $ln(1 + Y_{j(i)t})$, the expected value of the log-linearized error depends on the covariates, which makes OLS biased. Further, multiplicative models estimated using the non-linear least squares (NLS) method can be inefficient as it ignores heteroskedasticity.} The best way to estimate the parameters is through the Poisson Quasi Maximum Likelihood (QMLE) or Pseudo Maximum Likelihood Estimator (PMLE), which can very well account for zero values of $Y_{j(i)t}$ \citep{GourierouxMonfortTrognon1984, silvatenreyro2006}. The triple interaction DiD model in a generalized multiplicative form is specified as:

\vspace{-1em}
\begin{equation}
\begin{split}
Y_{j(i)t} = \exp\Biggl(&\underbrace{\lambda_{j(i)}}_{\text{county FE}}+\underbrace{\gamma_t}_{\text{time FE}}+\underbrace{\sum_{k=1}^{K} \beta_{1k} M_{ik} Post_{ct} B_{j(i)t}}_{\text{triple interaction (treated-post-broadband)}}+\underbrace{\sum_{k=1}^{K} \beta_{2k} M_{ik} Post_{ct}}_{\text{treat-post interaction}} \\
    &+ \underbrace{\sum_{k=1}^{K} \beta_{3k} M_{ik}B_{j(i)t}}_{\text{treated-broadband interaction}}+\underbrace{\beta_{4} Post_{ct}B_{j(i)t}}_{\text{post-broadband interaction}}+\underbrace{\beta_5' X_{j(i)t}}_{\text{controls}} 
\Biggr) \varepsilon_{j(i)t}
\end{split}
\label{eq:ppmltriple}
\end{equation}

$K$ is the total number of regulation policy types.\footnote{The state-level main effects \(M_{ik}\) are constant within groups and get subsumed into the county fixed effects \(\lambda_{j(i)}\), while the main \(Post_{ct}\) effects for each treatment type are subsumed into the time fixed effects \(\gamma_t\). The standalone broadband term is included in the estimation and removed by the estimator as an exact linear combination of the included terms. With never-adopting states assigned cohort zero, \(Post_{ct} = 1\) for these states in every year, which yields the identity \(B_{j(i)t} = Post_{ct} B_{j(i)t} + M_{i1} B_{j(i)t} - M_{i1} Post_{ct} B_{j(i)t}\), where \(k = 1\) denotes Payment Parity, the main treatment indicating TPL adoption. All three terms on the right appear in \textit{equation \ref{eq:ppmltriple}}, so the standalone term is an exact linear combination of included regressors, its role is carried by those terms, and no other coefficient is affected by its removal. The broadband term inside \(A_{j(i)t}\) of the data generating process therefore enters \textit{equation \ref{eq:ppmltriple}} through this combination. The decision to logarithmize a covariate or not was based on whether it was approximately log-normally distributed after taking natural logs or not \citep{BEYER2022}.} \textquote{$k$} is the type of treatment as determined by the framing of the TPL. \(X_{j(i)t}\) is composed of control variables.

If the sign of a coefficient is positive (negative), the respective policy environment is conducive (unfavorable) to service provision and physician practice. The coefficients capture the effect on service quantity and physician counts post-policy adoption, capturing the relative conduciveness of the policy-technology environment compared to areas without such a policy. A positive coefficient indicates a higher relative conduciveness score, implying that the respective policy-technology combination makes such areas more appealing relative to others that don't have the said policy environment. A positive (negative) coefficient implies an increase (decrease) in the equilibrium quantity of physician services in the adopting states, as predicted by the theoretical framework. This reallocation is a manifestation of the heterogeneous policy impacts of TPL, as it demonstrates how certain areas become relatively more conducive or unfavorable to physician practice or location based on their policy-technology environments as compared to those that don't have such policy-technology environment. If the sign of a coefficient in the vector $\beta_5'$ is positive (negative), the county-level control variable is an attractive (frictional) component. There are six different indicators, one indicating whether the state adopted the TPL or not, and five for the types of framing.

\subsubsection{\textbf{Partial Equilibrium Effects and Staggered Adoption}}
The policy-relevant question is what happens, on average, to physician supply and service provision when states adopt these regulations. We seek before-and-after comparisons that identify equilibrium shifts, not dynamic treatment effects that trace out adjustment paths over time. The identification strategy is difference-in-differences, estimated with PPML and county and year fixed effects.
Recent econometric literature has raised concerns about staggered difference-in-differences designs, arguing that heterogeneous treatment effects across cohorts can produce biased estimates when using two-way fixed effects estimators \citep{goodman2021, SNA2021, callaway2021}. However, as \cite{wooldridge2021} demonstrates, the problem lies not with the estimator itself but with restrictive model specifications. When the model correctly specifies treatment heterogeneity, the estimator consistently estimates the ATT. Several features of this setting support the use of PPML with county and year fixed effects for estimating partial equilibrium effects. First, the treatment is irreversible, i.e., once a state adopts parity laws, it remains treated. Second, county fixed effects absorb all time-invariant confounders, while year fixed effects control for secular trends affecting all counties. Third, the specification explicitly models treatment heterogeneity through the regulation type indicators $M_{ik}$, addressing the concern that motivates newer estimators.\footnote{The estimators proposed by \cite{callaway2021} and \cite{SNA2021} address treatment timing heterogeneity by producing cohort-time specific ATTs, $ATT(g,t)$, for adoption cohort $g$ at calendar time $t$. These are designed for settings where researchers want to trace dynamic effects over event time. However, they do not address regulatory design heterogeneity, which is the central feature of this study. All treated states receive the same treatment, a telehealth parity law. What differs is the statutory framing through which that law is implemented. No existing staggered difference-in-differences framework estimates separate ATTs by regulation type. Attempting to create cohorts by regulation type would produce cells with unreliable aggregation weights, and with multiple regulation types, post-treatment periods, and adoption cohorts, such an approach would yield dozens of coefficients, obscuring rather than clarifying the policy-relevant estimates. Furthermore, both \cite{callaway2021} and \cite{SNA2021} are developed for linear specifications and are not suited for count outcomes or nonlinear settings.}
\subsubsection{\textbf{Exogeneity of Treatment Adoption}}
Identification requires that parity law adoption is exogenous to county-level trends in service provision, conditional on covariates and fixed effects. This assumption is assessed using the event study framework of \cite{SNA2021} for the main treatment indicator, whether the state adopted a telehealth parity law or not, with no evidence of differential pre-trends in physician supply or service provision (\textit{Table~\ref{tab:pretrends}, Figure~\ref{fig:sunabraham}, Appendix}). Although \cite{SNA2021} is designed for linear specifications, it provides a reliable approximation for validating parallel trends in nonlinear settings. Pre-trends are also tested directly using PPML (\textit{Table~\ref{tab:pretrends}, Column (b)}). A placebo test assigning a fictitious treatment period confirms that pre-treatment coefficients are jointly indistinguishable from zero (\textit{Table~\ref{tab:placebo}, Appendix}). Pre-treatment covariate balance between treated and control counties is documented in \textit{Table~\ref{tab:sumstats}}. Covariates are statistically indistinguishable across groups at baseline with two exceptions, risk-adjusted per capita Medicare costs ($p = 0.038$) and the non-metro share ($p = 0.100$). Both are controlled throughout, the former as a time-varying covariate in every specification and the latter through the metro and non-metro split samples, and county fixed effects absorb baseline differences in levels.

In this setting, the relevant exogeneity condition applies to the adoption decision. The statutory framing, whether as Price Floor, Price Ceiling, Cost Ceiling, or Cost Parity, is a second-stage policy design choice reflecting how each state's legislature drafted the specific payment and cost-sharing provisions. Framing was not a response to anticipated trends in physician location or service provision. The appropriate strategy, which this paper follows, is to validate exogeneity for the adoption decision and then decompose the ATT by regulation type in a fully saturated specification. The framing itself is plausibly exogenous conditional on adoption.\footnote{\textit{Section \ref{sec:framingexo}} in the Appendix documents the origin of the framing language in national model legislation and standard drafting formulations and reports framing-specific pre-trend estimates.}

After ensuring satisfaction of\hspace{.2cm}\enquote{Conditional No Anticipation} and \enquote{Conditional Ratio (Proportional) Parallel Trends} assumptions, the latter implied by the exponential conditional mean function, $ATT_{k}$, where $k$ is the treatment type, is identified. We get the counterfactual percentage change in the mean outcome for the treated group (denoted by \enquote{\(M_{i1} = 1\)}) using the observed percentage change for the never treated group (denoted by \enquote{\(M_{i0} = 0\)}).\footnote{The identifying assumption is a conditional ratio (proportional) version of parallel trends implied by the exponential conditional mean:\newline
\(\dfrac{E[Y_{j(i)t}(0)\mid X_{it}, M_{1i} = 1, Post_{ct} = 1]}{E[Y_{j(i)t}(0)\mid X_{it}, M_{1i} = 1, Post_{ct} = 0]} =  \dfrac{E[Y_{j(i)t}(0)\mid X_{it}, M_{1i} = 0, Post_{ct} = 1]}{E[Y_{j(i)t}(0)\mid X_{it}, M_{1i} = 0, Post_{ct} = 0]}\), where \(X_{it}\) is the covariate set described above. This is parallel trends in ratios, equivalently in \(\ln E[Y(0)\mid \cdot]\), not in levels.} In this nonlinear estimation method, including unit-specific dummies does not lead to the incidental parameters problem. After incorporating covariates, the Poisson PMLE, given in \cite{CHENROTH2024} and formalized in \cite{WOOLDRIDGE2023}, gives a consistent estimate of $\theta_{ATT(k)}\%$, if the ratio version of parallel trends holds and $\varepsilon_{j(i)t}$ has a mean of 1, conditional on covariates, and $Y_{j(i)t}$ takes the exponential functional form.\footnote{This proportional treatment effect for a treatment type $k$ is given by $exp(\beta_{2k})-1$ and formulated as:\\ \(\theta_{ATT(k)}\% = \dfrac{E[Y_{j(i)t}(1) \mid M_{ki} = 1, Post_{ct} = 1] - E[Y_{j(i)t}(0) \mid M_{ki} = 1, Post_{ct} = 1]}{E[Y_{j(i)t}(0) \mid M_{ki} = 1, Post_{ct} = 1]}\).}  

The data sample spans from 2010 to 2019. States treated in or before 2011 were categorized as \textquote{always treated}, removed to avoid skewing the results, and treated as missing for those years, while states treated in 2018 or later were labeled \textquote{never treated.} A span of two years before the first treatment helps establish the baseline conditions, while a two-year period after the last treatment underscores the persistence or decay of the treatment effects over time. Consequently, the first treated cohort comprises states treated in 2012, and the last treated cohort includes states treated in 2017. The estimation is performed using the procedure in \cite{CGZ2020}. This ensures that our estimates reflect the policy-driven reallocation of physician services while accounting for underlying structural differences across regions.

\subsection{\textbf{Identification of the Average Causal Response on the Treated (ACRT)}}

The broadband variable $B$ is the standardized county volume of residential broadband connections, constructed from FCC Form 477 data and standardized on the full county-year panel (\textit{Section \ref{sec:bbddata}, Appendix}). One unit of $B$ is one standard deviation of that panel.

The $ATT_k$ at a specific broadband level ($B$) is given by $ATT_k(B) = \exp(\beta_{2k} + \beta_{1k} B) - 1$. The $ATT_k$ from the two-way interaction model do not give any idea about broadband's role in modifying the treatment effect itself, since these are $ATT_k$ at $B=0$ (mean broadband level), given by $\exp(\beta_{2k}) - 1$. The $ACRT$ reported in \textit{Tables~\ref{tab:withbbd}} and \textit{\ref{tab:main_acrt}} is identified from the coefficient on the triple interaction of the post-treatment indicator, the treatment type, and standardized broadband. This coefficient measures how the treatment effect changes with a one-standard-deviation increase in standardized broadband.\footnote{The increase in broadband, the effects of which are captured by $ACRT$, is spatio-temporal. It reflects both across counties and time variation in broadband levels within the treated group during the post-treatment period.} The $ACRT$ adapts the average causal response of the continuous-treatment difference-in-differences literature \citep{CGS2024}, where the continuous variable is the dose of the treatment itself, to this setting, where the treatment is the adoption of the law and broadband is a measured modifier of its effect.

Identification of the $ACRT$ requires that broadband variation is orthogonal to unobserved determinants of service provision, conditional on covariates and fixed effects. This is supported on both institutional and empirical grounds. Broadband deployment is governed by FCC spectrum allocation, universal service fund disbursements, and ISP capital-expenditure schedules tied to multi-year infrastructure plans. Parity-law adoption is determined by state legislative processes, interest group advocacy, and legislative calendars. These operate through distinct institutional channels with no direct feedback loop. County and year fixed effects absorb all time-invariant county characteristics and common annual shocks, while the full set of time-varying controls (income, poverty, unemployment, risk-adjusted Medicare costs, hospital admissions) accounts for the primary economic channels through which local development could jointly drive broadband expansion and parity-law adoption.\footnote{This approach is consistent with the broader literature treating broadband penetration as exogenous to local economic outcomes conditional on fixed effects and time-varying controls \citep{Amaral-Garcia2022, CANZIAN201987, hsls2018}. Results are also robust to alternative broadband specifications (\textit{Table~\ref{tab:lognormalandarsinh}}).} In the main specification's covariate set, the variance inflation factor for broadband is 1.68, with a mean of 2.76 across covariates, so broadband is not collinear with these controls. Household take-up on an existing network moves with these controlled factors, and the county fixed effects absorb each county's persistent adoption propensity. The remaining threat, changes correlated with broadband growth that coincide with the policy timing, is testable. Treatment effects attenuate below mean broadband (\textit{Table \ref{tab:below3SDandmean}}), are absent in specialties without telehealth scope (\textit{Table \ref{tab:lightusers}}), and the framing-specific pre-period coefficients are small and statistically indistinguishable from zero for the framings that carry the results (\textit{Section \ref{sec:framingexo}, Appendix}).

\vspace{1em}
\begin{table}[h!]  
\begin{small}
\caption{Summary Statistics}
\label{tab:sumstats}
\centering
\setlength{\tabcolsep}{4pt} 
\begin{spacing}{1.5}
\begin{tabular}{@{}p{4.5cm}*{3}{r}@{}} 
\cline{1-4}
\multicolumn{1}{r}{} &
  \multicolumn{1}{c}{Never Treated} &
  \multicolumn{1}{c}{Treated} &
  \multicolumn{1}{c}{Test} \\
\multicolumn{1}{r}{} &
  \multicolumn{1}{c}{$(49.4\%)$} &
  \multicolumn{1}{c}{$(50.6\%)$} &
  \multicolumn{1}{r}{} \\
\cline{1-4}
\multicolumn{1}{l}{NonFederal \& Federal MDs} &
  \multicolumn{1}{r}{$293.09 \, (1146.31)$} &
  \multicolumn{1}{r}{$312.85 \, (1187.04)$} &
  \multicolumn{1}{r}{} \\
\multicolumn{1}{l}{Radiology} &
  \multicolumn{1}{r}{$10.63 \, (41.60)$} &
  \multicolumn{1}{r}{$11.20 \, (44.91)$} &
  \multicolumn{1}{r}{} \\
\multicolumn{1}{l}{Psychiatry} &
  \multicolumn{1}{r}{$10.30 \, (47.57)$} &
  \multicolumn{1}{r}{$12.83 \, (70.03)$} &
  \multicolumn{1}{r}{} \\
\multicolumn{1}{l}{Emergency Medicine} &
  \multicolumn{1}{r}{$9.73 \, (36.71)$} &
  \multicolumn{1}{r}{$9.96 \, (34.68)$} &
  \multicolumn{1}{r}{} \\
\multicolumn{1}{l}{Cardiovascular Dis} &
  \multicolumn{1}{r}{$7.14 \, (29.30)$} &
  \multicolumn{1}{r}{$7.24 \, (28.96)$} &
  \multicolumn{1}{r}{} \\
\multicolumn{1}{l}{Gastroenterology} &
  \multicolumn{1}{r}{$3.95 \, (15.52)$} &
  \multicolumn{1}{r}{$4.17 \, (17.39)$} &
  \multicolumn{1}{r}{} \\
\multicolumn{1}{l}{Standardized Broadband} &
  \multicolumn{1}{r}{$-0.06 \, (0.67)$} &
  \multicolumn{1}{r}{$-0.05 \, (0.74)$} &
  \multicolumn{1}{r}{$0.925$} \\
\multicolumn{1}{l}{\% Persons in Poverty} &
  \multicolumn{1}{r}{$15.92 \, (5.63)$} &
  \multicolumn{1}{r}{$16.20 \, (6.17)$} &
  \multicolumn{1}{r}{$0.263$} \\
\multicolumn{1}{l}{Unemployment Rate, 16+} &
  \multicolumn{1}{r}{$8.99 \, (3.00)$} &
  \multicolumn{1}{r}{$9.04 \, (3.29)$} &
  \multicolumn{1}{r}{$0.691$} \\
\multicolumn{1}{l}{Area} &
  \multicolumn{1}{r}{} &
  \multicolumn{1}{r}{} &
  \multicolumn{1}{r}{} \\
\multicolumn{1}{l}{\hspace{1em}Non metro} &
  \multicolumn{1}{r}{$712 \, (63.3\%)$} &
  \multicolumn{1}{r}{$767 \, (66.6\%)$} &
  \multicolumn{1}{r}{$0.100$} \\
\multicolumn{1}{l}{\hspace{1em}Metro} &
  \multicolumn{1}{r}{$413 \, (36.7\%)$} &
  \multicolumn{1}{r}{$385 \, (33.4\%)$} &
  \multicolumn{1}{r}{} \\
\multicolumn{1}{l}{Age $ \geq 65$ w/o Health Insurance} &
  \multicolumn{1}{r}{$16710.52 \, (43661.04)$} &
  \multicolumn{1}{r}{$16512.61 \, (43627.61)$} &
  \multicolumn{1}{r}{0.914} \\
\multicolumn{1}{l}{Std Risk Adj. Per Capita Medicare Costs} &
  \multicolumn{1}{r}{$9571.65 \, (1171.83)$} &
  \multicolumn{1}{r}{$9466.81 \, (1235.12)$} &
  \multicolumn{1}{r}{$0.038$} \\
\multicolumn{1}{l}{Median Household Income} &
  \multicolumn{1}{r}{$43217.63 \, (8722.34)$} &
  \multicolumn{1}{r}{$43458.90 \, (11410.14)$} &
  \multicolumn{1}{r}{$0.572$} \\
\multicolumn{1}{l}{Hospital Admissions Per Population} &
  \multicolumn{1}{r}{$12128.73 \, (40188.68)$} &
  \multicolumn{1}{r}{$11602.18 \, (36143.94)$} &
  \multicolumn{1}{r}{$0.742$} \\
\multicolumn{1}{l}{Population} &
  \multicolumn{1}{r}{$93598.31 \, (2.4 \times 10^{5})$} &
  \multicolumn{1}{r}{$92489.77 \, (2.4 \times 10^{5})$} &
  \multicolumn{1}{r}{$0.914$} \\
\multicolumn{1}{l}{MD's, Non-Federal, Total} &
  \multicolumn{1}{r}{$286.20 \, (1127.05)$} &
  \multicolumn{1}{r}{$304.41 \, (1158.28)$} &
  \multicolumn{1}{r}{} \\
\cline{1-4}
\end{tabular}
\footnotesize
\begin{tabularx}{\textwidth}{@{}X@{}}
\vspace{0.5em}
\textit{Note:} The table reports means and standard deviations for continuous variables and counts and percentages for factor variables for year 2010, before the treatments rolled out in a staggered manner. ``NF" denotes ``Non-Federal". Total sample: N = 22,838. 
 The p-values are produced using \textit{test(pearson)} for metro and using \textit{test(regress)} across levels of \textit{Treated} for continuous variables. The higher p-values indicate comparability of the two groups at the baseline. 
\end{tabularx}
\end{spacing}
\end{small}
\end{table}

Two independent pieces of evidence confirm that the $ACRT$ reflects the telehealth mechanism discussed in the theory part. First, effects are attenuated precisely where telehealth cannot operate. \textit{Table~\ref{tab:below3SDandmean}} shows that treatment effects on physician supply are statistically insignificant when standardized broadband is below its mean, and even more so when counties more than 3~SD above the mean are excluded. If unobserved economic development were the driver, treatment effects would persist at low broadband levels where telehealth is not operationally feasible. The sharp dose-response pattern is consistent only with telehealth feasibility.

Second, the $ACRT$ is concentrated among specialties for which telehealth is feasible. \textit{Table~\ref{tab:heavyusers}} shows large and significant $ACRT$ for radiologists, psychiatrists, and emergency physicians, the three specialties with the highest pre-pandemic telehealth usage rates, while \textit{Table~\ref{tab:lightusers}} shows the Price Floor $ACRT$ is statistically indistinguishable from zero for cardiologists and gastroenterologists. Any economic confounder correlated with broadband growth would affect physician location across all specialties. Only the telehealth mechanism produces this selective pattern. These tests establish that the $ACRT$ identifies genuine regulation--technology interactions rather than spurious correlations with broadband rollout.


\section{Results}\label{sec:results}
\vspace{-1em}
\subsection{Treatment Effect Estimates}
\textit{Table \ref{tab:w/obbd}} \hspace{.1cm} provides $ATT$ estimates without broadband interaction, with dependent variables, CHSPI, which captures total realized service provision (Columns 1 to 3 without considering framing and columns 4 to 6 considering framing), and physician count, which isolates the extensive margin of provider location and suggests spatial market restructuring (Columns 7 to 9 without considering framing and columns 10 to 12 considering framing). When the framing effects are taken into account, estimates for CHSPI are mostly similar to those for physician counts.

When the framing effects are taken into account, the results indicate that Price Floor has a significant conducive effect for metro areas for both CHSPI and physician counts, and for the full sample physician count. The standard errors suggest more precise estimates for the metro subsample compared to the non-metro subsample. Price Ceiling has a significant unfavorable effect across most specifications, with the exception of the non-metro CHSPI estimate which is insignificant. Thus, at aggregate broadband levels, Price Floor has an opposite (positive) effect to that of Price Ceiling (negative). This aligns with the predictions of the supply chain framework. However, subsequent results show that these predictions don't hold at higher broadband levels. Additionally, Cost Parity shows significant positive coefficients at mean broadband for CHSPI and the full sample physician count (columns 4, 6, and 10). However, this reverses at higher broadband, where Cost Parity's unfavorable effect emerges. The negative $ACRT$ in \textit{Table \ref{tab:main_acrt}} and the profile in \textit{Figure \ref{fig:attprofiles}} carry this gradient over the framing's treated support. The estimates for Cost Ceiling are negligible and imprecise.

When the estimated effects of parity laws on CHSPI and physician counts exhibit the same sign or similar magnitude, it indicates that policy-induced changes in the number of providers are the dominant driver of overall service provision shifts, with limited countervailing adjustments on the intensive margin. For instance, positive effects on both metrics suggest that parity provisions, such as Price Floor, attract more physicians (extensive expansion), directly amplifying aggregate quantity without substantial reliance on productivity enhancements per provider. Conversely, when estimates differ in sign or magnitude, e.g., negative effects on physician counts alongside positive effects on CHSPI, this points to intensive margin offsets, where fewer providers achieve higher service output through increased workload or efficiency gains (e.g., telehealth adoption enabling more encounters per physician). This divergence highlights parity laws' role in fostering technology-driven productivity, particularly in underserved areas, allowing equilibrium quantity to rise despite supply-side constraints like physician shortages.\footnote{The Interstate Compact shows a positive effect across most specifications, with relatively precise estimates indicated by smaller standard errors. This shows that the states belonging to the compact saw increased service provisions and physician presence.}

Thus, the findings corroborate the fact that the effects of TPL on service quantity and physician count diverge not only by region, but also by how the laws are framed. However, these are $ATT_k$ aggregated at mean broadband level. The effects of Price and Cost Controls, which relate to physician reimbursement and consumer out-of-pocket costs for telehealth respectively, would be more pronounced at higher levels of broadband because broadband is indispensable for telehealth. The progression in $ATT$ with broadband gives a clearer picture about the impact of these policy types.

\setstretch{1.5}
\newgeometry{left=2cm,right=0.5cm,top=0cm,bottom=0cm} 
\begin{landscape}
\null\vfill 
\centering
\begin{small}
\def\sym#1{\ifmmode^{#1}\else\(^{#1}\)\fi}
\captionof{table}{ATT estimates}
\label{tab:w/obbd}
\setlength{\tabcolsep}{2pt}
\begin{tabular}{@{}l*{12}{c}@{}}
\hline\hline
& \multicolumn{3}{c}{CHSPI Without Considering Framing} & \multicolumn{3}{c}{CHSPI Considering Framing} & \multicolumn{3}{c}{Physician Count Without Considering Framing} & \multicolumn{3}{c}{Physician Count Considering Framing} \\
\cmidrule(lr){2-4}\cmidrule(lr){5-7}\cmidrule(lr){8-10}\cmidrule(lr){11-13}
& \multicolumn{1}{c}{(1)} & \multicolumn{1}{c}{(2)} & \multicolumn{1}{c}{(3)} & \multicolumn{1}{c}{(4)} & \multicolumn{1}{c}{(5)} & \multicolumn{1}{c}{(6)} & \multicolumn{1}{c}{(7)} & \multicolumn{1}{c}{(8)} & \multicolumn{1}{c}{(9)} & \multicolumn{1}{c}{(10)} & \multicolumn{1}{c}{(11)} & \multicolumn{1}{c}{(12)} \\
\cmidrule(lr){2-4}\cmidrule(lr){5-7}\cmidrule(lr){8-10}\cmidrule(lr){11-13}
& Full sample & Non-metro & Metro & Full sample & Non-metro & Metro & Full sample & Non-metro & Metro & Full sample & Non-metro & Metro \\
\hline
Post Price Floor & \_\_\_ & \_\_\_ & \_\_\_ & $0.0035$ & $-0.0045$ & $0.0198\sym{***}$ & \_\_\_ & \_\_\_ & \_\_\_ & $0.0270\sym{***}$ & $-0.0179$ & $0.0385\sym{***}$ \\
                    & & & & $(0.0021)$ & $(0.0061)$ & $(0.0053)$ & & & & $(0.0061)$ & $(0.0147)$ & $(0.0079)$ \\
[1em]
Post Price Ceiling & \_\_\_ & \_\_\_ & \_\_\_ & $-0.0105\sym{*}$ & $-0.0072$ & $-0.0200\sym{***}$ & \_\_\_ & \_\_\_ & \_\_\_ & $-0.0169\sym{***}$ & $-0.0511\sym{***}$ & $-0.0133\sym{***}$ \\
                    & & & & $(0.0060)$ & $(0.0070)$ & $(0.0063)$ & & & & $(0.0044)$ & $(0.0073)$ & $(0.0044)$ \\
[1em]
Post Cost Parity & \_\_\_ & \_\_\_ & \_\_\_ & $0.0352\sym{***}$ & $\_\_\_$ & $0.0370\sym{***}$ & \_\_\_ & \_\_\_ & \_\_\_ & $0.0106\sym{***}$ & $\_\_\_$ & $0.0035$ \\
                    & & & & $(0.0082)$ & & $(0.0077)$ & & & & $(0.0026)$ & & $(0.0025)$ \\
[1em]
Post Cost Ceiling & \_\_\_ & \_\_\_ & \_\_\_ & $0.0070$ & $0.0102$ & $-0.0010$ & \_\_\_ & \_\_\_ & \_\_\_ & $-0.0015$ & $-0.0060$ & $-0.0017$ \\
                    & & & & $(0.0056)$ & $(0.0083)$ & $(0.0053)$ & & & & $(0.0040)$ & $(0.0090)$ & $(0.0041)$ \\
\midrule
Post Payment Parity & $0.0038$ & $0.0102\sym{**}$ & $-0.0043$ & $-0.0027$ & $0.0014$ & $-0.0031$ & $-0.0003$ & $0.0097\sym{*}$ & $-0.0014$ & $-0.0003$ & $0.0188\sym{**}$ & $-0.0027$ \\
                    & $(0.0036)$ & $(0.0048)$ & $(0.0040)$ & $(0.0060)$ & $(0.0076)$ & $(0.0046)$ & $(0.0031)$ & $(0.0057)$ & $(0.0034)$ & $(0.0040)$ & $(0.0084)$ & $(0.0043)$ \\
\midrule
Broadband & $0.0288\sym{***}$ & $0.0523$ & $0.0153\sym{***}$ & $0.0271\sym{***}$ & $0.0482$ & $0.0146\sym{***}$ & $0.0047\sym{***}$ & $-0.0073$ & $0.0030\sym{*}$ & $0.0043\sym{***}$ & $-0.0080$ & $0.0024$ \\
                    & $(0.0058)$ & $(0.0761)$ & $(0.0041)$ & $(0.0055)$ & $(0.0752)$ & $(0.0037)$ & $(0.0017)$ & $(0.0848)$ & $(0.0017)$ & $(0.0016)$ & $(0.0831)$ & $(0.0016)$ \\
[1em]
Post Interstate Compact & $0.0073$ & $0.0085$ & $0.0120\sym{***}$ & $0.0054$ & $0.0049$ & $0.0132\sym{***}$ & $0.0089\sym{***}$ & $0.0212\sym{***}$ & $0.0101\sym{***}$ & $0.0076\sym{**}$ & $0.0195\sym{**}$ & $0.0076\sym{**}$ \\
                    & $(0.0049)$ & $(0.0067)$ & $(0.0042)$ & $(0.0047)$ & $(0.0061)$ & $(0.0044)$ & $(0.0034)$ & $(0.0075)$ & $(0.0035)$ & $(0.0034)$ & $(0.0077)$ & $(0.0031)$ \\
[1em]
\hline
Time Fixed Effects & Yes & Yes & Yes & Yes & Yes & Yes & Yes & Yes & Yes & Yes & Yes & Yes \\
County Fixed Effects & Yes & Yes & Yes & Yes & Yes & Yes & Yes & Yes & Yes & Yes & Yes & Yes \\
\hline
Observations & 20502 & 13312 & 7190 & 20502 & 13312 & 7190 & 22332 & 14354 & 7978 & 22332 & 14354 & 7978 \\
\hline\hline
\end{tabular}
\bigskip
\footnotesize
\begin{tabularx}{1.15\textwidth}{@{}X@{}}
\footnotesize \sym{*} \(p<0.10\), \sym{**} \(p<0.05\), \sym{***} \(p<0.01\)\\
\textit{Note \Rom{1}:} The table shows difference-in-difference PPML estimates. The dependent variables are CHSPI (columns 1-6) and physician counts (columns 7-12). The standard errors are in parenthesis and clustered at the state level. The first three columns for each dependent variable show $ATT$ without accounting for the framing of TPL, while the last three columns do account for the framing ($ATT_k$, where k is the treatment type). The controls used are (not shown): median household income, standardized risk adjusted per capita medicare costs, percentage poverty, unemployment rate for population aged 16 or more, population with age 65 or more without health insurance and broadband. For physician counts, additional controls used are---log transformed population and total hospital admissions. ``$\_\_\_$" indicates there are no non-metro observations for Cost Parity. The estimation sample spans 41 state clusters, 23 ever treated and 18 never treated. Treated state clusters by framing are 2 for Price Floor, 2 for Price Ceiling, 2 for Cost Parity, and 10 for Cost Ceiling. \\
\textit{Note \Rom{2}:} ``Post Payment Parity'' which represents the non-specific treatment or TPL adoption indicator is shown here, and has been included in all subsequent specifications but not shown. The purpose of showing it here is to demonstrate how incorporating the framing affects $ATT$ for main treatment. Additionally, ``Post Price Parity'' is included in all specifications but not shown, since our main interests are Price Floor and Price Ceiling. All subsequent regressions follow the same specification. \\
\end{tabularx}
\end{small}
\vfill 
\end{landscape}
\restoregeometry


\newgeometry{left=1cm,right=1cm,top=0.5cm,bottom=0.5cm} 
\begin{landscape}
\null\vfill 
\centering
\begin{small}
\def\sym#1{\ifmmode^{#1}\else\(^{#1}\)\fi}
\captionof{table}{Causal Response Estimates} \label{tab:withbbd}
\setlength{\tabcolsep}{5pt} 
\begin{tabular}{@{}l*{8}{c}@{}}
\toprule\toprule
& \multicolumn{3}{c}{Service Provision (CHSPI)} & \multicolumn{5}{c}{Physician Counts and Utilization} \\
\cmidrule(lr){2-4}\cmidrule(lr){5-9}
                    &\multicolumn{1}{c}{(1)} &\multicolumn{1}{c}{(2)} &\multicolumn{1}{c}{(3)} &\multicolumn{1}{c}{(4)} &\multicolumn{1}{c}{(5)} &\multicolumn{1}{c}{(6)} &\multicolumn{1}{c}{(7)} &\multicolumn{1}{c}{(8)} \\
                    &\multicolumn{1}{c}{\begin{tabular}[t]{@{}c@{}}CHSPI\\Full Sample\end{tabular}} &\multicolumn{1}{c}{\begin{tabular}[t]{@{}c@{}}CHSPI\\Non-metro\end{tabular}} &\multicolumn{1}{c}{\begin{tabular}[t]{@{}c@{}}CHSPI\\Metro\end{tabular}} &\multicolumn{1}{c}{\begin{tabular}[t]{@{}c@{}}Physician Count\\Full Sample\end{tabular}} &\multicolumn{1}{c}{\begin{tabular}[t]{@{}c@{}}Physician Count\\Non-metro\end{tabular}} &\multicolumn{1}{c}{\begin{tabular}[t]{@{}c@{}}Physician Count\\Metro\end{tabular}} &\multicolumn{1}{c}{\begin{tabular}[t]{@{}c@{}}Outpatient Visits\\Full Sample\end{tabular}} &\multicolumn{1}{c}{\begin{tabular}[t]{@{}c@{}}Radiologists\\Full Sample\end{tabular}} \\
\midrule
Post Price Floor $\times$ Broadband& $0.0277\sym{***}$& $0.3870\sym{***}$& $0.0167\sym{*}$& $0.0318\sym{***}$& $-0.3077\sym{*}$ & $0.0171\sym{***}$& $-0.0057$ & $0.0973\sym{***}$\\
                    & $(0.0096)$ & $(0.0746)$ & $(0.0091)$ & $(0.0013)$ & $(0.1857)$ & $(0.0020)$ & $(0.0071)$ & $(0.0033)$ \\[1em]
Post Price Ceiling $\times$ Broadband& $0.0249\sym{***}$& $0.2296\sym{**}$ & $0.0267\sym{***}$& $0.0350\sym{***}$& $-0.3189\sym{***}$& $0.0321\sym{***}$& $0.0158$ & $0.0372\sym{***}$\\
                    & $(0.0038)$ & $(0.0979)$ & $(0.0028)$ & $(0.0038)$ & $(0.0813)$ & $(0.0037)$ & $(0.0276)$ & $(0.0094)$ \\[1em]
Post Cost Parity $\times$ Broadband& $-0.0106\sym{***}$& --- & $-0.0138\sym{***}$& $-0.0319\sym{***}$& --- & $-0.0280\sym{***}$& $-0.1407\sym{***}$& $-0.0346\sym{***}$\\
                    & $(0.0039)$ & & $(0.0046)$ & $(0.0047)$ & & $(0.0047)$ & $(0.0377)$ & $(0.0052)$ \\[1em]
Post Cost Ceiling $\times$ Broadband& $-0.0040$ & $-0.1062$ & $-0.0033$ & $0.0007$ & $0.2221\sym{**}$ & $0.0011$ & $-0.0260$ & $0.0158\sym{*}$ \\
                    & $(0.0037)$ & $(0.1119)$ & $(0.0022)$ & $(0.0030)$ & $(0.0867)$ & $(0.0026)$ & $(0.0196)$ & $(0.0087)$ \\
                    [1em]
\midrule
Post Interstate Compact&      $0.0055$         &      $0.0047$         &      $0.0131\sym{***}$&      $0.0069\sym{**}$ &      $0.0200\sym{***}$&      $0.0070\sym{**}$ &      $0.0344$         &      $0.0177\sym{**}$ \\
                    &    $(0.0047)$         &    $(0.0060)$        &    $(0.0044)$        &    $(0.0034)$         &    $(0.0074)$        &    $(0.0031)$        &    $(0.0224)$         &    $(0.0084)$        \\
\midrule
Observations & $20502$ & $13312$ & $7190$ & $22332$ & $14354$ & $7978$ & $19077$ & $12188$ \\
\midrule
Time Fixed Effects & Yes & Yes & Yes & Yes & Yes & Yes & Yes & Yes \\
County Fixed Effects & Yes & Yes & Yes & Yes & Yes & Yes & Yes & Yes \\
\bottomrule\bottomrule
\end{tabular}
\bigskip
\footnotesize
\begin{tabularx}{\textwidth}{@{}X@{}}
\sym{*} \(p<0.10\), \sym{**} \(p<0.05\), \sym{***} \(p<0.01\) \\
\smallskip
\textit{Note:} The table shows difference-in-difference PPML estimates. The dependent variables are CHSPI (columns 1 to 3), physician counts (columns 4 to 6), outpatient visits (column 7) and radiologists (column 8). The standard errors are clustered at the state level. The estimation sample spans 41 state clusters, 23 ever treated and 18 never treated. Treated state clusters by framing are 2 for Price Floor, 2 for Price Ceiling, 2 for Cost Parity, and 10 for Cost Ceiling.\\
\end{tabularx}
\end{small}
\vfill 
\end{landscape}
\restoregeometry

\setstretch{2}

\subsection{ATT at Broadband Levels and Causal Response Estimates}
\textit{Table \ref{tab:withbbd}} presents the raw coefficients on the interaction terms between the post-treatment period, treatment type, and standardized broadband, which reflect how $ATT_k$ varies as broadband increases from $B=0$ (the mean) to $B=1$ (one unit above the mean), with each coefficient approximating the percentage change in the expected outcome for treated units per unit increase in $B$.\footnote{The estimates corresponding to Post Price Floor for Radiologists, who are among the highest telehealth users, are relatively more pronounced. This is because as more intensive users of telehealth, Radiologists are set to gain more surplus from binding Price Floor and set to lose from binding Cost Parity. The estimates for Radiologists are further discussed while discussing the results for specialty-wise estimates, \textit{Table \ref{tab:heavyusers}} in the Appendix.}

\textit{Table \ref{tab:main_acrt}} presents $ATT$ at standardized broadband levels (B=0, 1, 2, 4), with each framing tabulated over its treated post-period support as described in the table note, and the Average Causal Response on the Treated ($ACRT$) for four policy types: Post Price Floor, Post Price Ceiling, Post Cost Parity, and Post Cost Ceiling. $ACRT$ quantifies the marginal change in the treatment effect per unit increase in a standardized broadband, thereby indicating the effect's sensitivity (slope) to variations in broadband. Notably, both Post Price Floor and Post Price Ceiling exhibit a similar pattern in $ATT$ progression as broadband increases, which contradicts conventional theoretical models and the supply chain framework, which would predict different impacts. The finite difference between these levels, \(ATT(B=1) - ATT(B=0)\), approximates $ACRT$, which shows how $ATT$ vary with a one-standard-deviation increase in broadband.\footnote{While not exactly equal in general, the standardization of \(B\) and small \(\beta_1\) in this context (e.g., 0.0277 vs.\ 0.0282 (0.0333-0.0051) for ``Post Price Floor'', with the interpretation in PPML context being 2.78\% vs.\ 2.82\%) make them effectively equal (compare with \textit{Column 1, Table \ref {tab:withbbd}}), rendering the finite difference a good approximation of \(ACRT\). An additional explanation can be found in \textit{Section \ref{sec:acrtclarification}} in the Appendix.}

For Post Price Floor, $ATT$ is small but significant at $B=0$ and substantially larger at $B=1$, with a positive $ACRT$ indicating a conducive effect with broadband increase. Similarly, Post Price Ceiling shows a significant negative $ATT$ at $B=0$, which diminishes and turns positive at higher broadband levels, also accompanied by a positive $ACRT$. In contrast, Post Cost Parity starts with a positive $ATT$ at $B=0$ that falls to zero at $B=1$, with a negative $ACRT$ indicating a diminishing effect as broadband increases. Since these policies pertain to telehealth reimbursement and consumer costs, their effects become more pronounced when telehealth provision and utilization is substantial, i.e., at higher levels of broadband.\footnote{The standardized broadband variable is right-skewed, reflecting substantial spatial inequality in broadband across US counties. The 90th, 95th, and 99th percentiles correspond to $B = 0.34$, $B = 1.11$, and $B = 3.52$ respectively. Each framing in \textit{Table~\ref{tab:main_acrt}} is tabulated over its treated post-period support, with the support counts reported in the table note and the full distribution of the variable displayed in \textit{Figure~\ref{fig:bbddistribution}}.} Post Cost Ceiling shows no significant $ATT$ across broadband levels and an insignificant $ACRT$, suggesting no substantial impact of broadband on this policy's effect.

\setstretch{2}

\vspace{1em}
\begin{table}[h!]
\begin{small}
\caption{Estimates of $ATT$ at broadband levels and $ACRT$ (Aggregate Sample)}
\setlength{\tabcolsep}{12pt}
\label{tab:main_acrt}
\begin{spacing}{1.5}
\begin{tabular}{llccccc}
\toprule
\textbf{Policy type} & \textbf{Metric} & \textbf{Coefficient} & \textbf{Std. Error} & \textbf{Z-value} & \textbf{p-value} & \textbf{95\% Conf. Interval} \\
\midrule
\textbf{Post Price Floor} & ATT(B=0) & $0.0051$ & $(0.0022)$ & $2.2920$ & $[ 0.022]$ & $[0.0007, 0.0094]$ \\
 & ATT(B=1) & $0.0333$ & $(0.0099)$ & $3.3729$ & $[ 0.001]$ & $[0.0139, 0.0526]$ \\
 & ATT(B=2) & $0.0623^{\dagger}$ & -- & -- & -- & -- \\
 & ATT(B=4) & -- & -- & -- & -- & -- \\
\midrule
 & ACRT & $0.0277$ & $(0.0096)$ & $2.8726$ & $[ 0.004]$ & $[0.0088, 0.0466]$ \\
\midrule
\textbf{Post Price Ceiling} & ATT(B=0) & $-0.0120$ & $(0.0054)$ & $-2.2338$ & $[ 0.025]$ & $[-0.0226, -0.0015]$ \\
 & ATT(B=1) & $0.0129$ & $(0.0045)$ & $2.8414$ & $[ 0.004]$ & $[0.0040, 0.0218]$ \\
 & ATT(B=2) & $0.0384$ & $(0.0065)$ & $5.8622$ & $[ 0.000]$ & $[0.0256, 0.0512]$ \\
 & ATT(B=4) & -- & -- & -- & -- & -- \\
\midrule
 & ACRT & $0.0249$ & $(0.0038)$ & $6.4929$ & $[ 0.000]$ & $[0.0174, 0.0324]$ \\
\midrule
\textbf{Post Cost Parity} & ATT(B=0) & $0.0132$ & $(0.0059)$ & $2.2242$ & $[ 0.026]$ & $[0.0016, 0.0249]$ \\
 & ATT(B=1) & $0.0026$ & $(0.0025)$ & $1.0328$ & $[ 0.302]$ & $[-0.0023, 0.0075]$ \\
 & ATT(B=2) & $-0.0079^{\dagger}$ & -- & -- & -- & -- \\
 & ATT(B=4) & -- & -- & -- & -- & -- \\
\midrule
 & ACRT & $-0.0106$ & $(0.0039)$ & $-2.7251$ & $[ 0.006]$ & $[-0.0181, -0.0030]$ \\
\midrule
\textbf{Post Cost Ceiling} & ATT(B=0) & $0.0062$ & $(0.0055)$ & $1.1224$ & $[ 0.262]$ & $[-0.0046, 0.0171]$ \\
 & ATT(B=1) & $0.0022$ & $(0.0046)$ & $0.4651$ & $[ 0.642]$ & $[-0.0069, 0.0113]$ \\
 & ATT(B=2) & $-0.0019$ & $(0.0063)$ & $-0.2989$ & $[ 0.765]$ & $[-0.0143, 0.0105]$ \\
 & ATT(B=4) & $-0.0099$ & $(0.0125)$ & $-0.7929$ & $[ 0.428]$ & $[-0.0345, 0.0146]$ \\
\midrule
 & ACRT & $-0.0040$ & $(0.0037)$ & $-1.0898$ & $[ 0.276]$ & $[-0.0113, 0.0032]$ \\
\bottomrule
\end{tabular}
\end{spacing}
\footnotesize
\begin{justify}
\textit{Note:} The dependent variable is CHSPI. The standard errors are clustered at the state level. The ATT estimates at specific broadband levels are computed as nonlinear combinations of the regression coefficients using the delta method to derive standard errors and confidence intervals. Each framing is tabulated over its treated post-period support. Cells whose support spans at least two state clusters carry delta-method standard errors and confidence intervals. Cells within treated support that rest on a single state cluster are marked with $\dagger$ and reported without inference, for completeness, with the remaining columns of those rows dashed. Rows beyond a framing's maximum treated broadband are fully dashed. Maximum treated post-period broadband is $B = 2.03$ for Price Floor, $3.32$ for Price Ceiling and Cost Parity, and $12.69$ for Cost Ceiling. Treated post-period county-years (state clusters) at $B \geq 1$ are 11 (2) for Price Floor, 50 (2) for Price Ceiling, 39 (2) for Cost Parity, and 149 (9) for Cost Ceiling; at $B \geq 2$, 3 (1), 21 (2), 14 (1), and 96 (8); at $B \geq 4$, Cost Ceiling alone with 29 (4).
\end{justify}
\end{small}
\end{table}

\begin{figure}[htbp!]
    \centering
    \caption{Service-Provision (CHSPI) Treatment-Effect Profiles over Each Framing's Treated Broadband Support}
    \vspace{0.4em}
    \label{fig:attprofiles}
    \includegraphics[width=.95\textwidth]{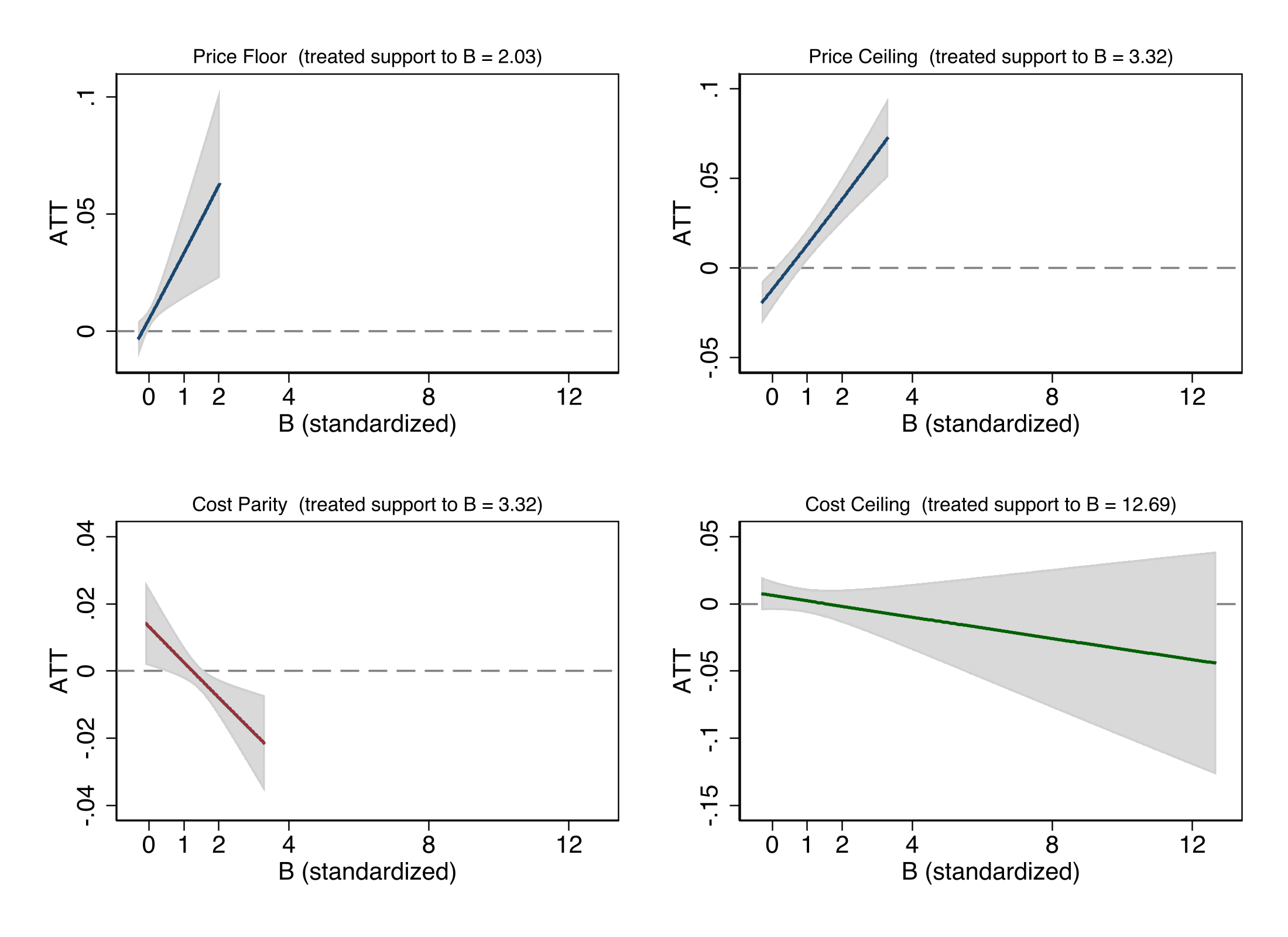}
    \vspace{0.6em}

    \noindent
    \begin{minipage}{\textwidth}
    \begin{spacing}{1.5}
    \footnotesize
    \justifying
    \textit{\textbf{Note}}: Model-implied treatment-effect profiles for the composite healthcare service provision index (CHSPI), $ATT_k(B) = \exp(\beta_{2k} + \beta_{1k} B) - 1$, from the specification of \textit{Table \ref{tab:main_acrt}}, with 95\% confidence bands computed by the delta method and standard errors clustered at the state level. Each framing's profile is drawn only over that framing's treated post-period support, whose end point is stated in each panel; the common horizontal axis makes the differences in treated support across framings visible. Treated support counts (county-years and state clusters) at $B \geq 1$, $B \geq 2$, and $B \geq 4$ are reported in the note to \textit{Table \ref{tab:main_acrt}}. The Cost Parity profile is confined to its support, with the slope carried by the $ACRT$ of $-0.0106$. The flat Cost Ceiling profile, statistically indistinguishable from zero at every broadband level, matches the model's prediction for a non-binding ceiling.
    \end{spacing}
    \end{minipage}
\end{figure}

\setstretch{2}

\vspace{1em}
\begin{table}[h!]
\begin{small}
\def\sym#1{\ifmmode^{#1}\else\(^{#1}\)\fi}
\caption{Causal Response Estimates for Lower Levels of Broadband}
\label{tab:below3SDandmean}
\setlength{\tabcolsep}{10pt}
\begin{spacing}{1.5}
\begin{tabular}{@{}l*{6}{c}@{}}
\hline\hline
& \multicolumn{3}{c}{Broadband Below 3SD Above Mean} & \multicolumn{3}{c}{Broadband Below Mean} \\
\cline{2-7}
& \multicolumn{1}{c}{(1)} & \multicolumn{1}{c}{(2)} & \multicolumn{1}{c}{(3)} & \multicolumn{1}{c}{(4)} & \multicolumn{1}{c}{(5)} & \multicolumn{1}{c}{(6)} \\
\cline{2-7}
&\multicolumn{1}{c}{Full sample} &\multicolumn{1}{c}{Non-metro} &\multicolumn{1}{c}{Metro} &\multicolumn{1}{c}{Full sample} &\multicolumn{1}{c}{Non-metro} &\multicolumn{1}{c}{Metro} \\
\hline
Post Price Floor $\times$ Broadband & $0.0106\sym{**}$ & $0.0163$ & $0.0048$ & $0.0166\sym{*}$ & $0.0140$ & $0.0128$ \\
& $(0.0047)$ & $(0.0113)$ & $(0.0041)$ & $(0.0094)$ & $(0.0111)$ & $(0.0127)$ \\
[1em]
Post Price Ceiling $\times$ Broadband & $0.0072\sym{*}$ & $0.0037$ & $0.0095\sym{**}$ & $0.0002$ & $0.0031$ & $0.0001$ \\
& $(0.0042)$ & $(0.0070)$ & $(0.0042)$ & $(0.0067)$ & $(0.0070)$ & $(0.0131)$ \\
[1em]
Post Cost Parity $\times$ Broadband & $0.0153\sym{***}$ & $\_\_\_$ & $0.0036\sym{*}$ & $-0.0022\sym{***}$ & $\_\_\_$ & $-0.0011\sym{**}$ \\
& $(0.0037)$ &  & $(0.0021)$ & $(0.0004)$ &  & $(0.0005)$ \\
[1em]
Post Cost Ceiling $\times$ Broadband & $-0.0046$ & $-0.0082$ & $-0.0048$ & $-0.0110$ & $-0.0060$ & $-0.0164$ \\
& $(0.0062)$ & $(0.0134)$ & $(0.0053)$ & $(0.0128)$ & $(0.0136)$ & $(0.0181)$ \\
\hline
Observations & 22004 & 14354 & 7650 & 18154 & 14245 & 3909 \\
\hline
Time Fixed Effects  & Yes & Yes & Yes & Yes & Yes & Yes \\
County Fixed Effects & Yes & Yes & Yes & Yes & Yes & Yes \\
\hline\hline
\end{tabular}
\bigskip
\footnotesize
\begin{tabularx}{\textwidth}{@{}X@{}}
\footnotesize \sym{*} \(p<0.10\), \sym{**} \(p<0.05\), \sym{***} \(p<0.01\) \\
\smallskip
\textit{Note:} The dependent variable is CHSPI. Standard errors in parentheses are clustered at the state level. The natural log of min-max normalized broadband is used because the standardized variable’s compressed range renders a 1-unit change impractical. The table reports the lower-broadband subsample, where broadband is insufficient to raise the telehealth supply elasticity and the effects vary little with broadband. \\
\end{tabularx}
\end{spacing}
\end{small}
\end{table}

\textit{Table~\ref{tab:below3SDandmean}} presents two progressively stricter sample restrictions to assess whether the main results are driven by the extreme right tail of the broadband distribution. The standardized broadband variable is right-skewed, so the below-3SD restriction serves as a robustness check that retains the vast majority of the sample while removing the outermost tail. The below-mean restriction further limits the sample to counties where broadband falls below the national average, representing the natural lower bound for telehealth feasibility.

When high broadband areas are excluded from the analysis, it essentially dilutes the conducive and unfavorable effects seen for the aggregate and metro sample which had high broadband counties. \textit{Table \ref{tab:below3SDandmean}} shows the results for counties at lower broadband levels. This shows how policy impacts on healthcare service quantity differ markedly in areas with lower broadband access, revealing their sensitivity to broadband levels, with the effects of policies getting diluted at lower broadband levels, with no economically meaningful effects at broadband levels below the mean.

In \textit{Table \ref{tab:withbbd}}, the results for non-metro areas are mostly opposite, while the aggregate impact is dominated by the effect in metro areas, despite the larger sample size of non-metro areas. Firstly, there is a scale effect owing to the higher baseline density of physicians in metro areas, so that large absolute changes result in smaller percentage changes. In contrast, in non-metro areas, due to the lower physician count and baseline demand, the small absolute changes result in large percentage changes. In non-metro areas, the demand for in-person services is likely to remain low or static, which may lead to status-quo response or provider relocation to different spatial market or exit, even when there is a Price Floor. Secondly, while both metro and non-metro areas have counties with broadband levels higher than mean, metro areas have more such counties, benefiting from existing infrastructure, stronger competition and better availability of resources. Consequently, metro areas can effectively leverage increased broadband to enhance telehealth offerings, so that policies such as Price Floor would guarantee a surplus to physicians. Finally, the implementation and support mechanisms for telehealth policies are often more robust in metro areas, which can benefit from targeted initiatives, training programs, and financial incentives that facilitate adoption. In non-metro areas, the absence or lesser extent of such support can hinder effective implementation. A leave-one-state-out jackknife across all 37 states with non-metro counties confirms that the non-metro estimates are not driven by any single state.\footnote{Full results are reported in \textit{Table~\ref{tab:jackknife}} in the Appendix.}

\begin{figure}[htbp!]
    \centering
    \caption{Physician-Count Treatment-Effect Profiles over Each Framing's Treated Broadband Support}
    \vspace{0.2em}
    \label{fig:attprofilesmd}
    \includegraphics[width=.95\textwidth]{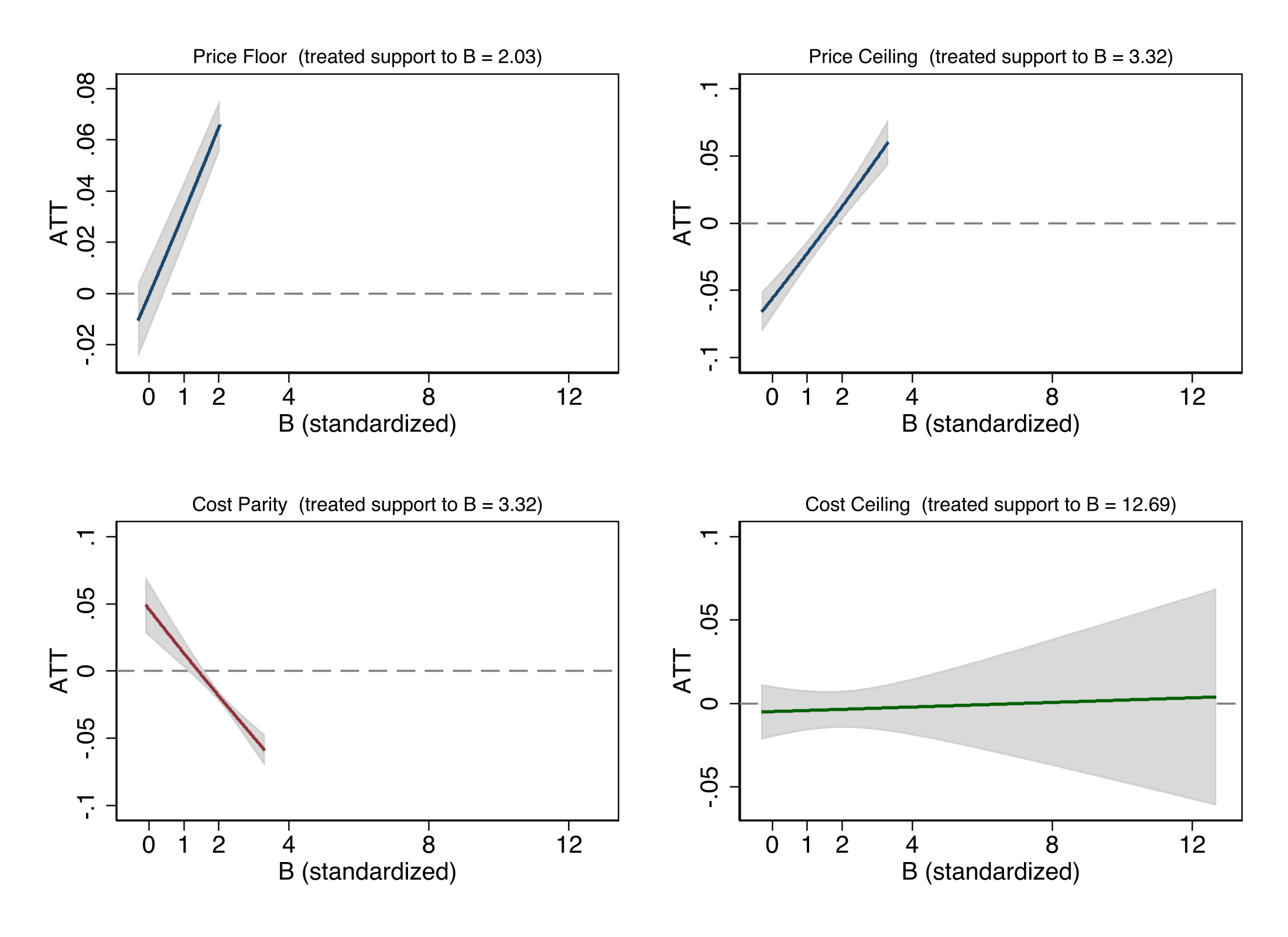}
    \vspace{0.6em}

    \noindent
    \begin{minipage}{\textwidth}
    \begin{spacing}{1.5}
    \footnotesize
    \justifying
    \textit{\textbf{Note}}: Model-implied treatment-effect profiles $ATT_k(B) = \exp(\beta_{2k} + \beta_{1k} B) - 1$ for the total number of physicians (MDs), from the full-sample physician-count specification underlying \textit{Tables \ref{tab:w/obbd}} (columns 7--12) and \textit{\ref{tab:withbbd}}, with 95\% confidence bands computed by the delta method and standard errors clustered at the state level. Each framing's profile is drawn only over that framing's treated post-period support, whose end point is stated in each panel, following the same rule as \textit{Table \ref{tab:main_acrt}} and \textit{Figures \ref{fig:attprofiles}} and \textit{\ref{fig:predictionplotscomb}}. Price Floor and Price Ceiling carry positive broadband gradients over their treated support. The Cost Parity profile carries a negative broadband gradient over its support. The flat Cost Ceiling profile is statistically indistinguishable from zero at every broadband level, consistent with the model's prediction for a non-binding ceiling.
    \end{spacing}
    \end{minipage}
\end{figure}

The results discussed focus on estimates of $ATT$ showing the heterogeneous effects of parity laws and $ACRT$ showing the spatiotemporal variation in $ATT$ with broadband. \textit{Figure \ref{fig:attprofiles}} traces these treatment-effect profiles over each framing's treated support, with Price Floor and Price Ceiling progressively conducive with broadband, Cost Parity carrying a negative broadband gradient, and Cost Ceiling statistically indistinguishable from zero at every level. The corresponding physician-count gradients appear in \textit{Table \ref{tab:withbbd}} and are traced in \textit{Figure \ref{fig:attprofilesmd}}. The predictions of physician counts from these regressions show which policy-technology combinations made certain areas more conducive or unfavorable to physicians after the parity laws were adopted.

\subsection{Spatial Restructuring of the Physician Market}

\vspace{1em}
\begin{figure}[htbp!]
    \caption{Predictions at Broadband Levels by Price Control-Cost Control combinations}
    \label{fig:predictionplotscomb}
    \centering
    \includegraphics[width=.95\textwidth]{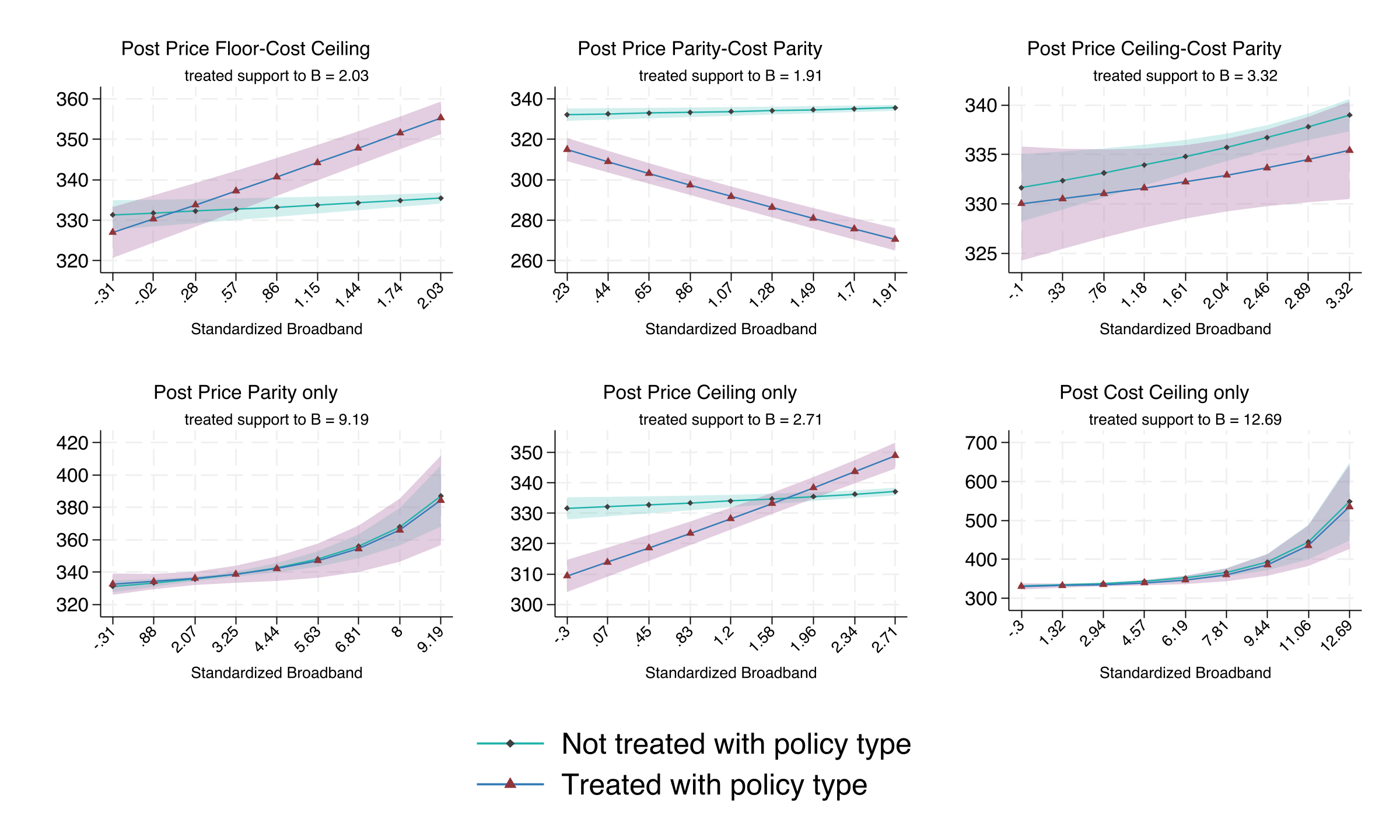}
    \vspace{0.6em}
\noindent
\begin{minipage}{\textwidth}
    \begin{spacing}{1.5}
    \footnotesize
    \justifying
    \textit{\textbf{Note}}: The figure presents predictions from a PPML regression of the total number of physicians (MDs) for the Price Control--Cost Control combinations as specified by the states, with 95\% confidence intervals. Each combination's predictions are drawn only over that combination's treated post-period support, whose end point is stated in each panel, following the same rule as \textit{Table \ref{tab:main_acrt}} and \textit{Figure \ref{fig:attprofiles}}. Maximum treated post-period broadband is $B = 2.03$ for Price Floor--Cost Ceiling, $1.91$ for Price Parity--Cost Parity, $3.32$ for Price Ceiling--Cost Parity, $9.19$ for Price Parity only, $2.71$ for Price Ceiling only, and $12.69$ for Cost Ceiling only. 
    \end{spacing}
\end{minipage}
\end{figure}

\textit{Figure \ref{fig:predictionplotscomb}}\hspace{.1cm} shows such predictions for the policy types as actually specified by the states---types of Price Control and the types of Cost Control, either in isolation, or in combination. Price Floor---Cost Ceiling combination turns out to be a conducive factor for aggregate physician count. Price Parity---Cost Parity combination turns out to be an unfavorable factor. This indicates that the unfavorable effect of Cost Parity dominates when used in combination with a Price Control. Thus, financial incentives to physicians may not necessarily increase care provision as in \cite{CG2014}. The regional variation in regulatory and technological environment and demand side factors play a significant role.

\vspace{1em}
\titlespacing*{\section}{0pt}{0ex plus 0ex minus 0ex}{0pt}
\titleformat{\subsection}{\centering\bfseries\normalsize}{\thesubsection}{0em}{}
\titlespacing{\subsection}{0pt}{0em plus 0ex minus 0ex}{0pt}

\section{Theoretical Framework}\label{sec:tf}

The empirical results in Section~\ref{sec:results} establish patterns that require a unified theoretical explanation. First, Price Floor and non-binding Price Ceiling produce effects on service quantity that converge with broadband, the Floor conducive throughout its treated support and the Ceiling turning from unfavorable at mean broadband to conducive at higher broadband. Second, Cost Parity produces an unfavorable effect on service quantity, with a negative broadband gradient. This section develops the equilibrium framework that generates these patterns. 

\subsection{\textbf{Rotations in Supply Curves: A Supply Chain Framework}}

\textbf{Adaptation of the Household Production Model}: \enquote{Telehealth (\(T\))} and \enquote{In-person (\( I \))} are considered as two factors in the spirit of the household production model \citep{Becker1965}. The impact on the physician count could depend on the effect on surplus of the physicians and the substitutability between the inputs. Price regulations distort the factor mix, create externalities and cause rotation of the marginal cost curve. Considering the supply chain framework similar to \cite{Mulligan2024}, where both physicians and patients are considered co-producers in the healthcare transaction, the production function is given by: \( Y = AF(T, I) \). \( T \) signifies telehealth resources predominantly controlled by providers and \( I \) symbolizes in-person inputs primarily at the discretion of consumers.\footnote{\(T\) and \(I\) inputs are analogous to capital \(K\) and labor \(L\) in traditional economic frameworks, respectively. \(T\) represents telehealth resources such as digital infrastructure, software, and machines predominantly controlled by providers, much like capital in other industries. \(I\)  corresponds to labor-intensive activities such as traveling, standing in queues, and waiting, which demand time and effort primarily from consumers. This distinction is used to reflect the primary control exerted by providers over telehealth technology and infrastructure \((T)\), whereas consumers generally govern their own time and effort \((I)\) in seeking in-person healthcare services. The setup highlights the industry norms where providers are equipped with and responsible for the technological aspects of care \((T)\), and patients navigate the logistics of in-person engagement \((I)\), underlining the separate but interdependent contributions of both parties to healthcare delivery.} The full price \(P_Y\) is expressed as: \( P_Y = \frac{r_T T}{Y} + \frac{r_I I}{Y} \), indicating how total revenue \( P_Y \cdot Y \) is allocated across telehealth and in-person services.  Consumers undertake dual roles: first, as suppliers of \( I \), they receive remuneration valued at \( r_I I/Y \) per unit of output \( Y \) consumed; second, as end-users, they incur the full price \( P_Y \), resulting in a net payment to providers of: \( r_T T/Y = P_Y - r_I I/Y \) per unit of output, where \textquote{\( r_T \)} and \textquote{\( r_I \)} are the inverse factor-supply functions derived from the strictly convex and increasing cost functions, \(\Gamma_T(T)\) and \(\Gamma_I(I)\), for telehealth and in-person services, respectively. The marginal costs associated with in-person and telehealth are captured by \( r_I = \Gamma_I'(I) \) and \( r_T = \Gamma_T'(T) \), respectively. To acknowledge the impacts of cost-sharing and broadband access, the market demand is now characterized by a curve \(Y_R = ZD(P_Y; \gamma)\). This adapts the unregulated demand function, \(Y_U = ZD(P_Y)\),  to incorporate the policy parameter \(\gamma\), thereby accounting for regulatory influences on consumer demand elasticity.  \( A \) and \( Z \) serve as shift parameters, which are each equal to 1 and remain constant with the absence of productivity and demand shock, respectively. Similarly, supply within this market is represented by the regulated market supply curve, \( Y_R = S(P_Y; \rho) \), where \(\rho\) incorporates the policy parameter. \( Y_U = S(P_Y) \) is the unregulated supply. Under the presence of an up-stream price regulation compliance condition ($PRC$): \( \frac{r_T T}{Y} = \rho \), the equilibrium in the telehealth space is not necessarily optimized solely around provider reimbursements, but is also attuned to the collective value derived from healthcare services, inclusive of patient's contributions. The production function defined as $F(T, I)$ relates to how demand for inputs escalates with increased service provision $Y$, and $C(r_I, r_T, Y)$ denotes the corresponding unregulated cost function. These considerations amalgamate to form an equilibrium that is now dynamically responsive to the policy environment and distinct consumer behaviors.\footnote{Payroll taxes funding Medicare can be incorporated into the model by introducing a one-time lump-sum tax \( \psi \), which reduces the total resources available for healthcare production. For simplicity, the main model assumes no taxes.} The equilibrium conditions are:

    \begin{multicols}{2}
        \begin{itemize}
            \item Full Price (\textit{FP}): \( P_Y = \frac{r_T T}{Y} + \frac{r_I I}{Y} \)
            \item Final Demand (\textit{FD}): \( Y = D(P_Y; \gamma) \)
            \item Production Function (\textit{PF}):\\ \( Y = F(T, I) \)
            \item In-person Supply (\textit{IS}): \( r_I = \Gamma_I'(I) \)
            \item Telehealth Supply (\textit{TS}): \( r_T = \Gamma_T'(T) \)
        \end{itemize}
        \columnbreak
        \begin{itemize}
            \item Price Regulation Compliance (\textit{PRC}):\\ \( \frac{r_T T}{Y} = \rho \)
            \item Cost Minimization (\textit{CMC}):\\ \( \frac{\tfrac{\partial F(T, I)}{\partial T}}{\tfrac{\partial F(T, I)}{\partial I}} = \frac{r_T}{r_I} \)
        \end{itemize}
    \end{multicols}

\textbf{\textit{Divergence from Conventional Models}}: The divergence arises when we introduce price regulation compliance ($PRC$). Cost minimization condition ($CMC$) would set relative marginal costs equal to the marginal rate of substitution derived from the production function $F(T, I)$. However, $PRC$, an upstream-price compliance condition, mandates that the price per unit of telehealth service $(r_T T/Y)$, as a policy parameter $\rho$, aligns with either a legislatively mandated Price Floor or Price Ceiling, transforming the allocation of resources within the telehealth and in-person services markets. This regulation-induced adjustment distorts the market allocation away from the cost-minimizing mix of telehealth and in-person services that would naturally arise in an unregulated market.

\textbf{Quality Adjustments Under Non-price Competition}: Price Controls cause changes in the product characteristics as market actors adapt to the imposed constraints. In case of a Price Ceiling, particularly when the supply is highly elastic, the markets do not simply collapse or trade does not disappear as shown in standard textbooks. Instead, there is an incentive for providers to modify the quality or nature of the product to maintain some level of market functionality, while avoiding the total disappearance of trade\textemdash{}a scenario supported by the elasticity of supply and a reluctance to forego potential gains from trade. This implies a rotation of the supply curve due to a change in the mix of the factors of production utilized. The price regulated markets are not slack, due to the incentive among providers to react to the willingness of the consumers to accept a lower quality product. Regulation deviates providers from cost-minimization towards regulatory compliance, which raises the overall costs. However, marginal costs may not rise due to adaptations. For instance, if excessive \(T\) and a minuscule \(I\) are being used instead of the cost minimizing factor mix, the marginal cost of producing an extra unit of \(Y\) would be lower since more of \(T\), which is cheaper, can be used.

For a unit output \(Y\), consumers are quoted the full price \(P_Y\) that reflects both the cost of healthcare services provided and the value of consumer input. When faced with a binding Price Ceiling, the consumers are compelled to find alternative ways to compete due to the inability to pay higher prices, a consequence of the regulatory cap. Such alternatives may involve increased personal effort, such as being more accommodating with appointment scheduling or willing to travel greater distances for care. In contrast, a Price Floor stimulates providers to enhance their offerings. For example, they may invest in expanding telehealth services by introducing additional features or capabilities. In this regulated landscape, where price cannot serve as the primary signaling mechanism, the quality and assortment of services become the focal points of communication between healthcare providers and consumers. Telehealth is a capital-intensive, higher-quality product that requires less time and effort from the consumer, whereas in-person or office-based healthcare is a low-quality product that demands more time and effort from the consumer.

\vspace{1em}
\begin{figure}[h!]
    \caption{Illustration of Price Controls created by Parity Laws}
    \centering
    \includegraphics[scale=.24]{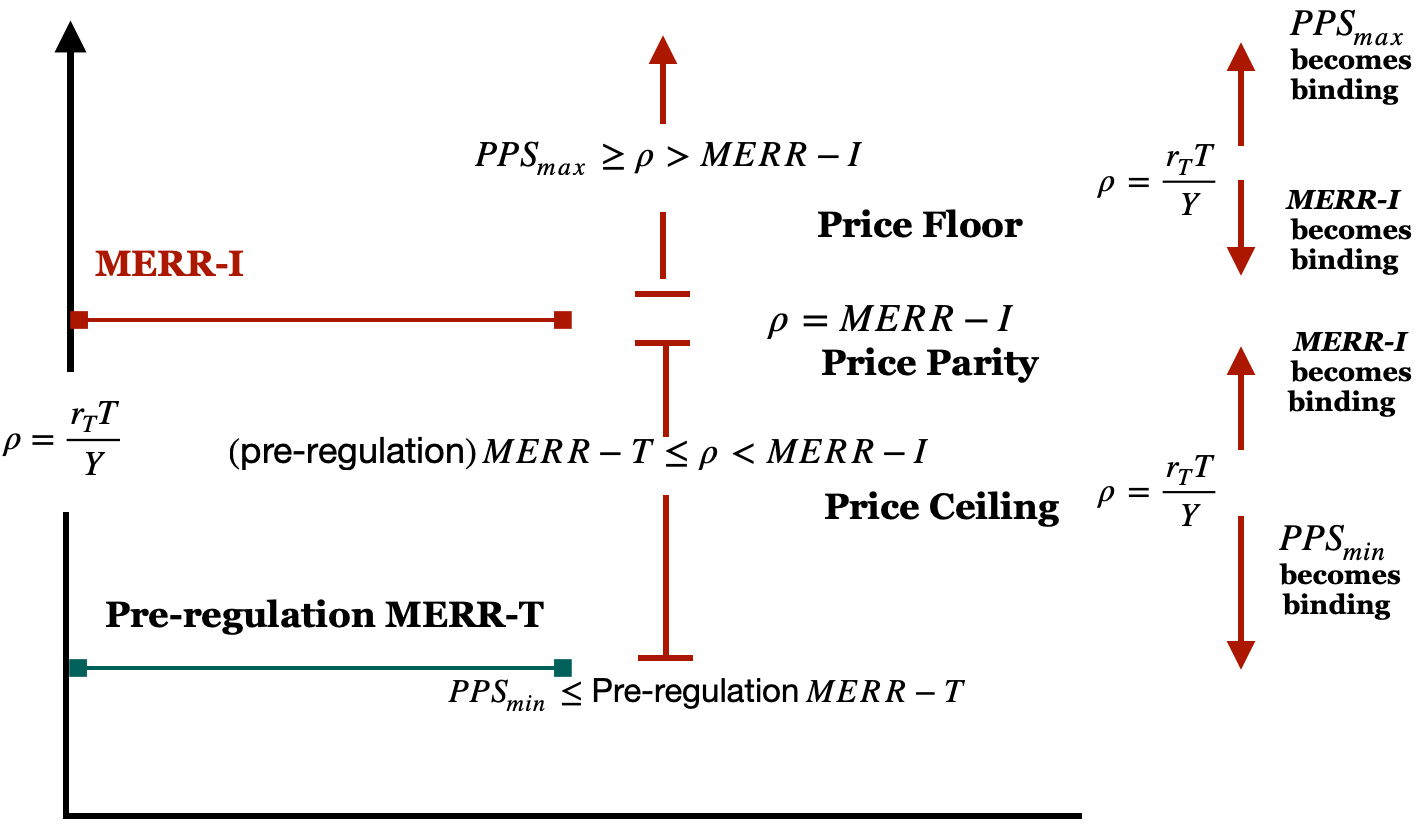}
    \label{fig:pcillustration}
\noindent
\begin{minipage}{\textwidth}
    \begin{spacing}{1.5}
    \footnotesize
    \justifying
    \vspace{2em}
    \textit{Note}: The Figure shows the design of Price Ceiling and Price Floor created by Telehealth Parity Laws (TPL). $MERR-I$ denotes Market Equilibrium Reimbursement Rate for in-person and $MERR-T$ denotes Market Equilibrium Reimbursement Rate for telehealth. \( \rho\) is the policy parameter. There is a point \(E(\rho\)) where \( \rho = \dfrac{r_T T}{Y}\). \(\dfrac{r_T T}{Y}\) signifies the ``physician reimbursement" for telehealth per unit of healthcare quantity \(Y\). 
    \vspace{0.5em}
    
    The design for Cost Controls is analogous to that of Price Controls, with \(\gamma\) as the cost regulation policy parameter, and $MERR-I$ and $MERR-T$ replaced by ``Market Equilibrium Cost Rate for in-person" $MECR-I$ and ``Market Equilibrium Cost Rate for telehealth" $MECR-T$, respectively. The Cost is the out-of-pocket costs paid by consumers in the form of ``deductibles, copay and coinsurance", with \(E(\gamma\)) as a point such that \(\gamma = E_{oop-T}\), where \(E_{oop-T}\) denote out-of-pocket costs for telehealth per unit of healthcare quantity, \(Y\). 
    \vspace{0.5em}
    
    \(PPS\) denotes the physician pay schedule with \(PPS_{min}\) denoting its lower limit and \(PPS_{max}\) denoting its upper limit.
    \end{spacing}
\end{minipage}
\end{figure}

\vspace{1em}
\textbf{\textit{Policy Parameters and Regulatory Conditions}}: $E(\rho)$ exists as a point where $\rho = \frac{r_T T}{Y}$ such that $\rho \in [\rho_{min}, \rho_{max}]$. When Price Floor is specified by a state through TPL, post regulation, $E(\rho)$ lies above $MERR-I$ shown in \textit{Figure \ref{fig:pcillustration}}. As one goes above  $E(\rho)$, the upper limit of Physician Pay Schedule (\(PPS_{max})\) becomes binding and takes the role as a non-binding Price Ceiling. As one goes below  $E(\rho)$, $MERR-I$ becomes binding and takes the role as a Price Floor. On the contrary, when Price Ceiling is specified by a state through TPL, post regulation, $E(\rho)$ lies below $MERR-I$ shown in \textit{Figure \ref{fig:pcillustration}}. As one goes above $E(\rho)$, $MERR-I$ becomes binding and takes the role as a Price Ceiling. As one goes below $E(\rho)$, the lower limit of Physician Pay Schedule (\(PPS_{min})\) becomes binding and takes the role as a Price Floor. A similar analogy follows for $E(\gamma)$ and Cost Ceiling or Cost Parity with $\gamma \in [\gamma_{min}, \gamma_{max}]$. $E(\rho)$ and $E(\gamma)$ are different from equilibrium points. $E(\rho)$ is a point where regulatory parameter for physician reimbursement is equal to physician reimbursement for telehealth such that \(\rho = \frac{r_T T}{Y}\), and $E(\gamma)$ is a point where regulatory parameter for consumer out-of-pocket costs is equal to out-of-pocket costs for telehealth such that \(\gamma = E_{oop-T}\). $E(\rho)$ might coincide with the equilibrium point only if the demand curve passes through it, while $E(\gamma)$ might coincide with the equilibrium point only if the supply curve passes through it. Holding constant tastes ($B$, $D()$) and technology ($A$, $F()$, $\Gamma_I$, $\Gamma_T$), both the unregulated equilibrium $U$ and any point on the unregulated supply curve are on the same demand curve in the $[Y,P_Y]$ plane, indicating that regulation effects stem from supply rotations. Pre-regulation, $E(\rho)$ satisfies unregulated equilibrium conditions---$FP$, $PF$, $IS$, $TS$, and $CMC$. Post-regulation, $E(\rho)$ satisfies regulated equilibrium conditions---$FP$, $PF$, $IS$, $TS$, and $PRC$. 

Let $\rho$ be the regulatory parameter for Price Controls: $\rho \in [\rho_{\text{min}}, \rho_{\text{max}}]$. \(MERR-I\) and \(MERR-T\) are the Market Equilibrium Reimbursement Rate for in-person and telehealth, respectively. Then:

\begin{itemize}
    \item Price Floor (PF): Sets $\rho=\dfrac{r_T T}{Y} > MERR-I$, increasing $\dfrac{r_T T}{Y}$, since $ MERR-T < MERR-I$ pre-regulation.
    \item Price Ceiling (PC): Caps $\rho=\dfrac{r_T T}{Y}< MERR-I$, but $\dfrac{r_T T}{Y}$ may rise if $MERR - T < \rho_{\text{max}}$.
\end{itemize}
 
\vspace{1em}
\textbf{Factor Distortions and Production Inefficiency-Price Floor vs Price Ceiling:} Irrespective of whether pre-regulation $MERR-T$ is close to $MERR-I$ or not, the introduction of a binding Price Floor reallocates spending towards the upstream input $T$ at the expense of the downstream input $I$, which ultimately causes an uptick in $r_T$ and a decline in $r_I$. If  \(T\) is more elastically supplied than $I$ ($\varepsilon_{\Gamma T} > \varepsilon_{\Gamma I}$), it would lead to counter-clockwise rotation in the Marginal Cost (MC) curve, and vice versa. Conversely, a binding Price Ceiling would reduce $T$, hence $r_T$, while increasing $I$, and therefore $r_I$. If $T$ is more elastically supplied than $I$ ($\varepsilon_{\Gamma T} > \varepsilon_{\Gamma I}$), it would lead to clockwise rotation in the Marginal Cost (MC) curve, and vice versa. In the case of telehealth, pre regulation $MERR-T$ is much lower than $MERR-I$, i.e., the pre-regulation equilibrium physician reimbursement rate is much lower than the post-regulation rate. Thus, owing to such a unique case, even in the face of a Price Ceiling, providers still reallocate spending towards the upstream input $T$ at the expense of the downstream input $I$, which ultimately causes an increase in $r_T$ and a decline in $r_I$. Thus, the effect could be similar to that of a Price Floor. The feasibility of these mechanisms depend on the spatial area. Typically, in areas with high broadband, $T$ has enough supply elasticity to make such adjustments feasible. In the areas with low broadband, where supply elasticity of $T$ is much less such that ($\varepsilon_{\Gamma T} < \varepsilon_{\Gamma I}$), there is more tendency to reallocate spending from $T$ towards $I$.\footnote{Both metro and non-metro areas have counties with high or low levels of broadband, even though metro areas have more counties with high broadband. The supply elasticity of telehealth is more contingent on the level of broadband than on the degree of urbanity, as internet access is crucial for telehealth services.} In cases where $I$ is also not supplied elastically enough, it could lead to reduction in service quantity, and possibly relocation or exit of the upstream input supplier or physician.

When the model operates near an efficient factor mix, no incremental costs arise, given the capability to substitute between inputs to maintain the desired output level. Inputs $T$ and $I$ can be traded off along the isoquant at an exchange rate of $r_I / r_T$ (from the Cost Minimization Condition, CMC), so there is no price effect if $r_I$ and $r_T$ are kept constant. However, changes in $r_I$ and $r_T$ have repercussions for $P_Y$, contingent on the relative elasticity of supply for $T$ and $I$. The provider's ability to adjust depends on these elasticities, and the Marginal Cost (MC) curve rotates in response. Therefore, the overall impact of a price regulation on $P_Y$, whether it leads to an increase or decrease in $P_Y$ at a specific quantity, depends on the elasticity of supply for the inputs $T$ and $I$. Let $\varepsilon_{\Gamma T}$ and $\varepsilon_{\Gamma I}$ be the elasticities of supply for the inputs \(T\) and \(I\), respectively. The relation between the regulated and unregulated equilibrium quantities, respectively, depend upon the position of the regulated demand curve or the demand curve post regulation. 

\proposition{prop:1}:
With $\varepsilon_{\Gamma T}$ and $\varepsilon_{\Gamma I}$ as input supply elasticities for \(T\) and \(I\), respectively, and $\tilde{Q} \equiv \varepsilon_{\Gamma I} + \sigma\left[1 + s_I\left(\varepsilon_{\Gamma T} - \varepsilon_{\Gamma I}\right)\right]$, where $\sigma$ is the elasticity of substitution between the inputs and $s_I$ is the cost share of \(I\)---    
\begin{enumerate}
    \item[(a)] When \textsc{$MERR-I$} acts as a binding Price Floor or a non-binding Price Ceiling, the supply curve rotates counter-clockwise around $E(\rho)$ if $\varepsilon_{\Gamma T} - \varepsilon_{\Gamma I}$ is positive, in which case $\tilde{Q} > 0$ holds automatically. The supply curve rotates clockwise if $\varepsilon_{\Gamma T} - \varepsilon_{\Gamma I}$ is negative, provided $\tilde{Q} > 0$, for which $s_I\left(\varepsilon_{\Gamma I} - \varepsilon_{\Gamma T}\right) \leq 1$ is a sufficient condition that holds for any $\sigma$. 
    \item[(b)] For each type of regulation policy, the position of the regulated equilibrium quantity $E_R$ with respect to the unregulated equilibrium quantity $E_U$ depend on the position of the regulated demand curve. 
\end{enumerate}

The proof of \textit{Proposition \ref{prop:1} (a)} is in \textit{Section \ref{subsec:proof1a}} in the Appendix. For \textit{\ref{prop:1} (a)} to hold, specifically, for the Price Ceiling and Price Floor to exhibit similar behavior—it is necessary that pre-regulation telehealth reimbursement is lower than $MERR-I$, which ensures that the Price Ceiling is non-binding. As previously noted, this scenario is indeed the case with pre-regulation $MERR-T$ being considerably lower than $MERR-I$. $PPS_{min}$ and $PPS_{max}$ remain similar across states due to the need for uniformity and private insurers following in footsteps of Medicare. This scenario allows physician reimbursement to rise even under a Price Ceiling ($PC$), resulting in the regulated supply curve ($Y_R = S^{}_{PC}(P_Y; \rho)$) exhibiting behavior similar to the regulated supply curve ($Y_R = S^{}_{PF}(P_Y; \rho)$) of a binding Price Floor ($PF$), as shown in \textit{Figure \ref{fig:floorceiling}}. Conversely, if the Price Ceiling were binding, $\rho$ would cap $\frac{r_T T}{Y}$ below its unregulated level, and providers would reallocate spending away from the upstream input $T$ towards the downstream input $I$, reducing $r_T$ and raising $r_I$. This case lies outside the hypothesis of \textit{Proposition \ref{prop:1}}~(a) and outside the data. With $\varepsilon_{\Gamma T} - \varepsilon_{\Gamma I} > 0$, the regulated supply curve would then be the light green curve, a clockwise rotation from the unregulated dark green curve, rather than the red-black curve that is the regulated supply curve under the non-binding Price Ceiling. With the demand curve crossing from below in \textit{Panel (b), Figure \ref{fig:3b}}, the regulated equilibrium quantity would fall rather than rise. The sign of $\varepsilon_{\Gamma T} - \varepsilon_{\Gamma I}$ is determined by the level of broadband.

\vspace{2em}
\newcommand{\assumeref}[1]{\hyperref[ass:#1]{#1 Assumption}}

\textbf{Broadband Enhanced Telehealth Supply Elasticity Assumption (BETSEA)}\label{ass:betsea}: At broadband levels one standard deviation or more above the mean, the telehealth supply elasticity is greater than the supply elasticity of in-person input such that, $\varepsilon_{\Gamma T} - \varepsilon_{\Gamma I} > 0$. 

The supply elasticity of telehealth depends on multiple factors, among which broadband is a crucial one. An increase in broadband increases the supply elasticity of telehealth. This assumption is backed by empirical evidence. The findings from the ``Telehealth Broadband Pilot Program" suggest that broadband is indispensable for telehealth. Lower broadband and poor connections disrupt telehealth access. Poor internet limits the provision and adoption of telehealth, creating healthcare gaps. On the other hand, strong broadband connectivity supports video consultations, remote monitoring and other telehealth services \citep{Bogulski2024}. Broadband not only increases telehealth adoption and provision but also utilization or demand \citep{Amaral-Garcia2022, Okoye2021, Pandit2025}.\footnote{As shown in \textit{Figure \ref{fig:bivariate}}, stronger broadband supports physician numbers. This fosters increased investment in telehealth capital, especially for those specialties which use telehealth the most, similar to how movement of people fosters movement of capital. In addition, broadband supports digital-literacy programs to increase access to telehealth and ramp up ancillary services such as digital health data management, analytics and population healthcare management.}\footnote{The empirical results provide direct validation of this assumption: the $ACRT$ is 
concentrated among radiologists, psychiatrists, and emergency physicians --- the 
specialties with the highest pre-pandemic telehealth usage --- while remaining 
statistically indistinguishable from zero for cardiologists and gastroenterologists 
(\textit{Tables~\ref{tab:heavyusers}--\ref{tab:lightusers}}). Any confounder 
correlated with broadband growth would affect all specialties uniformly; the selective 
pattern is consistent only with the telehealth supply elasticity mechanism that BETSEA 
formalizes.}

$ E_{oop} $ represents the average out-of-pocket cost per healthcare service, considering that the total healthcare services consumed are a mix of telehealth and in-person services. Let $\gamma$ be the regulatory parameter for Cost Controls: $\gamma \in [\gamma_{\text{min}}, \gamma_{\text{max}}]$. \(MECR-I\) and \(MECR-T\) be the Market Equilibrium Cost Rate for in-person and telehealth, respectively. Consumer out-of-pocket costs are given by $E_{oop}$. $E_{oop}$ is a weighted average of the out-of-pocket costs of the two types of services, telehealth (\(T\)) and in-person (\(I\)), weighted by their proportions such that at regulation compliance ($\gamma=E_{oop-T}).$ Then:

\begin{itemize}
    \item Cost Parity (CP): Sets $\gamma=E_{oop-T} = MECR-I$, increasing $E_{oop}$ since $MECR-T < MECR-I$ pre-regulation.
    \item Cost Ceiling (CC): Caps $\gamma=E_{oop-T} \leq MECR-I$. Pre-regulation $MECR-T < MECR-I$, so $E_{oop}$ rises if $MECR - T < \gamma_{\text{max}}$.\footnote{This ensues when the insurer passes on the costs to the consumer making the ceiling non-binding unless $MECR - T \geq MECR-I$ or difference between pre-regulation $MECR-I$ and $MECR-T$ is $\approx 0$.}
\end{itemize}

\subsection{\textbf{Rotations in Demand Curves}}\label{subsec:rotations_in_demand_curves}
\textbf{Cost Controls as Demand Modulators}: A binding consumer Cost Ceiling prevents consumer cost for telehealth from rising above that for in-person services. However, consumer cost for telehealth can still increase to that of in-person if pre-regulation difference between out-of-pocket costs for telehealth and in-person is substantial and the insurer passes on additional burden of expenditure to the consumer. In that case, a Cost Ceiling is non-binding. Cost Parity equalizes consumer costs for telehealth services with those for in-person services, increasing the full price with certainty (\textit{Section \ref{subsec:costregfullprice}, Appendix}). Thus, Cost Parity is necessarily binding. If only a type of Price Control is specified, it may affect demand if the insurer changes the cost sharing structure, which is reflected through the Medicare costs. However, as shown in \textit{Table \ref{tab:withbbd}}, when only Price Controls are deployed without Cost Controls, the effect on outpatient visits is negligible and insignificant. This shows that for Price Controls only, the cost-sharing structure is not altered meaningfully in a way that would affect the demand. Thus, Price Controls affect the supply only, in consonance with the supply chain framework. The demand is modified by Cost Controls.

\textbf{Demand for Healthcare Services (Time Price vs Money Price)}: Within this framework, the approach similar to that in \cite{acton1975} informs the dynamic nature of demand. The utility function of individuals incorporates medical services, denoted by \( Y \), and a composite good represented by \( X \). The constraint is the full income equation that combines money price and time price for both medical services and the other composite good. \( P_Y \) is the money price, which is also the full price paid by the consumer in the supply chain approach. If \( w\) is the wage and \(\tau \) is time required for accessing medical services, \( w \times \tau \) becomes the time price per unit of medical services. The total price \(\Pi_Y\), which is the sum of money price and time price, becomes: \( \Pi_Y = P_Y + w \times \tau \). Both money price and time price influence the demand for medical service \( Y \), depending on money price elasticity, \( \varepsilon_Y^{P_Y}\), and time price elasticity, \( \varepsilon_Y^{\tau} \), of demand for medical services, such that: \( \varepsilon_Y^{P_Y} = \dfrac{P_Y}{\Pi_Y}  \varepsilon_Y^{\Pi_Y}\), and \( \varepsilon_m^{\tau} = \dfrac{w \tau}{\Pi_Y}  \varepsilon_m^{\Pi_Y}\). This yields a prediction from the model: \begin{equation}
   \varepsilon_m^{w\tau} \mathrel{\substack{<\\>}} \varepsilon_m^{P_Y} \text{ as } w\tau \mathrel{\substack{<\\>}} P_Y \tag{B1} \label{eq:elasticityrelative}
\end{equation}

This implies that the relation between money price and time price elasticities is determined by the relation between the respective prices. The proof is in \textit{Section \ref{sec:proofb1}} in the Appendix.
  
\textbf{Changes in Consumption Mix Owing to Expected Change in Cost}: The higher wages in metro areas make the time price of medical services higher compared to the money price. Thus, at the same level of broadband, the demand for medical services is more sensitive to changes in time price in metro areas, as opposed to non-metro areas, which have relatively lower wages,  as implied by (\ref{eq:elasticityrelative}). Moreover, the demand for medical services is relatively more sensitive to money prices in non-metro areas than in metro areas, making the demand curve more elastic for the former. Thus, at $E(\gamma)$, where the out-of-pocket costs for telehealth becomes the same as $MECR-I$ and where broadband is sufficient so that telehealth usage is considerable, a Cost Parity would cause a counter-clockwise rotation in the demand curve. This suggests that Cost Parity raises \(E_{oop-T}\) such that quantity demanded post regulation is lesser at the same pre-regulation \(P_Y\). A non-binding Cost Ceiling has similar effect to that of Cost Parity and cause a counter-clockwise rotation but with much lesser magnitude than Cost Parity, since Cost Parity is binding. These effects are more pronounced in areas with higher broadband levels with higher telehealth usage. 
  
\textbf{Income and Substitution Effects of Broadband}: An increase in broadband may have an income effect such that it could increase the demand for both medical good as well as composite good.\footnote{Broadband inevitably boosts e-commerce and healthcare utilization owing to better product and health consciousness, access to information and enhanced access through home delivery of goods and remote delivery of healthcare through telehealth.} It would increase the opportunity cost of time, thus raising the time price. Whether this change results in a positive substitution effect causing an increase in demand for medical services depends on the relative proportions of the time price to the total price of medical services and the composite good. With \( q \) as the money price and \( w \times s \) as the time price per unit of the composite good \( X \), the substitution effect of broadband on the demand for medical services is positive if:
\vspace{-0.5em}
\begin{equation}
    \frac{ws}{ws + q} > \frac{w\tau}{w\tau + P_Y}, \tag{B2} \label{eq:substitution}
\end{equation}
that is, if the time price is a larger proportion of the total price for the composite good, \( X \), than it is for medical services, \( Y \). The proof is given in \textit{Section \ref{sec:proofb2}} in the Appendix.

\begin{figure}[htbp!]
    \centering
    \caption{Supply and Demand Rotations under Price and Cost Controls}
    \vspace{0.2em}
    \label{fig:floorceiling}
    \begin{minipage}{.1\textwidth}
        {(a) \medium \mbox{MERR\texttt{-}I} Floor}
    \end{minipage}%
    \begin{minipage}{.87\textwidth}
        \includegraphics[width=\textwidth, height=0.38\textheight, keepaspectratio]{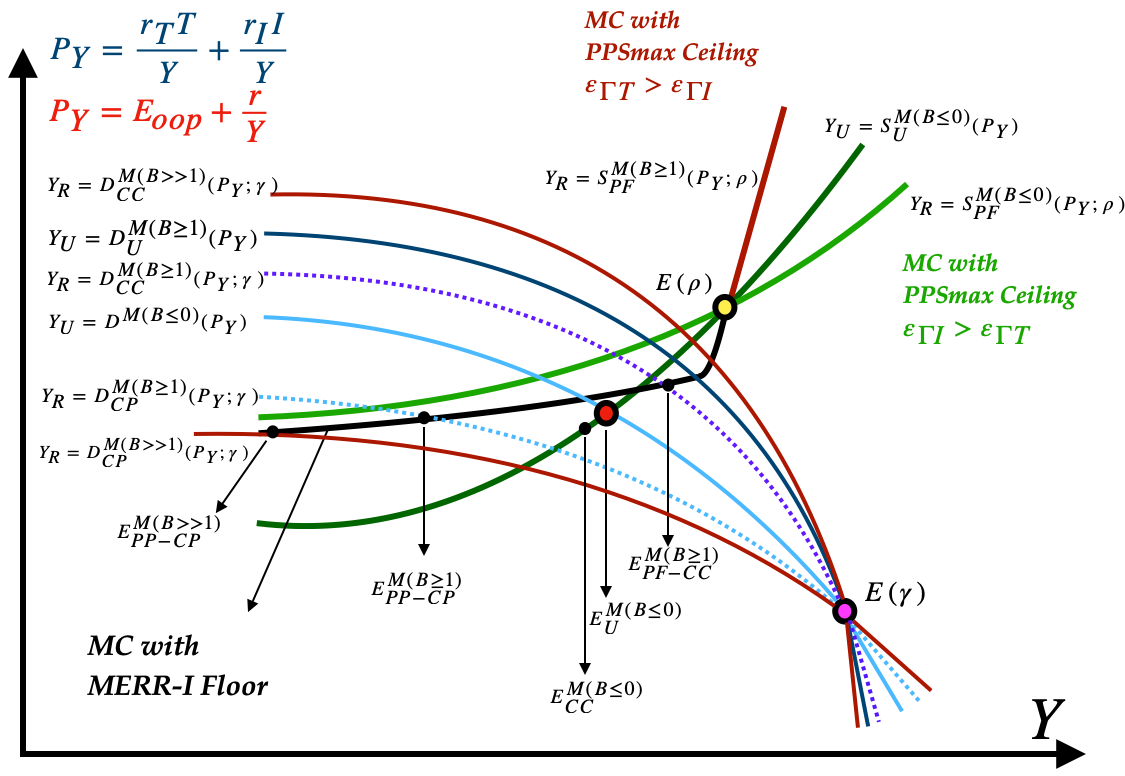}
        \label{fig:3a}
    \end{minipage}

    \begin{minipage}{.1\textwidth}
        {(b) \medium \mbox{MERR\texttt{-}I} Ceiling}
    \end{minipage}%
    \begin{minipage}{.85\textwidth}
        \includegraphics[width=\textwidth, height=0.38\textheight, keepaspectratio]{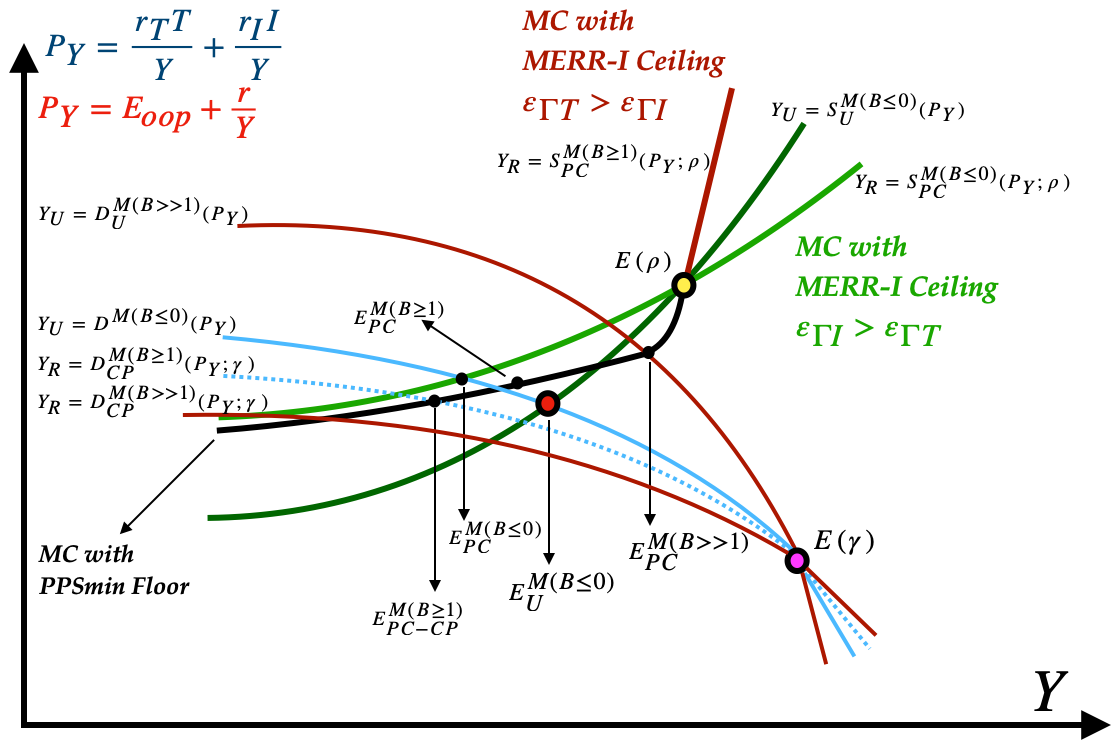}
        \label{fig:3b}
    \end{minipage}
       \vspace{1em}
        \footnotesize
        \justifying
        
\textit{\textbf{Note}}: \textbf{(a): $\boldsymbol{MERR-I}$ as Price Floor ($\boldsymbol{PPS_{max}}$ as Ceiling)}. 
\textbf{(b): $\boldsymbol{MERR-I}$ as Price Ceiling ($\boldsymbol{PPS_{min}}$ as Floor)}.
        
        \vspace{0.8em}
        \textbf{Superscripts:} M: Metro. B: Standardized Broadband. $B\leq0$, $B\geq1$ and $B>>1$---standardized broadband at the mean, one unit above the mean, and well above one unit above the mean.

        \textbf{Subscripts denote policy type:} CP--Cost Parity, CC--Cost Ceiling, PC--Price Ceiling, PF--Price Floor, and PP--Price Parity, respectively. 
        
        R: Regulated. U: Unregulated. 
        
        The light green supply curve ($Y_R = S^{M(B\leq0)}_{PF}(P_Y; \rho)$) denotes the regulated supply at $B\leq0$, with a Price Control specified. The dark green supply curve ($Y_U = S^{M(B\leq0)}_U(P_Y)$) denotes the unregulated supply at $B\leq0$ : ($\varepsilon_{\Gamma T} - \varepsilon_{\Gamma I} < 0$), with no Price Control specified. The maroon---black supply curve  ($Y_R = S^{M(B\geq1)}_{PF}(P_Y; \rho)$) denotes the regulated supply at $B\geq1$: ($\varepsilon_{\Gamma T} - \varepsilon_{\Gamma I} > 0$), with a Price Control specified. 
        
        The yellow and pink circles with black rings denote $E(\rho)$ and $E(\gamma)$, respectively. The red circle with black ring denotes the unregulated equilibrium for metro ($E^{M}_{U}$). The equilibrium quantities for each policy type are denoted by black dots. The solid maroon demand curves represent demand curves where $B>>1$.  As the red supply curve crosses and goes below $E(\rho)$, in Panel (a), $MERR-I$ becomes binding taking the role of Price Floor, and in Panel (b), $PPS_{min}$ becomes binding taking the role of Price Floor, so the supply curve becomes almost horizontal.
\end{figure}

If there is increase in broadband causing a rise in the opportunity cost of time, and if the time intensity of composite good is more than that of medical services, the substitution effect is positive, leading to an increased demand for medical services. The increase in demand is more pronounced in metro areas where the demand for medical services is more sensitive to changes in time price. In metro areas, a Cost Parity, which causes telehealth costs to rise since it is binding, has moderately higher effect than a non-binding Cost Ceiling regulation. Thus, for counties with higher broadband levels above the mean, the demand curves fan out progressively as shown in \textit{Figure \ref{fig:floorceiling}}.\footnote{For free or highly subsidized services, where \( P \approx 0 \), the right-hand side of (\ref{eq:substitution}) becomes greater than the left-hand side, causing the substitution effect to be negative. For a high \( P \), the right-hand side becomes less than the left-hand side, making the substitution effect positive. The switch in the substitution effect would happen around \( E(\gamma) \), where neither Cost Parity nor Cost Ceiling is binding, and the substitution effect is negligible.} The regulated demand curve takes into account this effect of broadband and the effect of Cost Control. The quantity corresponding to the intersection of the regulated demand and supply curves gives the regulated equilibrium quantity.

\vspace{1em}
\proposition{prop:2}: For the supply and demand curves in the \([Y, P_Y]\) plane as the locus of pairs \(\{Y, P_Y\}\) and broadband at least one unit above the mean ($B \geq 1$):

\begin{itemize}
  \item[(a)] A Cost Parity causes the demand curve to rotate counter-clockwise since it is binding, with higher broadband amplifying this effect. 
  \item[(b)] A non-binding Cost Ceiling has similar effect as Cost Parity causing the demand curve to rotate counter-clockwise. However, this effect is of lesser magnitude since Cost Parity is binding, while Cost Ceiling is not.
  \item[(c)] Broadband increase nullifies the effect in (b) because higher broadband rotates the demand curve clockwise due to non-binding nature of Cost Ceiling. 
\end{itemize}

\vspace{1em}

The proof is given in \textit{Section \ref{sec:prop2proof}} in the Appendix. A Cost Parity in areas with mean or lesser levels of broadband may not be sufficient to have an effect on demand. In areas with sufficient broadband, and hence with telehealth utilization, a Cost Parity has a significant effect since it effectively increases the out-of-pocket costs of accessing telehealth. Due to higher telehealth supply elasticity in higher broadband areas and the dis-utility of in-person access, the consumer is not readily able to substitute the former with the latter, causing reduction in healthcare quantity demanded. This implies that quantity demanded at a price would be lesser than before, which leads to subsequent counter-clockwise rotation. Since the magnitude of difference between $MECR-I$ and pre-regulation $MECR-T$ was considerable, Cost Parity induces a robust and certain negative effect since \(E_{oop-T}\) is forced to rise up to $MECR-I$. This increases as standardized broadband rises further above $B \geq 1$, such that the negative effect of the unfavorable Cost Parity deepens, as reflected in the negative $ACRT$ over the framing's treated support (\textit{Figure \ref{fig:attprofiles}}). 

The regulated equilibrium depends on the type of regulation in effect—a type of Price Control, or Cost Control, or a combination of a type of Price Control with that of a Cost Control. If only a Price Control is in effect, then the regulated equilibrium is at the intersection of regulated supply and unregulated demand. If only a Cost Control is in effect, then the regulated equilibrium is at the intersection of unregulated supply and regulated demand. If a Price Control-Cost Control combination is present, then the regulated equilibrium is at the intersection of regulated supply and regulated demand.\footnote{The focus of the study is restricted to service reallocation and market restructuring. Welfare analysis requires integrating the supply and demand curve rotations across multiple regulation types simultaneously, each with distinct effects on provider 
surplus, consumer out-of-pocket costs, and insurer cost sharing --- a separate undertaking beyond the scope of this paper. For welfare effects of such regulations, see \cite{Bulow2012}, \cite{BJ2024} and \cite{DKA2021}.}

\textit{Figure \ref{fig:floorceiling}} shows the rotations in supply curves as per \textit{Proposition  \ref{prop:1}} and the rotations in demand curves as per \textit{Proposition  \ref{prop:2}}. The estimates of $ATT_k$ for service quantity (CHSPI) are analogous to the distances between the projections of unregulated equilibrium point ($E^{M(B\leq0)}_{U}$) and the projections of regulated equilibrium points ($E^{M(B)}_{k}$) corresponding to each policy type, $k$, on the quantity axis (B is the standardized broadband). Higher levels of broadband progressively amplify the conducive effects of policies such as Price Floor or Price Ceiling (further counter-clockwise rotation in regulated supply curve). Higher broadband also deepens the unfavorable effects of policies such as Cost Parity (further counter-clockwise rotation in regulated demand curve). Thus, Cost Parity's negative effect dominates when it is used in combination with a type of Price Control, more so at higher broadband levels within the treated support. The effects are diluted at lower broadband levels, especially below mean broadband levels, as shown in \textit{Table \ref{tab:below3SDandmean}}.

The effect of Cost Ceiling is ambiguous. Cost Ceiling, which is non-binding, may have a similar effect as Cost Parity in areas with low broadband, since insurers might be able to pass higher burden of cost sharing with the consumers. This can cause a counter-clockwise rotation, implying a reduction in quantity as shown in \textit{Figure \ref{fig:floorceiling}} (point $E^{M(B\leq0)}_{CC})$. However, in this case, the role of an increased broadband is ameliorative rather than aggravating as in case of Cost Parity. Since Cost Ceiling is not binding, it can still allow the demand to rise owing to clockwise rotation due to higher broadband. That would nullify the unfavorable tendency of Cost Ceiling. As broadband further increases, a further clockwise rotation may cause a higher positive effect. If Cost Ceiling is combined with a conducive component such as Price Floor, that can have an overall positive effect ($E^{M(B\geq1)}_{PF-CC}$ in \textit{Figure \ref{fig:floorceiling}}). The progressive effect of or modification of treatment effect by broadband for aggregate sample becomes clearer while discussing the results for $ATT_k$ at levels and ``Average Causal Response on Treated" ($ACRT_k$) in \textit{Table \ref{tab:main_acrt}}. The switch in sign ($\varepsilon_{\Gamma_T} - \varepsilon_{\Gamma_I} < 0$ to $\varepsilon_{\Gamma_T} - \varepsilon_{\Gamma_I} > 0$) mimics the transition from short-run to long-run behaviour when telehealth becomes relatively more elastic with time. 

\textbf{Lower Broadband Levels:} For lower broadband levels, the conducive effects of Price Floor and non-binding Price Ceiling on service quantity (CHSPI) get diluted, since the elasticity of input supply of telehealth remains very low. Similarly, Cost Parity may not be able to exert an unfavorable effect on service quantity (CHSPI) due to lower dependence on telehealth, while Cost Ceiling may still show an ambiguous or negligible impact.

Telehealth offers a quality premium due to its convenience and effectiveness similar to outpatient in-person services as observed empirically, and this premium grows with broadband. At low broadband the premium is minimal and telehealth is not operationally feasible for most patients, so consumers have little scope to substitute between the two services. Pre-regulation, when \(E_{oop-T}^{\text{pre}} < E_{oop-I}\), telehealth is less costly, but its adoption is constrained by limited broadband access. The pre-regulation full price is: \(P_Y^{\text{pre}} = \frac{E_{oop-T}^{\text{pre}} \cdot Y_T + E_{oop-I} \cdot Y_I}{Y} + \frac{r}{Y}\). Post-regulation, Cost Parity enforces \(E_{oop-T} = E_{oop-I}\), raising telehealth costs and yielding a full price of \(P_Y^{\text{post}} = E_{oop-I} + \frac{r}{Y}\). For the same pre-regulation quantity \(Y\), \(P_Y^{\text{post}} > P_Y^{\text{pre}}\) because the average out-of-pocket cost increases. With little scope for substitution at low broadband, consumers cannot easily shift their service mix, resulting in a less elastic demand curve, characterized by clockwise rotation. This causes a higher equilibrium quantity for Cost Parity at lower broadband levels (through a clockwise rotation) as suggested by the estimates for $ATT(B=0)$ in \textit{Table \ref{tab:main_acrt}} and estimates for ``broadband below mean" in \textit{Table \ref{tab:below3SDandmean}}. \textit{Row 4, Column 7, Table \ref{tab:withbbd}} shows that Cost Ceiling fails to exert a significant impact on outpatient visits and hence demand. Thus, a non-binding Cost Ceiling would simply have negligible effects.

\vspace{1em}
\section{Robustness Checks}\label{sec:rc}

County and year fixed effects were included in all specifications to control for unobserved, time-invariant characteristics and common time trends. To ensure balance and comparability of treatment and control groups, logistic regression-based Propensity Score Matching (PSM) and model diagnostics were performed. Multicollinearity was evaluated, and sensitivity analyses were conducted. To ensure the correctness of the functional form, the heteroskedasticity-robust Ramsey's Regression Specification Error Test (RESET) was applied \citep{ramsey1969}. The test showed no signs of misspecification (\textit{Table \ref{tab:reset}, Appendix}).

\begin{figure}[htbp!]
    \caption{Dynamic Treatment Effects for Physician Count}
    \centering
    \includegraphics[scale=.3]{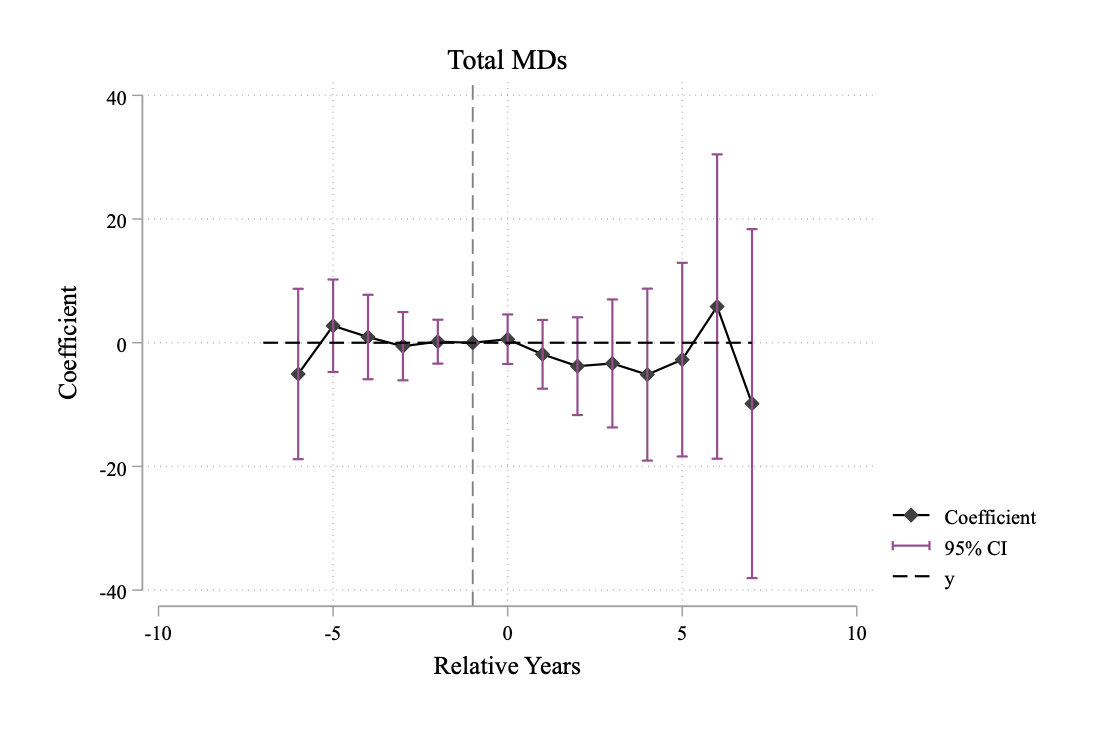}
    \label{fig:sunabraham}
\noindent
\begin{minipage}{\textwidth}
    \begin{spacing}{1.5}
    \footnotesize
    \justifying
    \textit{Note}: These plots show dynamic treatment effects on number of non-federal and federal physicians using \cite{SNA2021}, with a baseline at relative year = $-1$ (coefficient = 0). The figure confirms two things. First, at rel\_year = $-1$ (95\% CI includes 0), there is no significant effect. This supports No Anticipation and Parallel Trends in Ratio assumptions. Second, post treatment effects for the main treatment are negligible, which show that the framing effects need to be taken into account. For count outcomes with large means, the linear SA method approximates dynamic effects reasonably well. 
    \end{spacing}
\end{minipage}
\end{figure}

The ratio version of the parallel trends assumption was tested by conducting a PPML event study, where \enquote{relative time from treatment} was interacted with treatment indicator and standardized broadband variable. The results support the assumption (\textit{Column (a), Table \ref{tab:pretrends}, Appendix}). To test the no-anticipation assumption, PPML model with broadband interaction on the lag of count of MDs was estimated (\textit{Column (b), Table \ref{tab:pretrends}, Appendix}). In \textit{Figure \ref{fig:sunabraham}, Appendix}, the estimated dynamic treatment effects are presented using the linear \cite{SNA2021}, which approximates well for count outcomes, despite the main specification being nonlinear. The results support the assumption that future treatment does not affect current outcomes. Additionally, a placebo test for pre-treatment trends, conducted by assigning a fictitious treatment period, showed no significant pre-treatment differences between treated and control groups (\textit{Table \ref{tab:placebo}, Appendix}). Finally, PPML is robust to high variability and over-dispersion, ensuring unbiased estimates even with large deviations. While precision may decrease (e.g., wider confidence intervals), this does not affect the validity of our results. 

Some estimates for non-metro show higher standard errors. Non-metro counties differ significantly in their infrastructure, causing greater inherent variability in responses to policy changes in these areas. If baseline demand and physician count are low, even a small change in absolute numbers could cause higher treatment effect estimates with low precision. Since non-metro areas have both urban and rural counties, there is large variability in the estimates for non-metro areas, as shown in the results. Nevertheless, the use of cluster-robust standard errors at the state level accounts for heteroskedasticity and autocorrelation within states, and intragroup correlation due to the state-level treatment, giving more accurate standard errors.  

As an additional robustness check, I used a log-transformed min-max normalized and an arcsinh transformed broadband variable, both normally distributed, to address skewness in the original standardized variable (\textit{Table \ref{tab:lognormalandarsinh}, Appendix}). Despite differing magnitudes due to scaling, sign patterns and significance persisted, showing stable directional effects across all samples and areas, confirming that skewness did not unduly affect results and that findings are robust to alternative broadband specifications.


\vspace{2em}
\section{Conclusion}\label{sec:conclusion}

This study advances the price regulation literature by incorporating non-price and technological factors into the analysis of the impact of Telehealth Parity Laws (TPL) on physician counts, before the pandemic. The effects of these laws varied across states due to differences in regulatory framing and were significantly influenced by broadband availability. Challenging traditional price regulation theories, the study identifies technology-driven quality adjustments as a critical factor shaping outcomes, with policy effects differing between metro and non-metro areas. Empirical evidence underscores that these heterogeneous effects depend on regulatory design, local infrastructure, and the degree of reliance of physician specialties on telehealth. As telehealth expands post-pandemic, these insights call for policymakers to craft a balanced consumer- and provider-centric regulatory environment, considering consumers' role as primary input providers, while accounting for the interplay of regulations and local broadband, to enhance healthcare access and provision.


\vspace{1em}
\begin{spacing}{1.0}
\bibliographystyle{aea}
\bibliography{thbbdreferences}
\end{spacing}


\newpage
\setstretch{1.15}
\section{Appendix}

\subsection{Proof of Proposition I.A}
\label{subsec:proof1a}

Consider a production function \( Y = F(T, I) \) that is homogeneous of degree one, with inputs \( T \) (upstream, regulated input) and \( I \) (downstream input). The input supply functions are upward-sloping, with input prices \( r_T = \Gamma_T'(T) \) and \( r_I = \Gamma_I'(I) \), where \( \Gamma_T(\cdot) \) and \( \Gamma_I(\cdot) \) are convex. The full (competitive) price of output is
\[
P_Y = \frac{r_T T + r_I I}{Y}.
\]

Cost shares are \( s_T = \frac{r_T T}{r_T T + r_I I} \) and \( s_I = \frac{r_I I}{r_T T + r_I I} \), so \( s_T + s_I = 1 \). The input supply elasticities are
\[
\varepsilon_{\Gamma T} = \frac{\Gamma_T'(T)}{T \Gamma_T''(T)}, \qquad \varepsilon_{\Gamma I} = \frac{\Gamma_I'(I)}{I \Gamma_I''(I)},
\]
so that a larger \( \varepsilon \) denotes a more elastically supplied input. Let \( \nu_T \equiv 1/\varepsilon_{\Gamma T} \) and \( \nu_I \equiv 1/\varepsilon_{\Gamma I} \) denote the corresponding inverse supply elasticities, which govern the log-differentials of input prices through \( \hat{r}_T = \nu_T \hat{T} \) and \( \hat{r}_I = \nu_I \hat{I} \).
The elasticity of substitution between \( T \) and \( I \) is \( \sigma > 0 \). Logarithmic differentials are denoted \( \hat{X} = dX/X \).

\subsubsection*{Pivot Lemma}
Define $E(\rho)$ as the point on the unregulated supply curve at which physician reimbursement for telehealth equals the regulatory parameter, $\frac{r_T T}{Y} = \rho$. Such a point exists for $\rho \in [\rho_{min}, \rho_{max}]$, as stated in Section V, since $\frac{r_T T}{Y}$ varies continuously along the curve. $E(\rho)$ lies on the unregulated supply curve by this definition and satisfies the unregulated equilibrium conditions $FP$, $PF$, $IS$, $TS$, and $CMC$. At $E(\rho)$ the price regulation compliance condition $PRC$ also holds, so $E(\rho)$ satisfies the regulated equilibrium conditions $FP$, $PF$, $IS$, $TS$, and $PRC$, and the regulated supply locus passes through it as well. Rotation around $E(\rho)$ is therefore well defined.

\subsubsection*{Unregulated Equilibrium}

Firms minimize cost \( C = r_T T + r_I I \) subject to \( Y = F(T, I) \). The first-order condition is
\[
\frac{\partial F / \partial T}{\partial F / \partial I} = \frac{r_T}{r_I}.
\]
By Euler's theorem (since \( F \) is homogeneous of degree one),
\[
Y = \frac{\partial F}{\partial T} T + \frac{\partial F}{\partial I} I,
\]
so output elasticities equal cost shares: \( \theta_T = s_T \) and \( \theta_I = s_I \).

Zero profits imply \( P_Y \cdot Y = r_T T + r_I I \). Log-differentiating gives
\[
\hat{P}_Y + \hat{Y} = s_T (\hat{r}_T + \hat{T}) + s_I (\hat{r}_I + \hat{I}).
\]
From production,
\[
\hat{Y} = s_T \hat{T} + s_I \hat{I}.
\]
Input supply conditions are \( \hat{r}_T = \nu_T \hat{T} \) and \( \hat{r}_I = \nu_I \hat{I} \).

Input substitution satisfies
\[
\hat{T} - \hat{I} = \sigma (\hat{r}_I - \hat{r}_T) = \sigma (\nu_I \hat{I} - \nu_T \hat{T}).
\]
Rearranging yields
\[
\hat{T} (1 + \sigma \nu_T) = \hat{I} (1 + \sigma \nu_I),
\]
so
\[
\alpha \equiv \frac{\hat{I}}{\hat{T}} = \frac{1 + \sigma \nu_T}{1 + \sigma \nu_I}.
\]

Express output in terms of \( \hat{T} \):
\[
\hat{Y} = (s_T + s_I \alpha) \hat{T}.
\]
Substitute into the zero-profit equation:
\[
\hat{P}_Y + \hat{Y} = s_T (1 + \nu_T) \hat{T} + s_I (1 + \nu_I) \alpha \hat{T}.
\]
Thus,
\[
\hat{P}_Y = (s_T \nu_T + s_I \alpha \nu_I) \hat{T}.
\]
The unregulated supply elasticity is
\[
\eta_{\text{unreg}} = \frac{\hat{Y}}{\hat{P}_Y} = \frac{s_T + s_I \alpha}{s_T \nu_T + s_I \alpha \nu_I}.
\]

\subsubsection*{Regulated Equilibrium}

Regulation fixes the upstream payment per unit of output: \( \frac{r_T T}{Y} = \rho \). Log-differentiating gives
\[
\hat{r}_T + \hat{T} = \hat{Y}.
\]
With \( \hat{r}_T = \nu_T \hat{T} \),
\[
\hat{Y} = (1 + \nu_T) \hat{T}.
\]

From production,
\[
\hat{Y} = s_T \hat{T} + s_I \hat{I} \implies s_I \hat{I} = (\nu_T + s_I) \hat{T},
\]
so
\[
\hat{I} = \frac{\nu_T + s_I}{s_I} \hat{T}.
\]

Substitute into zero profits:
\[
\hat{P}_Y + \hat{Y} = s_T (\hat{r}_T + \hat{T}) + s_I (\hat{r}_I + \hat{I})
\]
\[
\hat{P}_Y + (1 + \nu_T) \hat{T} = s_T (1 + \nu_T) \hat{T} + s_I (1 + \nu_I) \frac{\nu_T + s_I}{s_I} \hat{T}.
\]
The right-hand side simplifies to
\[
s_T (1 + \nu_T) \hat{T} + (1 + \nu_I) (\nu_T + s_I) \hat{T}.
\]
Hence,
\[
\hat{P}_Y = \bigl[ \nu_T (1 - s_I + \nu_I) + \nu_I s_I \bigr] \hat{T}.
\]
The regulated supply elasticity is
\[
\eta_{\text{reg}} = \frac{\hat{Y}}{\hat{P}_Y} = \frac{1 + \nu_T}{\nu_T (1 - s_I + \nu_I) + \nu_I s_I}.
\]

\subsubsection*{Comparison of Elasticities}

Define
\[
A = s_T + s_I \alpha, \quad B = s_T \nu_T + s_I \alpha \nu_I,
\]
\[
C = 1 + \nu_T, \quad D = \nu_T (1 - s_I + \nu_I) + \nu_I s_I.
\]
Then
\[
\eta_{\text{unreg}} - \eta_{\text{reg}} = \frac{A}{B} - \frac{C}{D} = \frac{AD - BC}{BD}.
\]
Under the maintained assumptions (\( 0 < s_I < 1 \), \( \nu_T > 0 \), \( \nu_I > 0 \), \( \sigma > 0 \)), the denominator satisfies \( BD > 0 \). Direct expansion of the numerator yields the exact factorization
\[
AD - BC = \left(\nu_I - \nu_T\right)\left(1 - s_I\right)\frac{Q}{1 + \sigma \nu_I}, \qquad Q = \nu_T\left(1 - \sigma s_I\right) + \sigma \nu_I \left(\nu_T + s_I\right).
\]
Since \( 1 - s_I > 0 \) and \( 1 + \sigma\nu_I > 0 \), the sign of the difference is
\[
\operatorname{sign}\left(\eta_{\text{unreg}} - \eta_{\text{reg}}\right) = \operatorname{sign}\left(\nu_I - \nu_T\right)\cdot \operatorname{sign}\left(Q\right).
\]
Converting to supply elasticities, \( \operatorname{sign}(\nu_I - \nu_T) = \operatorname{sign}(\varepsilon_{\Gamma T} - \varepsilon_{\Gamma I}) \), and multiplying \( Q \) by \( \varepsilon_{\Gamma T}\varepsilon_{\Gamma I} > 0 \) gives \( \operatorname{sign}(Q) = \operatorname{sign}(\tilde{Q}) \) with
\[
\tilde{Q} = \varepsilon_{\Gamma I} + \sigma\left[1 + s_I\left(\varepsilon_{\Gamma T} - \varepsilon_{\Gamma I}\right)\right].
\]
Therefore
\[
\operatorname{sign}\left(\eta_{\text{unreg}} - \eta_{\text{reg}}\right) = \operatorname{sign}\left(\varepsilon_{\Gamma T} - \varepsilon_{\Gamma I}\right)\cdot \operatorname{sign}\left(\tilde{Q}\right).
\]
Two cases follow. First, if \( \varepsilon_{\Gamma T} \geq \varepsilon_{\Gamma I} \), then \( s_I(\varepsilon_{\Gamma T} - \varepsilon_{\Gamma I}) \geq 0 \), so \( \tilde{Q} \geq \varepsilon_{\Gamma I} + \sigma > 0 \) holds automatically, and \( \eta_{\text{unreg}} > \eta_{\text{reg}} \). The regulated locus is steeper than the unregulated supply curve, a counter-clockwise rotation around \( E(\rho) \). Second, if \( \varepsilon_{\Gamma T} < \varepsilon_{\Gamma I} \), then \( \eta_{\text{unreg}} < \eta_{\text{reg}} \) and the rotation is clockwise provided \( \tilde{Q} > 0 \). A sufficient condition is \( s_I\left(\varepsilon_{\Gamma I} - \varepsilon_{\Gamma T}\right) \leq 1 \), since the bracketed term is then non-negative and \( \tilde{Q} \geq \varepsilon_{\Gamma I} > 0 \) for any \( \sigma \). \quad QED.

\textit{Remark}: The clockwise branch can fail only when \( s_I\left(\varepsilon_{\Gamma I} - \varepsilon_{\Gamma T}\right) \) substantially exceeds one, which requires the in-person input to be supplied far more elastically than the telehealth input while simultaneously dominating the cost share. In the present setting the in-person input consists of patient travel, waiting, and time, which the telehealth literature characterizes as the principal burden of access rather than an elastically supplied factor, so the empirically relevant region satisfies the condition.

{\centering \textbf{\large Substitution Effects of Broadband and Demand for Healthcare Services}\par}

The consumers maximize utility \( U(Y, X) \), where \( Y \) is healthcare services and \( X \) is a composite good, subject to the full-income constraint: \( (P_Y + w \tau) Y + (q + w s) X \leq M \), with \( M = y + w T \) as full income, \( y \) as unearned income, \( T \) as total time, \( \Pi_Y = P_Y + w \tau \) as the total price of \( Y \), \( \Pi_X = q + w s \) as the total price of \( X \), \( P_Y \) and \( q \) as money prices, \( w \tau \) and \( w s \) as time prices, and \( w \) as the wage rate. Broadband increases the opportunity cost of time, raising \( w \tau \).

Utility Maximization: \( \mathcal{L} = U(Y, X) + \lambda \left[ M - (P_Y + w \tau) Y - (q + w s) X \right]. \)

FOCs: \(\dfrac{\partial \mathcal{L}}{\partial Y} = U_Y - \lambda (P_Y + w \tau) = 0 \Rightarrow U_Y = \lambda \Pi_Y, \)\\
\(\dfrac{\partial \mathcal{L}}{\partial X} = U_X - \lambda (q + w s) = 0 \Rightarrow U_X = \lambda \Pi_X, \quad \dfrac{\partial \mathcal{L}}{\partial \lambda} = M - \Pi_Y Y - \Pi_X X = 0. \)

Thus, the marginal rate of substitution equals the relative price:
\( \dfrac{U_Y}{U_X} = \dfrac{\Pi_Y}{\Pi_X} = \dfrac{P_Y + w \tau}{q + w s}. \)

\subsection{\textbf{Equation B1 Proof}}\label{sec:proofb1}

Given: $\max_{Y, X} U(Y, X)$, subject to: $\Pi_Y Y + \Pi_X X \leq M$, where: $\Pi_Y = P_Y + w \tau$, $\Pi_X = q + w s$, $M = y + w T$. The elasticities with respect to money price (\(P_y\)), time price (\({w \tau}\)) and full price (\(\Pi_Y\)) are given by: \(\varepsilon_Y^{P_Y} = \dfrac{\partial Y}{\partial P_Y} \cdot \dfrac{P_Y}{Y},\quad \varepsilon_Y^{w\tau} = \dfrac{\partial Y}{\partial (w \tau)} \cdot \dfrac{w \tau}{Y}, \quad \varepsilon_Y^{\Pi_Y} = \dfrac{\partial Y}{\partial \Pi_Y} \cdot \dfrac{\Pi_Y}{Y}.\)\\

Since, \(\Pi_Y = P_Y + w \tau, \quad \dfrac{\partial \Pi_Y}{\partial P_Y} = 1, \quad \dfrac{\partial \Pi_Y}{\partial (w \tau)} = 1\). Thus, we get: \(\dfrac{\partial Y}{\partial P_Y} = \dfrac{\partial Y}{\partial \Pi_Y}, \quad \dfrac{\partial Y}{\partial (w \tau)} = \dfrac{\partial Y}{\partial \Pi_Y}.\)

Re-writing the elasticities, we get: \(\varepsilon_Y^{P_Y} = \dfrac{\partial Y}{\partial \Pi_Y} \cdot \dfrac{P_Y}{Y}, \quad \varepsilon_Y^{w\tau} = \dfrac{\partial Y}{\partial \Pi_Y} \cdot \dfrac{w \tau}{Y}.
\)

Thus: \(\varepsilon_Y^{\Pi_Y} = \dfrac{\partial Y}{\partial \Pi_Y} \cdot \dfrac{\Pi_Y}{Y} \implies  \dfrac{\partial Y}{\partial \Pi_Y} = \varepsilon_Y^{\Pi_Y} \cdot \dfrac{Y}{\Pi_Y}\). Substituting: \(\varepsilon_Y^{P_Y} = \varepsilon_Y^{\Pi_Y} \cdot \dfrac{P_Y}{\Pi_Y}\) and \(\varepsilon_Y^{w\tau}= \varepsilon_Y^{\Pi_Y} \cdot \dfrac{w \tau}{\Pi_Y}\), we get: \[\varepsilon_Y^{w\tau} \mathrel{\substack{<\\>}} \varepsilon_Y^{P_Y} \quad \text{as} \quad \dfrac{w \tau}{\Pi_Y} \mathrel{\substack{<\\>}} \dfrac{P_Y}{\Pi_Y} \implies  \varepsilon_Y^{w\tau} \mathrel{\substack{<\\>}} \varepsilon_Y^{P_Y} \quad \text{as} \quad  w \tau \mathrel{\substack{<\\>}} P_Y.\] QED.

\subsection{\textbf{Equation B2 Proof:}}\label{sec:proofb2}
Utility maximization: \( \max_{Y, X} U(Y, X) \quad \text{s.t.} \quad (P_Y + w \tau) Y + (q + w s) X \leq M \), \\
where \( M = y + w T \), with \( y \) as non-earned income and \( T \) as total time. \\

Total prices: \( \Pi_Y = P_Y + w \tau \), \( \Pi_X = q + w s \). FOC: \(\dfrac{U_Y}{U_X} = \dfrac{\Pi_Y}{\Pi_X} = \dfrac{P_Y + w \tau}{q + w s}. \)\\

Optimal demand: \( Y^* = Y(\Pi_Y, \Pi_X, M) \). Further, \(\dfrac{\partial}{\partial w} \left( \dfrac{\Pi_Y}{\Pi_X} \right) = \dfrac{\tau q - s P_Y}{(q + w s)^2}. \) \\

If \( \dfrac{w s}{w s + q} > \dfrac{w \tau}{w \tau + P_Y} \), then \( \dfrac{s}{q} > \dfrac{\tau}{P_Y} \). Thus: \( \tau q - s P_Y < 0 \) $\implies$ \( \dfrac{\Pi_Y}{\Pi_X} \) decreases.\\

Since \( \dfrac{U_Y}{U_X} = \dfrac{\Pi_Y}{\Pi_X} \), a decrease in \( \dfrac{\Pi_Y}{\Pi_X} \) implies the consumer substitutes toward \( Y \), increasing its demand as \( X \) is more time-intensive. QED.

\subsection{\textbf{Proof of Proposition 2}}\label{sec:prop2proof}

\textit{Setup:} Consumers maximize utility \( U(Y, X) \), where \( Y \) is healthcare services and \( X \) is a composite good, subject to the full-income constraint: \((P_Y + w \tau) Y + (q + w s) X \leq M,\)

where: \( M = y + w T \) is full income, with \( y \) as unearned income and \( T \) as total time; \( \Pi_Y = P_Y + w \tau \) is the full price of healthcare, with \( P_Y \) as  money price and \( w \tau \) as time price, \( w \) being the wage rate and \( \tau \) the time per unit of healthcare; \( \Pi_X = q + w s \) is the full price of \( X \), with \( q \) as money price and \( w s \) as time price. The money price is: \(P_Y = E_{oop} + \frac{r}{Y},\)
where \( E_{oop} = \frac{D}{Y} + C_o \) (deductibles plus copayments), and \( r \) is the insurance premium. Thus, the full price is:
\(\Pi_Y = P_Y + w \tau\).
Broadband (\( B \)) reduces \( \tau \) by enabling telehealth, and Cost Parity (\( \gamma = 1 \)) increases \( P_Y \) by raising \( E_{oop-T}\).

\textbf{Demand:} The aggregate healthcare service \(Y\) is a composite of telehealth (\(Y_T\)) and in-person (\(Y_I\)) services, aggregated by a linearly homogeneous function. The consumer chooses the optimal quantities of \(Y\) and \(X\) to maximize utility, while the mix of \(Y_T\) and \(Y_I\) is determined by cost minimization for a given level of \(Y\).

The consumer maximizes utility subject to the budget constraint: \( (P_Y + w \tau) Y + (q + w s) X \leq M\),
where \(P_Y\) is the money price of the composite healthcare good, \(w \tau\) is the time price of healthcare, \(q\) is the money price of \(X\), \(w s\) is the time price of \(X\), and \(M\) is the consumer's full income. The price \(P_Y\) is derived from the cost-minimizing combination of \(Y_T\) and \(Y_I\), given their respective prices \(P_{YT}\) and \(P_{YI}\). For a fixed level of \(Y\), the consumer minimizes the cost \(P_{YT} Y_T + P_{YI} Y_I\), and the resulting minimum cost defines \(P_Y\), which is independent of the level of \(Y\) by linear homogeneity. However, for utility maximization, \(P_Y\) is treated as a given parameter.
The first-order conditions (FOCs) with respect to \(Y\) and \(X\), are:\(\dfrac{\partial U}{\partial Y} = \lambda (P_Y + w \tau), \quad \dfrac{\partial U}{\partial X} = \lambda (q + w s),\) where \(\lambda\) is the Lagrange multiplier. The marginal rate of substitution (MRS) between \(Y\) and \(X\) is: \(\dfrac{\partial U / \partial Y}{\partial U / \partial X} = \dfrac{P_Y + w \tau}{q + w s}.\)

Consider a quasilinear utility where broadband and Cost Controls affect curvature, with the full price of \(Y\) denoted \(\Pi_Y = P_Y + w \tau\) and the full price of \(X\) normalized to one, \(\Pi_X = q + w s = 1\), so that the demand below is exact. \[U = \dfrac{Y^{1 - \eta(B, \gamma)}}{1 - \eta(B, \gamma)} + X, \quad 0 < \eta < 1, \quad Y = \Pi_Y^{-1 / \eta(B, \gamma)}\]

The full-price elasticity of demand is exact, \(\varepsilon_Y^{\Pi_Y} = -\dfrac{1}{\eta(B, \gamma)}\), while the money-price elasticity carries the factor \(P_Y/\Pi_Y\), \(\varepsilon_Y^{P_Y} = -\dfrac{1}{\eta(B, \gamma)}\dfrac{P_Y}{\Pi_Y}\). The comparative statics below are stated for \(\varepsilon_Y^{\Pi_Y}\), which equals \(-1/\eta\) under the quasilinear form. Since \(P_Y = \Pi_Y - w\tau\) at every \(Y\) and \(w\tau\) does not vary with \(Y\), slopes at any given \(Y\) are identical, so the steeper and flatter comparisons below apply unchanged in the \([Y, P_Y]\) plane of \textit{Figure \ref{fig:floorceiling}}.\\

\textbf{Demand Curvature Assumption:} The elasticity parameter \(\eta(B,\gamma)\) is assumed to be differentiable, to take values in \((0,1)\) for every \(B\) and \(\gamma\), and to satisfy \(\partial \eta/\partial B > 0\) when \(\gamma = 0\) and \(\partial \eta/\partial B < 0\) when \(\gamma = 1\). All results below use only these range and monotonicity properties. A parametrization with these properties for every value of \(B\) is
\[
\eta(B, \gamma) = \Bigl[1 + e^{-\left(\eta_0 + \eta_1 B - \eta_2 \gamma B\right)}\Bigr]^{-1}, \quad \eta_0 > 0, \quad \eta_2 > \eta_1 > 0,
\]
where \( B \) is broadband access, \( \gamma \) is a Cost Parity indicator (\( \gamma = 1 \) if active, \( 0 \) otherwise), and the index \(\eta_0 + \eta_1 B - \eta_2 \gamma B\) is increasing in \(B\) without Cost Parity and decreasing in \(B\) with Cost Parity.

\textbf{Functional Form Justification:} Keeping in mind \textit{equation \ref{eq:elasticityrelative}}: 
\begin{itemize}
    \item \( +\eta_1 B \): Broadband reduces the time price of access, and demand is modeled as less sensitive to the full price at higher broadband, with a positive coefficient.
    \item \( -\eta_2 \gamma B \): Cost Parity increases monetary costs, and demand is modeled as more sensitive to the full price under Cost Parity, especially in high-\( B \) areas, with a negative interaction term.
\end{itemize}

\textbf{Broadband’s Effect}: Without Cost Parity (\(\gamma = 0\)): \\
\[\dfrac{\partial \eta(B,0)}{\partial B} > 0 \implies \dfrac{\partial |\varepsilon_Y^{\Pi_Y}|}{\partial B} = -\dfrac{\partial \eta / \partial B}{\eta(B,0)^2} < 0\]

reducing \( |\varepsilon_Y^{\Pi_Y}| \), steepening the demand curve (\(  Y_U = D^{M(B\geq1)}_{U}(P_Y)\), \textit{Figure \ref{fig:floorceiling}}).\\

\textbf{With Cost Parity}: (\(\gamma = 1\)):\label{subsec:cpeffects} Since \( \eta_2 > \eta_1 \), the index is decreasing in \(B\), so \(\partial \eta(B,1)/\partial B < 0\) and \(\eta(B,1) < \eta(B,0)\) for every \(B > 0\). Hence

\[\dfrac{\partial |\varepsilon_Y^{\Pi_Y}|}{\partial B}\Big|_{\gamma=1} = -\dfrac{\partial \eta/\partial B}{\eta(B,1)^2} > 0 \quad \text{and} \quad |\varepsilon_Y^{\Pi_Y}|_{\gamma=1} > |\varepsilon_Y^{\Pi_Y}|_{\gamma=0}\]

increasing \( |\varepsilon_Y^{\Pi_Y}| \), flattening the demand curve, with a stronger effect at higher \( B \).

\begin{itemize}
\item Unregulated Case at Mean $ B = 0 $, $ \gamma = 0 $: The demand curve has elasticity $ \varepsilon_Y^{\Pi_Y} = -\dfrac{1}{\eta(0,0)} $.

\item Higher Broadband ($ B > 0 $), no Cost Parity ($ \gamma = 0 $): $ \eta(B,0) > \eta(0,0) $, so $ |\varepsilon_Y^{\Pi_Y}| = \dfrac{1}{\eta(B,0)} < \dfrac{1}{\eta(0,0)} $, leading to a steeper demand curve (less elastic, clockwise rotation).

\item Higher Broadband ($ B \geq 1 $), with Cost Parity ($ \gamma = 1 $): the index $ \eta_0 + (\eta_1 - \eta_2) B $ lies below both $ \eta_0 $ and $ \eta_0 + \eta_1 B $ (where $ \eta_2 > \eta_1 > 0 $), so $ \eta(B,1) < \eta(0,0) $ and $ \eta(B,1) < \eta(B,0) $. 

Thus, $ |\varepsilon_Y^{\Pi_Y}| > |\varepsilon_Y^{\Pi_Y}|_{\gamma=0} $, resulting in a flatter curve than without Cost Parity.

Since $ \eta(B,1) < \eta(0,0) $, $ |\varepsilon_Y^{\Pi_Y}| > \dfrac{1}{\eta(0,0)} $, making the curve flatter than the unregulated baseline (more elastic, counter-clockwise rotation), as illustrated by \(Y_R = D^{M(B>>1)}_{CP}(P_Y; \gamma)\) in \textit{Figure \ref{fig:floorceiling}}.
\end{itemize}

The empirical results in \textit{Table \ref{tab:main_acrt}} confirm the theoretical predictions of the model. Under Cost Parity, the $ATT$ declines as broadband (\( B \)) rises within the framing's treated support, from 0.0132 at \( B = 0 \) to 0.0026 at \( B = 1 \), statistically indistinguishable from zero, with the negative $ACRT$ of $-0.0106$ capturing the continued decline. This can be confirmed by the Cost Parity profile in \textit{Figure \ref{fig:attprofiles}}. This aligns with the model's implication that higher broadband amplifies the counterclockwise rotation of the demand curve, reducing the equilibrium quantity demanded more significantly.

\textbf{With Cost Ceiling:}\label{subsec:cceffects} For Cost Ceiling, let's denote \(\gamma = \gamma_{CC}\), where \(0 < \gamma_{CC} < 1\). 

Then the index becomes \(\eta_0 + (\eta_1 - \eta_2 \gamma_{CC}) B\), so \(|\varepsilon_Y^{\Pi_Y}| = \dfrac{1}{\eta(B, \gamma_{CC})}\) with \(\eta(B, \gamma_{CC})\) governed by that index. \\

Since \(\gamma_{CC} < 1\), the reduction, \(-\eta_2 \gamma_{CC} B\), under Cost Ceiling, is smaller in magnitude than the reduction, \(-\eta_2 B\), under Cost Parity. Thus, \(\eta_1 - \eta_2 \gamma_{CC} > \eta_1 - \eta_2\), making \(\eta\) larger than under Cost Parity, and \(|\varepsilon_Y^{\Pi_Y}|\) smaller. The demand curve flattens less than under Cost Parity. If \(\gamma_{CC}\) is small (close to 0), the term \(-\eta_2 \gamma_{CC} B\) becomes negligible, and the index approaches \(\eta_0 + \eta_1 B\), resembling the unregulated case, where elasticity decreases with \(B\).

As illustrated in \textit{Figure \ref{fig:floorceiling}}, Cost Ceiling (\(\gamma < 1\)) introduces a milder counter-clockwise rotation (flattening) of the demand curve compared to Cost Parity (\(\gamma = 1\)), with the effect diminishing as \(\gamma\) approaches 0. Broadband’s tendency to steepen the curve (clockwise rotation) at low \(\gamma_{CC}\) (curve \(Y_R = D^{M(B\geq1)}_{CC}(P_Y; \gamma) \) in \textit{Figure \ref{fig:floorceiling}}), resulting in a nearly neutral impact on demand elasticity. For a non-binding Cost Ceiling, the $ATT$ remains near zero across higher levels of \( B \), consistent with broadband potentially offsetting the effect through a clockwise rotation.

These findings align with the model’s prediction of a nearly neutral effect on demand elasticity, demonstrated in \textit{Table \ref{tab:main_acrt}} with near-zero $ATT$, insensitive to \(B\), for Cost Ceiling, illustrated in \textit{Figure \ref{fig:attprofiles}}. The coefficients are not statistically significant and have high standard errors, reflecting the policy’s unpredictable and mild effect. This is due to small net impact, offset by broadband’s influence and variability in $\gamma_{CC}$, making consistent detection challenging. This affirms the theoretical and empirical coherence of our analysis.

\newpage 

\section{Online Appendix}
\subsection{Additional Figures}

\vspace{2em}
\begin{figure}[h!]
    \caption{Textbook Treatment of Price Controls}
    \vspace{1em}
    \centering
    \includegraphics[scale=.23]{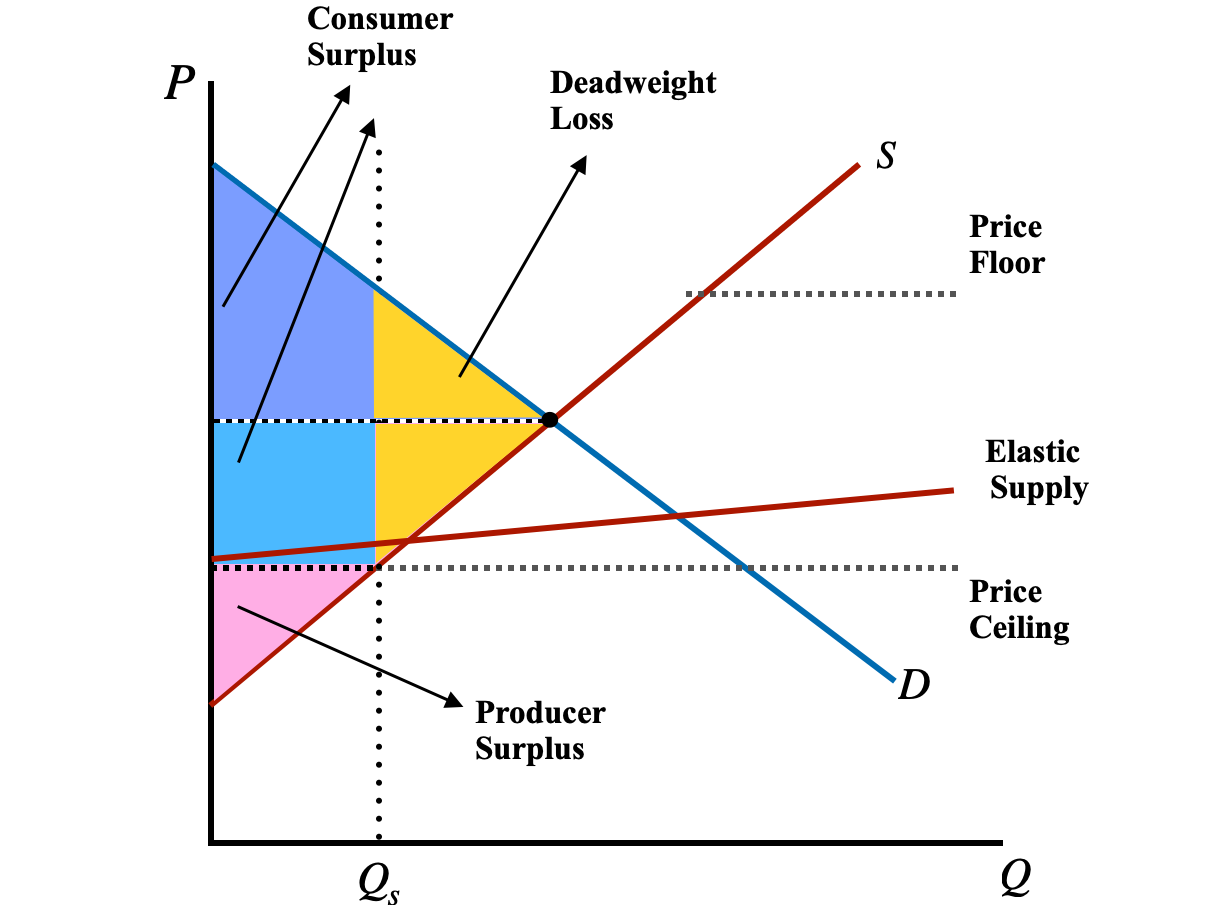}
    \label{fig:conventional_limitation}
\noindent
\begin{minipage}{\textwidth}
    \begin{spacing}{1.5}
    \footnotesize
    \justifying
    \vspace{1em}
    \textit{Note:} The figure illustrates the impact of Price Controls on quantity as per conventional models in standard textbooks \citep{all2018, KrugmanWells2020}, which are inadequate in dealing with the situation portrayed in \textit{Figure \ref{fig:pcillustration}}. In these models, Price Ceiling is necessarily below the unregulated equilibrium price. In case of TPLs, both Price Floor and Price Ceiling are above the unregulated equilibrium telehealth price, since pre-regulation $MERR-T < MERR-I$. According to these models, Price Ceiling necessarily causes excess demand, and Price Floor causes excess supply. Thus, these models do not account for quality adjustments, which could eliminate excess demand or supply. Both Price Ceiling and Price Floor cause reduced quantity. This runs in contrast to \cite{Mulligan2024}, where Floor and Ceiling have opposite effect. I show that the effect depends on whether the control is binding or not---the effects of binding Price Floor and non-binding Price Ceiling are similar. As per these models, Price Controls do not affect supply and demand curves. However, Price Controls do rotate supply curves and the Cost Controls rotate demand curves. These models cannot account for the role of technology, which affects input supply elasticity and opportunity cost of time. I show how broadband as a technological mediator, influences rotation of supply and demand curves. As per the conventional models, demand increase only adds to the shortage and sellers don’t react to it. Thus, demand is rendered irrelevant. However, in this study, the rotations in demand are a determinant of equilibrium quantity. Lastly, these models cannot account for implications of the third party insurer and the wedge created between prices posted and costs incurred by consumers. This study recognizes this wedge between prices posted and costs and their respective regulations.  
    \end{spacing}
\end{minipage}
\end{figure}

\vspace{1em}
\begin{figure}[h!]
    \centering
        \caption{Comparison with Supply Chain Framework}
    \label{fig:comparison}
    \begin{minipage}[t]{0.45\textwidth}
        \centering
\includegraphics[width=\textwidth,height=0.25\textheight]{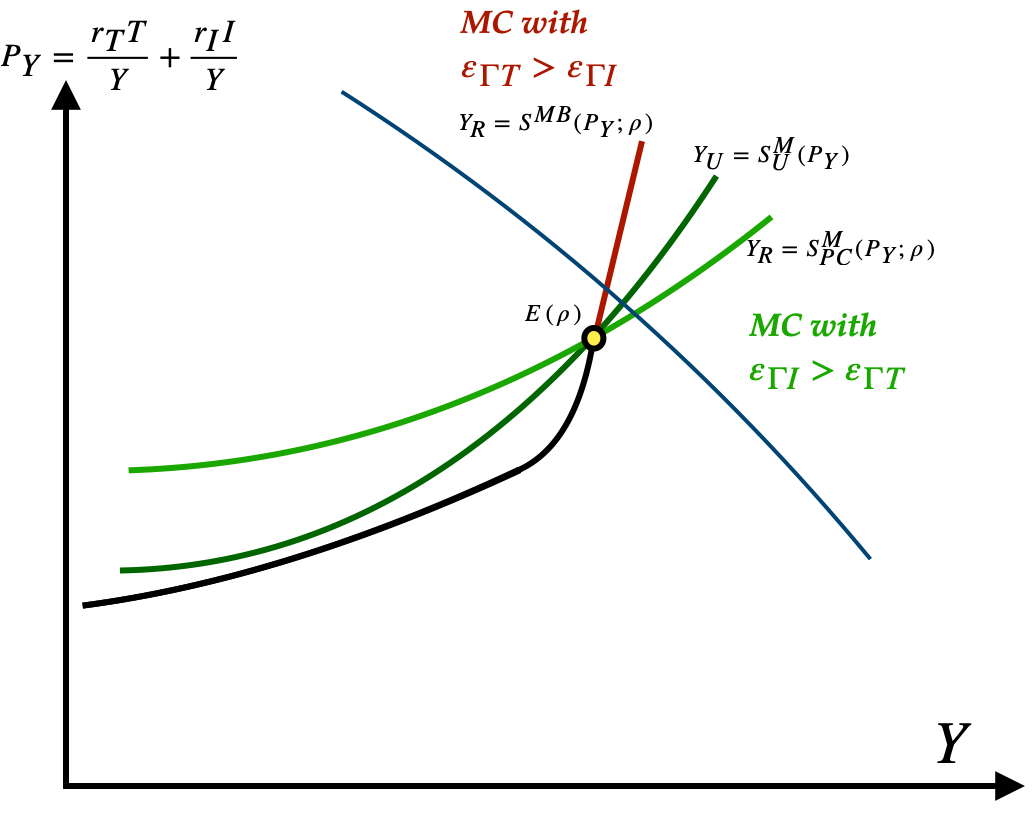}
        \subcaption[]{}
    \end{minipage}
    \hspace{0.01\textwidth} 
    \begin{minipage}[t]{0.48\textwidth}
        \centering
        \includegraphics[width=\textwidth,height=0.25\textheight]{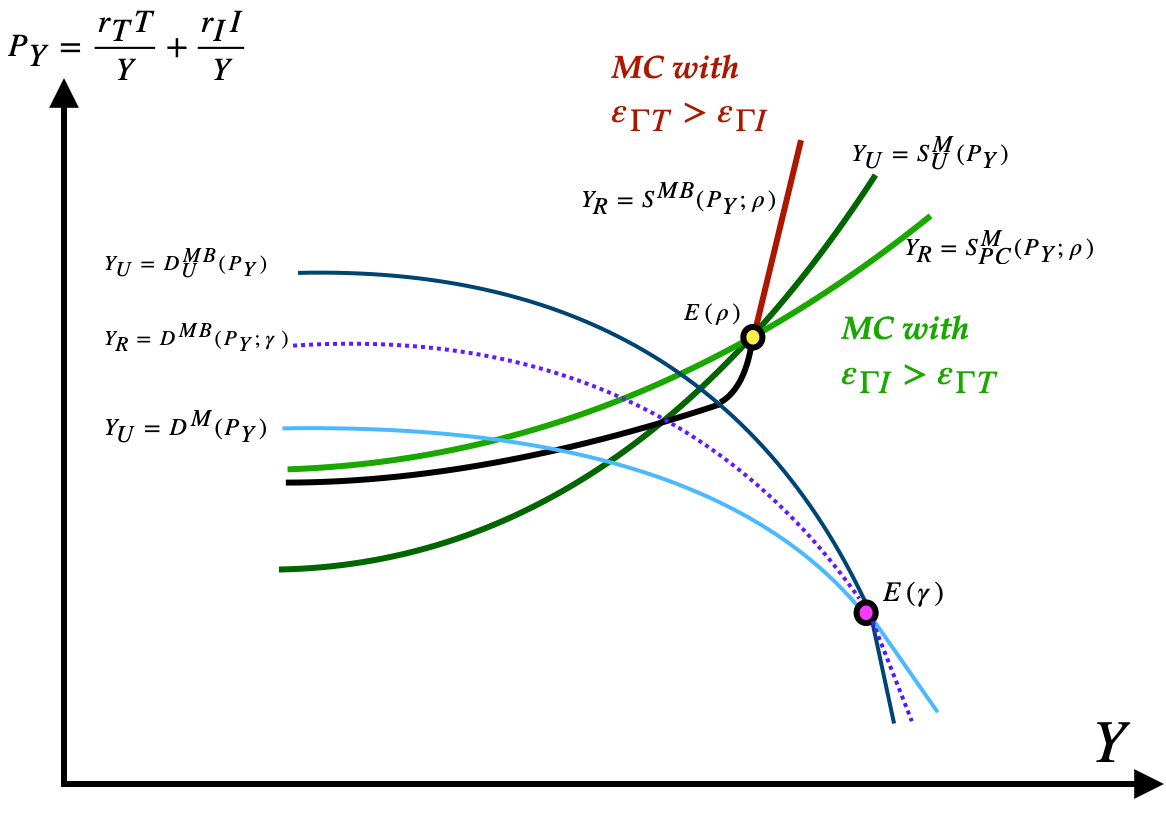}
        \subcaption[]{}
    \end{minipage}
    \noindent
    \begin{minipage}{\textwidth}
        \begin{spacing}{1.5}
        \footnotesize
        \justifying
        \textit{Note}: Panel (a) shows the set-up in \cite{Mulligan2024}. The emphasis is on the supply chain approach and the rotations of the supply curve along a static unregulated demand curve $ Y_U = D(P_Y)$, which is assumed to lie above $E(\rho)$. Panel (b) illustrates the basic approach in this study and the contrast with Mulligan's approach: 1) The demand curve is not static but rotates around $E(\gamma)$, depending on broadband level and the type of Cost Control specified. 2) The original position of the unregulated demand curve doesn't have to be above $E(\rho)$. In our case, since pre-regulation out-of-pocket costs for telehealth is less than that of in-person, the unregulated demand curve lies below $E(\rho)$. 3) Cost Controls rotate the demand curve. 4) Broadband facilitates access to healthcare. Thus, regulation, along with higher broadband, rotates the demand curve clockwise relative regulated demand curve at mean broadband. 5) Broadband (denoted by superscript B) increases the elasticity of telehealth such that $\varepsilon_{\Gamma T} > \varepsilon_{\Gamma I}$ (see \hyperref[ass:betsea]{\textit{BETSEA}}), causing the previously unregulated supply curve (dark green) to rotate counter-clockwise (red-black curve). Lack of broadband would lead to $\varepsilon_{\Gamma I} > \varepsilon_{\Gamma T}$, causing the previously unregulated supply curve (dark green) to rotate clockwise (light green curve). 6) The demand curve is non-linear. It becomes more elastic as $P_Y$ rises. 
        \end{spacing}
    \end{minipage}
\end{figure}

\clearpage

\subsection{Cost Regulation Parameter and Full Price}\label{subsec:costregfullprice} The full monetary price \( P_Y \) that consumers face is not just direct expenditures such as out-of-pocket costs, but also premium. The Premiums \( r \) are divided into the provider share \( S_p \), the insurer's profit share \( S_n \), and administrative costs share \( S_A \), where \( S_p + S_n + S_A = 1 \), shape the full price \( P_Y \) that consumers encounter. The full price \( P_Y \) is a sum of the out-of-pocket expenditure \( E_{oop} \) and the insurance premium per healthcare service unit, given by: \( P_Y = E_{oop} + \frac{r}{Y} \). The premium per service \( \frac{r}{Y} \) is allocated as \( \frac{S_p \cdot r}{Y} \) towards providers' reimbursements, \( \frac{S_n \cdot r}{Y} \) as the insurer's profit, and \( \frac{S_A \cdot r}{Y} \) for administrative costs.\footnote{The third-party insurer could be public or private. The reimbursement policies of private insurer follows the footsteps of Medicare \citep{CG2017}.} Here the focus is on these direct expenditures, excluding the value of consumer inputs. However, since telehealth and in-person services incorporate elements managed by both consumers and providers, the full price \(P_Y\) can also be analyzed through the value produced using both inputs.

Here, \( E_{oop} = \frac{D}{Y} + C_o \), where \( D \) denotes the annual deductible and \( C_o \) represents the copayment and coinsurance for each unit of healthcare service.\footnote{If coinsurance is established as 20\% of the total premium \( r \), the out-of-pocket cost per visit is expressed as \( E_{oop} = \frac{D}{Y} + C_o \), with \( D \) as the annual deductible and \( C_o \) as the total cost-sharing per service. Here, \( C_o = c_{\text{fixed}} + 0.2 S \), where \( S \) is the service cost per visit, \( c_{\text{fixed}} \) is the fixed copayment (a set dollar amount per service), and \( 0.2 S \) is the coinsurance (20\% of the service cost \( S \)).} The out-of-pocket cost is broken down as \(E_{oop} = \frac{D}{Y} + c_{\text{fixed}} + 0.2 S \), where \(D=\) is the annual deductible spread across all services, \( c_{\text{fixed}} \) is the fixed copayment per service, and \( 0.2 S \) is the coinsurance, which is 20\% of the service cost \(S\).\footnote{While \(S\) encompasses the total cost of the medical service, including physician fees, facility costs, equipment, and administrative overhead, physician reimbursement is solely the amount paid to the doctor for their professional services. These two are not directly tied due to factors such as insurance negotiations, government fee schedules, and overhead costs. Therefore, changes in physician reimbursement do not directly affect \( S \), ensuring that our model of Cost Controls and demand remain robust to variations in physician reimbursement rates.} Consumers choose between telehealth services (\(Y_T\)) and in-person services (\(Y_I\)), with total healthcare consumption defined as \(Y = Y_T + Y_I\). The total out-of-pocket expenditure is given by \(E_{oop}^{Total} = E_{oop-T} \cdot Y_T + E_{oop-I} \cdot Y_I\), where \(E_{oop-T}\) and \(E_{oop-I}\) represent the out-of-pocket costs for telehealth and in-person services, respectively. The average out-of-pocket cost per unit is then \(E_{oop} = \frac{E_{oop-T} \cdot Y_T + E_{oop-I} \cdot Y_I}{Y}\). Under Cost Parity, where \( \gamma = MECR-I \), telehealth expenses are aligned with the Market Equilibrium Cost Rate for in-person services (\( MECR-I \)), and under Cost Ceiling, where \( \gamma \leq MECR-I \), telehealth consumer costs are capped. Before regulation, telehealth out-of-pocket costs for telehealth were lower than those for in-person (\( E_{oop-T} \) < \( E_{oop-I} \)). Cost Parity raises telehealth out-of-pocket costs such that, \( E_{oop-T} = E_{oop-I} \). At lower broadband levels, telehealth is less elastically supplied already and a Cost Parity is unlikely to cause a reduction in demand. This aligns with the supply chain framework, where telehealth input is predominantly controlled by physicians and in-person input relies on consumer effort, where consumer bears the dis-utility. In settings with stable premiums, \( \gamma \) closely mirrors \( P_Y \), streamlining demand analysis.

\vspace{2em}
\section{\textit{Estimates from Table IV} and \textit{Table III}}\label{sec:acrtclarification}

\vspace{1em}
The $ACRT$ estimates from \textit{Table \ref{tab:main_acrt}} are similar to triple interaction coeffcient estimates from \textit{Table \ref{tab:withbbd}}, though not identical. In the PPML model from \textit{equation \ref{eq:ppmltriple}}, the $ATT$ at a specific broadband level \( B \) is defined as:
\[
ATT(B) = \exp(\beta_2 + \beta_1 B) - 1,
\]
where \( \beta_2 \) denotes the baseline effect, and \( \beta_1 \) represents the interaction coefficient modulating the effect of the standardized broadband variable \( B \). The Average Causal Response on the Treated (ACRT) measures the instantaneous rate of change of the ATT with respect to \( B \):
\[
 ACRT = \dfrac{\partial{[\exp(\beta_2 + \beta_1 B) - 1]}}{\partial{B}}  = \beta_1 \exp(\beta_2 + \beta_1 B).
\]
This derivative shows the sensitivity of the treatment effect to variations in \( B \) at a given level of \( B \). Given that \( B \) is standardized on the full construction panel, \( B = 0 \) corresponds to the mean broadband level, and \( B = 1 \) indicates one unit (one construction-panel standard deviation) above the mean. \(\text{ATT}(B=1) = \exp(\beta_2 + \beta_1) - 1, \quad \text{ATT}(B=0) = \exp(\beta_2) - 1\). Therefore, \(\text{ATT}(B=1) - \text{ATT}(B=0) = \exp(\beta_2 + \beta_1) - \exp(\beta_2).\)

The ACRT, being dependent on \( B \), is evaluated at \( B = 0 \) (the mean broadband level) as a reference point: \(ACRT|_{B=0} = \beta_1 \exp(\beta_2).\)
In general, the two expressions are not equivalent: \(\exp(\beta_2 + \beta_1) - \exp(\beta_2) \neq \beta_1 \exp(\beta_2),\)
since the former is a finite difference and the latter a marginal effect. However, for small values of \( \beta_1 \), a first-order Taylor expansion provides an approximation:
\(\exp(\beta_2 + \beta_1) \approx \exp(\beta_2) + \beta_1 \exp(\beta_2),\) \(\implies
ATT(B=1) - ATT(B=0) \approx \beta_1 \exp(\beta_2) = ACRT|_{B=0}. \)
This suggests that, when \( \beta_1 \) is sufficiently small, the finite difference approximates the ACRT at the mean broadband level.

As an example, take \enquote{Post Price Floor} in \textit{Table \ref{tab:main_acrt}}: $ATT(B=0) = 0.0051$ and $ATT(B=1) = 0.0333$, so the finite difference is $0.0333 - 0.0051 = 0.0282$, while the reported $ACRT$ is $0.0277$, a discrepancy of only $0.0005$. This illustrates that the finite difference \(ATT(B=1) - ATT(B=0)\) closely approximates the $ACRT$, though the two are not identical.

\vspace{1em}
\section{Additional Results}\label{sec:adres}

\vspace{2em}
\subsection{\textbf{Specialty-wise Estimates for Heavy Telehealth Users}}

\textbf{\textit{Telehealth modalities and usage}}\label{sec:modalities}: Telehealth entails direct, electronic patient-provider interactions and the use of medical devices to collect and transmit health information as well as to manage chronic conditions.\footnote{Telehealth incorporates telemedicine, which is a  bilateral, interactive health communications with clinicians on both ends of the exchange (e.g. videoconferenced grand rounds, x-rays transmitted between radiologists or consultations where a remote practitioner presents a patient to a specialist).} Specialties differ not just in their degree of usage of telehealth, but also the purpose or modality of usage, and the type of interaction. The type of interaction could be physician-patient or physician-health care professional. The responses of providers to the regulations depend on their respective specialties, since specialties differ in their modalities. Currently, there are three main modalities of telehealth. \textit{Synchronous or Live Video}, which involve video conferences. \textit{Asynchronous or Store-and-Forward (SFT)} refers to the transmission of diagnostic information, videos, and digital images such as x-rays, CT scans, and EEG printouts, collected at the patient's site of care, to a specialist in another location. \textit{Remote Patient Monitoring (RPM)}, used for the management of chronic illness, employs devices such as Holter monitors to transmit personal medical data and vital statistics (e.g., blood pressure, blood oxygen levels), to clinicians. \footnote{As per \cite{kane2018telemedicine}, Radiology (39.5\%) and Psychiatry (27.8\%) used telehealth the most to interact with patients. Emergency Medicine topped the list for interacting with other healthcare professionals at 38.8\%. For video-conferencing, Emergency Medicine (31.6\%) and Radiology (25.8\%) led the way. Radiology also stood out in using store-and-forward data at 42.7\%. Lastly, Cardiology, was the highest user of Remote Patient Monitoring (RPM) at 17.9\%. Thus, Radiologists are, in general, the leading users of telehealth.}

The results discussed so far, present valuable insights by discriminating the treatment effects by treatments types and geography. However, since telehealth usage differs in scope and nature according to specialties, it is expected that the effect of TPL interacted with broadband should differ according to specialties.

\vspace{2em}
\begin{table}[h!]
\begin{small}
\def\sym#1{\ifmmode^{#1}\else\(^{#1}\)\fi}
\caption{Specialty-wise PPML Estimates for Heavy Telehealth (Except RPM) Users}
\vspace{0.5em}
\label{tab:heavyusers} 
\setlength{\tabcolsep}{2pt} 
\begin{spacing}{1.5}  
\begin{tabular}{@{}l*{9}{c}@{}} 
\hline\hline
& \multicolumn{3}{c}{Psychiatrists Count} & \multicolumn{3}{c}{Radiologists Count} & \multicolumn{3}{c}{Emergency Physicians Count} \\
\cline{2-10}
& \multicolumn{1}{c}{(1)} & \multicolumn{1}{c}{(2)} & \multicolumn{1}{c}{(3)} & \multicolumn{1}{c}{(4)} & \multicolumn{1}{c}{(5)} & \multicolumn{1}{c}{(6)} & \multicolumn{1}{c}{(7)} & \multicolumn{1}{c}{(8)} & \multicolumn{1}{c}{(9)} \\
\cline{2-10}
& Full sample & Non-metro & Metro & Full sample & Non-metro & Metro & Full sample & Non-metro & Metro \\
\hline
Post Price Floor=1 $\times$ Broadband & $ 0.0807\sym{***} $ & $ 0.1249^{\dagger} $ & $ 0.0950\sym{***} $ & $ 0.0973\sym{***} $ & $ -3.7324^{\dagger} $ & $ 0.0566\sym{***} $ & $ 0.1325\sym{***} $ & $ 2.6131^{\dagger} $ & $ 0.1281\sym{***} $ \\
                    & ( $ 0.0250 $ ) & -- & ( $ 0.0050 $ ) & ( $ 0.0033 $ ) & -- & ( $ 0.0048 $ ) & ( $ 0.0323 $ ) & -- & ( $ 0.0238 $ ) \\
[1em]
Post Price Ceiling=1 $\times$ Broadband & $ 0.0116 $ & $ 0.0051^{\dagger} $ & $ 0.0272\sym{***} $ & $ 0.0372\sym{***} $ & $ 0.0069^{\dagger} $ & $ 0.0540\sym{***} $ & $ 0.0271\sym{***} $ & $ -0.7074^{\dagger} $ & $ 0.0156\sym{*} $ \\
                    & ( $ 0.0087 $ ) & -- & ( $ 0.0103 $ ) & ( $ 0.0094 $ ) & -- & ( $ 0.0094 $ ) & ( $ 0.0102 $ ) & -- & ( $ 0.0094 $ ) \\
[1em]
Post Cost Parity=1 $\times$ Broadband & $ 0.0303\sym{***} $ & $ \_\_\_ $ & $ 0.0153\sym{***} $ & $ -0.0346\sym{***} $ & $ \_\_\_ $ & $ -0.0485\sym{***} $ & $ -0.0115 $ & $ \_\_\_ $ & $ 0.0003 $ \\
                    & ( $ 0.0033 $ ) & & ( $ 0.0049 $ ) & ( $ 0.0052 $ ) & & ( $ 0.0045 $ ) & ( $ 0.0230 $ ) & & ( $ 0.0222 $ ) \\
[1em]
Post Cost Ceiling=1 $\times$ Broadband & $ 0.0111 $ & $ 0.9092^{\dagger} $ & $ 0.0098 $ & $ 0.0158\sym{*} $ & $ 0.5766^{\dagger} $ & $ 0.0166\sym{*} $ & $ 0.0057 $ & $ 0.0021^{\dagger} $ & $ 0.0074 $ \\
                    & ( $ 0.0095 $ ) & -- & ( $ 0.0110 $ ) & ( $ 0.0087 $ ) & -- & ( $ 0.0091 $ ) & ( $ 0.0088 $ ) & -- & ( $ 0.0066 $ ) \\
[1em]
\hline
Observations        &       12398         &        5890         &        6508         &       12188         &        5890         &        6298         &       14771         &        7903         &        6868         \\
\hline\hline
\multicolumn{10}{l}{\footnotesize Standard errors in parentheses}\\
\multicolumn{10}{l}{\footnotesize \sym{*} \(p<0.10\), \sym{**} \(p<0.05\), \sym{***} \(p<0.01\)}\\
\end{tabular}
\end{spacing}
\end{small}
\footnotesize
\begin{justify}
\textit{Note:} Non-metro cells of the broadband-interaction rows are marked with $\dagger$ and reported without inference, with their standard errors dashed. The within-cell treated post-period broadband span in these cells is below one standard deviation of the construction panel (0.17 to 0.46 units across framings, against 2.22 or more in every metro cell), so these coefficients are per-unit-$B$ slopes estimated off a sub-unit broadband range. The same rule binds nowhere in \textit{Table \ref{tab:main_acrt}}, where every framing's treated span exceeds one unit. Post Cost Parity has no non-metro treated observations ($\_\_\_$), and the non-metro Price Ceiling arm comprises a single state cluster (Tennessee). The non-metro estimation samples retain the observations shown against 14,816 non-metro county-years in the panel, the remainder dropped as separated by the county fixed effects, so the split estimates are identified within counties that ever employ the specialty.
\end{justify}
\end{table}

\textit{Table \ref{tab:heavyusers}} shows the specialty-wise break up of our estimates for Emergency Physicians (leading in video conferencing and interacting with other healthcare professionals), Psychiatrists (second only to Radiology in interacting with patients), and Radiologists who lead in the telehealth modality which involves storing and forwarding of data and interacting with patients.\footnote{The usage of telehealth prior to the COVID-19 pandemic in 2016, differed by specialty and by area \citep{kane2018telemedicine}. The modalities considered were: Videoconferencing, Remote Patient Monitoring (RPM), or Storing and Forwarding Data.} The ``effects" here would mean how $ATT_k$ vary with broadband as it increases from the mean to one unit above the mean. The Price Floor shows positive effects for all three specialties in both the full sample and the metro subsample. The effect is the strongest for Emergency Physicians, followed by Psychiatry, and then Radiology in metro areas. In metro areas with higher broadband and typically higher demand, Price Floor allows physicians to get reimbursed more than they otherwise would. The relatively precise estimates for the metro subsample suggests that Price Floor is a significantly conducive component for metro areas, irrespective of the specialty. For the non-metro subsample, the estimates for Price Floor, particularly for Psychiatrists and Emergency Physicians, indicate less precision, underscoring variability in telehealth usage patterns. It is important to note that these specialties rely heavily on telehealth, which reduces the scope for substitution to in-person services. This reliance implies stronger rotations in the supply curves and more pronounced estimates.

For non-metro areas, the broadband-interaction cells carry no inference under the support rule stated in the table notes, so the specialty conclusions are drawn from the full-sample and metro estimates. Price Ceiling shows a positive effect for Radiology in metro areas. Cost Parity is a strongly unfavorable element for Radiologists and aggregate physician count in metro areas. In contrast, the Cost Parity estimate for Psychiatrists is positive. This divergence is explained by the pre-existing federal regulatory environment for mental health services. The Mental Health Parity and Addiction Equity Act (MHPAEA, 2008) already requires that cost sharing for mental health services cannot be more restrictive than for medical and surgical services. As a result, when state-level TPL impose Cost Parity on telehealth, the incremental cost burden on psychiatric telehealth consumers is minimal because their in-person cost sharing was already indirectly constrained by federal parity requirements, and telehealth Cost Parity simply equalizes to that already-constrained rate. For Radiology, no comparable federal cost-sharing protection exists, so TPL Cost Parity creates a genuine new cost increase that deters utilization and reduces physician counts. Conversely, Cost Ceiling is expected to make telehealth cheaper for consumers, acting as a conducive component. The effects of Cost Parity are particularly pronounced for Radiology, with relatively robust estimates. The estimates for Price Floor in \textit{Table \ref{tab:heavyusers}} appear more amplified and show more conduciveness than those for the Price Floor in \textit{Table \ref{tab:withbbd}}, while the estimates for the Price Ceiling in \textit{Table \ref{tab:heavyusers}} appear more subdued compared to those in \textit{Table \ref{tab:withbbd}}. Specialties that use telehealth intensively for patient interactions may use more telehealth to capitalize on higher reimbursement opportunities when a Price Floor is present, or they may relocate to areas with more favorable policy environments with better technological infrastructure when faced with a Price Ceiling, thus moderating the conducive effect of a Price Ceiling observed in the aggregate sample in \textit{Table \ref{tab:withbbd}}.

\vspace{2em}
\subsection{\textbf{Specialty-wise Estimates for a Light Telehealth User and a RPM User}}

\begin{table}[h!]
\begin{small}
\def\sym#1{\ifmmode^{#1}\else\(^{#1}\)\fi}
\caption{Specialty-wise PPML Estimates for a Heavy Telehealth (RPM) User and a Light Telehealth User}
\vspace{1em}
\label{tab:lightusers} 
\setlength{\tabcolsep}{2pt} 
\begin{spacing}{1.5}  
\begin{tabular}{@{}l*{6}{c}@{}} 
\hline\hline
& \multicolumn{3}{c}{Cardiologists Count} & \multicolumn{3}{c}{Gastroenterologists Count} \\
\cline{2-7}
& \multicolumn{1}{c}{(1)} & \multicolumn{1}{c}{(2)} & \multicolumn{1}{c}{(3)} & \multicolumn{1}{c}{(4)} & \multicolumn{1}{c}{(5)} & \multicolumn{1}{c}{(6)} \\
\cline{2-7}
& Full sample & Non-metro & Metro & Full sample & Non-metro & Metro \\
\hline
Post Price Floor=1 $\times$ Broadband & $ -0.0400 $ & $ 0.6499^{\dagger} $ & $ -0.0166 $ & $ -0.0037 $ & $ 5.3501^{\dagger} $ & $ 0.0157 $ \\
& ($0.0637$) & -- & ($0.0717$) & ($0.0042$) & -- & ($0.0097$) \\
[1em]
Post Price Ceiling=1 $\times$ Broadband & $ 0.0083 $ & $ 0.6503^{\dagger} $ & $ 0.0022 $ & $ 0.0510\sym{***} $ & $ 1.2735^{\dagger} $ & $ 0.0280\sym{**} $ \\
& ($0.0165$) & -- & ($0.0185$) & ($0.0084$) & -- & ($0.0126$) \\
[1em]
Post Cost Parity=1 $\times$ Broadband & $ -0.0065 $ & $\_\_\_$ & $ -0.0037 $ & $ 0.0014 $ & $\_\_\_$ & $ 0.0194 $ \\
& ($0.0183$) & & ($0.0211$) & ($0.0089$) & & ($0.0135$) \\
[1em]
Post Cost Ceiling=1 $\times$ Broadband & $ -0.0038 $ & $ -0.1072^{\dagger} $ & $ -0.0058 $ & $ 0.0059 $ & $ -1.0452^{\dagger} $ & $ 0.0026 $ \\
& ($0.0080$) & -- & ($0.0080$) & ($0.0078$) & -- & ($0.0103$) \\
\hline
Observations        &        9687         &        3839         &        5848         &        7828         &        2440         &        5388         \\
\hline\hline
\multicolumn{7}{l}{\footnotesize Standard errors in parentheses}\\
\multicolumn{7}{l}{\footnotesize \sym{*} \(p<0.10\), \sym{**} \(p<0.05\), \sym{***} \(p<0.01\)}\\
\end{tabular}
\end{spacing}
\end{small}
\footnotesize
\begin{justify}
\textit{Note:} Non-metro cells of the broadband-interaction rows are marked with $\dagger$ and reported without inference, with their standard errors dashed. The within-cell treated post-period broadband span in these cells is below one standard deviation of the construction panel (0.17 to 0.46 units across framings, against 2.22 or more in every metro cell), so these coefficients are per-unit-$B$ slopes estimated off a sub-unit broadband range. The same rule binds nowhere in \textit{Table \ref{tab:main_acrt}}, where every framing's treated span exceeds one unit. Post Cost Parity has no non-metro treated observations ($\_\_\_$), and the non-metro Price Ceiling arm comprises a single state cluster (Tennessee). The non-metro estimation samples retain the observations shown against 14,816 non-metro county-years in the panel, the remainder dropped as separated by the county fixed effects, so the split estimates are identified within counties that ever employ the specialty.
\end{justify}
\end{table}

\textit{Table \ref{tab:lightusers}}\hspace{.08cm} shows the estimates for Cardiologists—the biggest RPM (Remote Patient Monitoring) users—and for Gastroenterologists—who are among the specialties that use telehealth the least. Most of the coefficients exhibit considerable variability and imprecision, indicating that the physician count for specialties with either extensive RPM usage or minimal telehealth use is not significantly affected by the TPL, as compared to the specialties with heavy telehealth usage or to the specialties which use modalities other than RPM. RPM requires relatively much lower patient input. This indicates that the scope for using patient input and it’s impact on supply and demand is crucial while determining the effect of TPL. This further bolsters this study's theoretical predictions, which rely on the mechanism of Price Controls distorting the input mix, and Cost Controls changing the consumption mix, implying that for specialties using telehealth less or lacking scope for consumer input, the TPL would not have a significant impact.

\newpage
\section{Exogeneity of Framing}\label{sec:framingexo}

\subsection{Model Legislation and Drafting Conventions}
The statutory language of the parity laws originated in national model legislation and standard drafting formulations, not in state healthcare markets. The American Telemedicine Association circulated model state legislative language for parity laws in its State Telemedicine Toolkit, whose consumer cost provision is the ceiling wording itself, a deductible, copayment, or coinsurance amount that ``may not exceed'' the amount applicable to the same service delivered in person \citep{ATAtoolkit2015}. The American Medical Association distributed its own model Telemedicine Act, which pairs reimbursement parity with the same ceiling wording on consumer cost sharing \citep{AMAmodel2017}. The National Conference of State Legislatures presents the reimbursement formulations, ``the same amount,'' ``at the same rate,'' and ``on the same basis,'' as standard drafting alternatives in its toolkit for state legislators \citep{Pitsor2021}. Research on policy diffusion finds that statutory text of this kind typically originates in interest-group model legislation \citep{GarrettJansa2015}, and 41 percent of the states with private payer parity laws share one identical, undefined phrase, a marker of common drafting sources \citep{CCHPMilbank2017}.

The framing wording also carried no strategic weight at adoption. The American Telemedicine Association's fifty-state report cards, which graded every state's telemedicine policy annually and drove much of the legislative activity of the period, classified floor, ceiling, and same-rate statutes together as full parity and assigned states in each category the same top grade \citep{ThomasCapistrant2015}. A state could not improve its grade by choosing one reimbursement wording over another. The interests whose bargaining shaped the wording are the same nationwide, with providers supporting parity mandates and insurers opposing them \citep{dills2021}, so the enacted formulation reflects the relative strength of the same two groups in each legislature rather than conditions in the state's healthcare market.

\subsection{Framing-Specific Pre-Trends}\label{subsec:framingpretrends}
\begin{figure}[htbp!]
    \centering
    \caption{Framing-Specific Event Study for Service Provision (CHSPI)}
    \vspace{0.2em}
    \label{fig:framingeventstudy}
    \includegraphics[width=.95\textwidth]{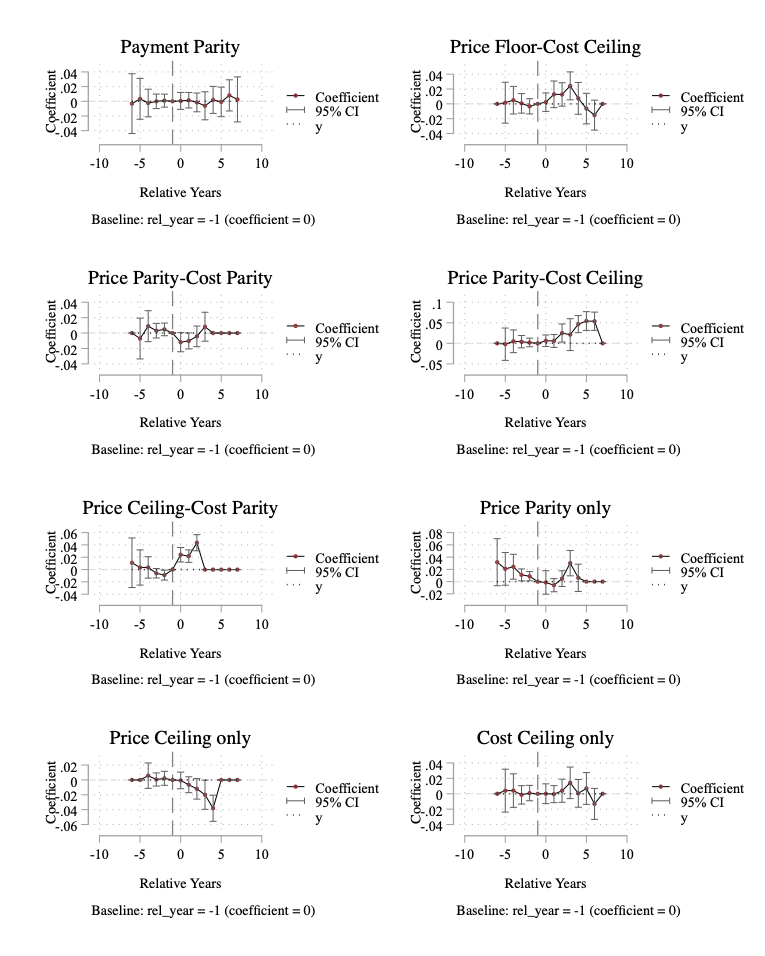}
    \vspace{0.6em}

    \noindent
    \begin{minipage}{\textwidth}
    \begin{spacing}{1.5}
    \footnotesize
    \justifying
    \textit{\textbf{Note}}: Each panel reports event-study coefficients for one framing combination from a single event-study regression of CHSPI on relative-year indicators interacted with each framing indicator, with county and year fixed effects and the covariate set of the main specification. Standard errors are clustered at the state level. The baseline is relative year $-1$, and bands are 95\% confidence intervals.
    \end{spacing}
    \end{minipage}
\end{figure}

The pre-trend evidence is framing specific. \textit{Figure \ref{fig:framingeventstudy}} reports an event study of CHSPI in which relative time from treatment is interacted with each framing combination separately, estimated with county and year fixed effects, the covariate set of the main specification, and standard errors clustered at the state level, with the baseline at relative year $-1$. Across the framings that carry the paper's results, the pre-period coefficients are small and statistically indistinguishable from zero, with one exception, a coefficient of $-0.009$ two years before adoption under Price Ceiling--Cost Parity, opposite in sign to that framing's positive post-period estimates. The ``Price Parity only'' arm shows small positive pre-period coefficients, between 0.009 and 0.024. ``Price Parity only'' is a secondary framing outside the paper's mechanism and results. Price Floor states, Price Ceiling states, Cost Parity states, and the never-treated comparison counties were on similar service-provision trends before the laws took effect, and the post-period patterns emerge only after adoption.

\newpage
\vspace{1em}
\section{Additional Robustness}

\vspace{3em}
\subsection{PCA Loadings for the Composite Healthcare Service Provision Index}

Table~\ref{tab:pca_loadings} reports the eigenvalues and eigenvectors for the principal component analysis underlying the CHSPI. The table is provided to document that physician density per capita does not mechanically dominate the index: it is the third-highest loading on PC1 (0.410), behind hospital admissions per capita (0.472) and hospitals per capita (0.463), confirming that PC1 reflects supply-side healthcare infrastructure broadly rather than physician counts specifically. Three components are retained based on the Kaiser criterion (eigenvalue $\geq 1$), collectively explaining 73.4 percent of total variance. The first principal component captures supply-side healthcare infrastructure; the second captures beneficiary enrollment intensity; the third captures emergency utilization.

\vspace{3em}
\begin{table}[htbp]
\centering
\caption{Principal Component Loadings for the Composite Healthcare Service Provision Index (CHSPI)}
\label{tab:pca_loadings}
\begin{threeparttable}
\begin{tabular}{lccc}
\toprule
\textbf{Variable} & \textbf{PC1} & \textbf{PC2} & \textbf{PC3} \\
\midrule
Hospital Admissions per Capita & 0.472 & 0.165 & $-$0.063 \\
Hospitals per Capita & 0.463 & $-$0.057 & 0.064 \\
Physician Density per Capita & 0.410 & $-$0.036 & $-$0.269 \\
Outpatient Visits per Capita & 0.407 & 0.240 & 0.081 \\
Inpatient Days per Capita & 0.365 & 0.298 & $-$0.155 \\
ED Visits per 1,000 Beneficiaries & 0.129 & 0.067 & 0.937 \\
FFS Beneficiaries per Capita & $-$0.223 & 0.637 & $-$0.093 \\
Aged and Disabled Enrollment per Capita & $-$0.170 & 0.642 & 0.045 \\
\midrule
Eigenvalue & 3.149 & 1.707 & 1.018 \\
Proportion of Variance Explained & 0.394 & 0.213 & 0.127 \\
Cumulative Variance Explained & 0.394 & 0.607 & 0.734 \\
\bottomrule
\end{tabular}
\begin{tablenotes}
\small
\item \textit{Notes:} All variables are arcsinh-transformed and standardized prior to PCA. Components are retained based on the Kaiser criterion (eigenvalue $\geq 1$). CHSPI is the first principal component rescaled to a 0--100 range.
\end{tablenotes}
\end{threeparttable}
\end{table}

\newpage
\subsection{Alternate Broadband Specifications}

\begin{figure}[htbp!]
    \centering
    \vspace{1.5em}
    \caption{Distribution of Broadband: Standardized and Log Min--Max Normalized}
    \vspace{0.2em}
    \label{fig:bbddistribution}
    \includegraphics[width=.95\textwidth]{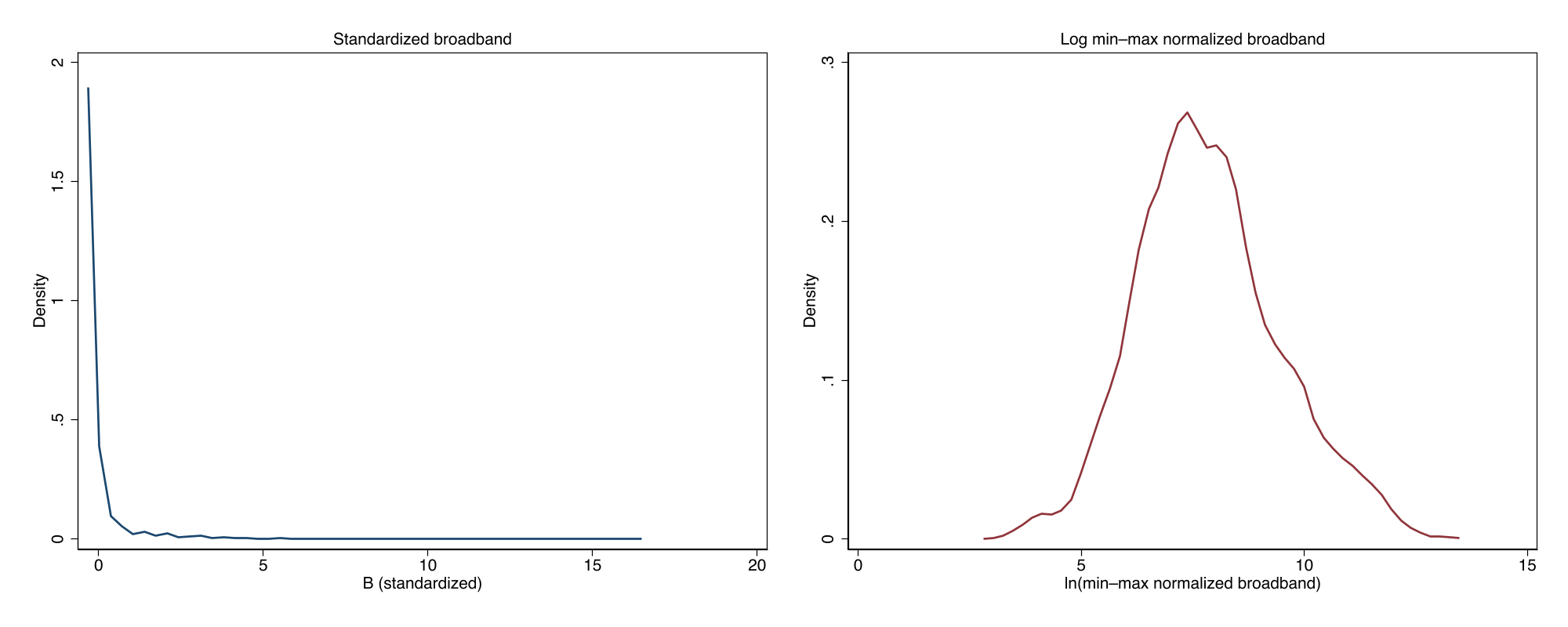}
    \vspace{0.6em}

    \noindent
    \begin{minipage}{\textwidth}
    \begin{spacing}{1.5}
    \footnotesize
    \justifying
    \textit{\textbf{Note}}: Kernel densities of the two broadband measures over the full county panel (N = 22,828). The left panel shows the standardized household-weighted Tier 1 broadband variable ($B$), standardized using the full-panel mean and standard deviation. The distribution is right-skewed (skewness 8.0), with the 90th, 95th, and 99th percentiles at $B = 0.34$, $B = 1.11$, and $B = 3.52$, and a sample maximum of $B = 16.48$, which is also the maximum within the estimation sample. The right panel shows the natural log of the min--max normalized variable used in \textit{Table \ref{tab:lognormalandarsinh}}, which is approximately normally distributed.
    \end{spacing}
    \end{minipage}
\end{figure}

\textit{Figure \ref{fig:bbddistribution}} displays the distribution of the standardized broadband variable and of the log min--max normalized variable used in the robustness estimates below.

\vspace{3em}
\begin{table}[h!]
\begin{small}
\def\sym#1{\ifmmode^{#1}\else\(^{#1}\)\fi}
\caption{PPML Estimates With Broadband Interaction}
\label{tab:lognormalandarsinh}
\setlength{\tabcolsep}{1.2pt}
\begin{spacing}{1.5}
\begin{tabular}{@{}l*{6}{c}@{}}
\hline\hline
& \multicolumn{3}{c}{log-normal broadband} & \multicolumn{3}{c}{asinh broadband} \\
\cline{2-7}
& \multicolumn{1}{c}{(1)} & \multicolumn{1}{c}{(2)} & \multicolumn{1}{c}{(3)} & \multicolumn{1}{c}{(4)} & \multicolumn{1}{c}{(5)} & \multicolumn{1}{c}{(6)} \\
\cline{2-7}
&\multicolumn{1}{c}{Full sample} &\multicolumn{1}{c}{Non-metro} &\multicolumn{1}{c}{Metro} &\multicolumn{1}{c}{Full sample} &\multicolumn{1}{c}{Non-metro} &\multicolumn{1}{c}{Metro} \\
\hline
Post Price Floor $\times$ Broadband & $0.0230\sym{***}$ & $-0.0340\sym{**}$ & $0.0199\sym{***}$ & $0.0230\sym{***}$ & $-0.0340\sym{**}$ & $0.0199\sym{***}$ \\
& ($0.0040$) & ($0.0140$) & ($0.0045$) & ($0.0040$) & ($0.0141$) & ($0.0045$) \\
[1em]
Post Price Ceiling $\times$ Broadband & $0.0239\sym{***}$ & $-0.0403\sym{***}$ & $0.0271\sym{***}$ & $0.0239\sym{***}$ & $-0.0404\sym{***}$ & $0.0271\sym{***}$ \\
& ($0.0030$) & ($0.0104$) & ($0.0041$) & ($0.0030$) & ($0.0105$) & ($0.0041$) \\
[1em]
Post Cost Parity $\times$ Broadband & $-0.0237\sym{***}$ & $\_\_\_$ & $-0.0252\sym{***}$ & $-0.0237\sym{***}$ & $\_\_\_$ & $-0.0253\sym{***}$ \\
& ($0.0069$) &  & ($0.0076$) & ($0.0069$) &  & ($0.0076$) \\
[1em]
Post Cost Ceiling $\times$ Broadband & $0.0017$ & $0.0185\sym{*}$ & $0.0007$ & $0.0017$ & $0.0184\sym{*}$ & $0.0007$ \\
& ($0.0030$) & ($0.0111$) & ($0.0040$) & ($0.0030$) & ($0.0111$) & ($0.0040$) \\
\hline
Observations & 22332 & 14354 & 7978 & 12188 & 5890 & 6298 \\
\hline
Time Fixed Effects  & Yes & Yes & Yes & Yes & Yes & Yes \\
County Fixed Effects & Yes & Yes & Yes & Yes & Yes & Yes \\
\hline\hline
\end{tabular}
\bigskip
\footnotesize
\begin{tabularx}{\textwidth}{@{}X@{}}
\footnotesize \sym{*} \(p<0.10\), \sym{**} \(p<0.05\), \sym{***} \(p<0.01\) \\
\smallskip
\textit{Note:} Standard errors in parentheses are clustered at the state level. The outcome variable is physician count. All the controls are included but not shown. ``$\_\_\_$" indicate there are no non-metro observations for Cost Parity.  \\
\end{tabularx}
\end{spacing}
\end{small}
\end{table}

As an additional robustness check, we estimated the model using a log-normalized (natural log of min-max normalization of original household-weighted Tier 1 variable) and arcsinh-transformed broadband variable (of original household-weighted Tier 1 variable), to address potential concerns about the skewness of the original standardized broadband variable. The results demonstrate that the results for both types of transformations are similar, and the signs and significance of the coefficients remain consistent with those from the original specification. Specifically, the interactions between post-policy indicators and broadband retain their directional effects across the full sample, non-metro, and metro areas for both Fed \& Non-Fed MDs and Radiologists. Although the magnitudes differ due to the distinct scaling of the broadband variables, the overall patterns and statistical significance are preserved. This consistency suggests that the skewness of the original broadband variable does not unduly influence our findings, confirming that our results are robust to alternative specifications of the broadband measure.

\newpage
\subsection{Pre-trends and No Anticipation}

\vspace{2em}
\begin{table}[h!]
\begin{small}
\centering
\caption{Combined Analysis of Pre-trends}
\label{tab:pretrends}
\begin{spacing}{1} 
\begin{tabular}{>{\def\arraystretch{1.5}}l>{\def\arraystretch{1.5}}c>{\def\arraystretch{1.5}}c>{\def\arraystretch{1.5}}c} 
\hline
&& (a) & (b)\\
\hline
&& Non Fed \& Fed MDs & Lag Non Fed \& Fed MDs \\
&& Full sample & Full sample \\
\hline
Treated $\times$ 7 years pre-treatment && $0.0008$ & $\_\_\_$ \\
&& ($0.0099$)  & \\
Treated $\times$ 6 years pre-treatment && $-0.0070$ & $0.0076$ \\
&& ($0.0092$)  & ($0.0108$) \\
Treated $\times$ 5 years pre-treatment && $-0.0015$ & $-0.0026$ \\
&& ($0.0088$)  & ($0.0096$) \\
Treated $\times$ 4 years pre-treatment && $-0.0027$ & $0.0032$ \\
&& ($0.0082$)  & ($0.0084$) \\
Treated $\times$ 3 years pre-treatment && $-0.0047$ & $0.0006$ \\
&& ($0.0075$)  & ($0.0074$) \\
Treated $\times$ 2 years pre-treatment && $-0.0057$ & $0.0000$ \\
&& ($0.0070$)  & ($0.0069$) \\
Treated $\times$ 1 years pre-treatment && $-0.0071$ & $-0.0015$ \\
&& ($0.0062$)  & ($0.0063$) \\
\hline
Observations && $11554$ & $10074$ \\
\hline
\multicolumn{4}{l}{Standard errors in parentheses} \\
\multicolumn{4}{l}{\sym{*} \(p<0.10\), \sym{**} \(p<0.05\), \sym{***} \(p<0.01\)} \\
\end{tabular}
\end{spacing} 
\begin{spacing}{1.5} 
\footnotesize
\noindent
Note: Specifications include county and year fixed effects. The results show that the assumptions pre-treatment analogue to the ratio version of conditional parallel trends and conditional no anticipation hold. 
\end{spacing}
\end{small}
\end{table}
\clearpage 


\begin{table}[htbp!]
\centering
\caption{RESET Test Results}
\label{tab:reset}
\begin{spacing}{1} 
\begin{adjustbox}{max width=\textwidth, center}
\begin{tabular}{>{\def\arraystretch{1.5}}l>{\def\arraystretch{1.5}}c>{\def\arraystretch{1.5}}c} 
\hline
& Model (4), Table 2 & Model (4), Table 3 \\
\hline
Null Hypothesis (H0) & \multicolumn{2}{c}{$(\hat{y})^2 = 0$} \\
Test Statistic & \multicolumn{2}{c}{chi2(1)} \\
p-value & $0.30$ & $0.63$ \\
\hline
\end{tabular}
\end{adjustbox}
\newline
\noindent
\vspace{1em}
{\footnotesize \parbox{\textwidth}{\textit{Note}: The null hypothesis (H0) in our context is that the model is correctly specified, such that the functional form of the model is appropriate. Specifically, it tests whether the model can be improved by adding higher-order terms of the predicted values (squared predicted values in this case). If the null hypothesis is not rejected, there is no evidence from the data to suggest that the model is misspecified. A low p-value indicates that we reject the null hypothesis, signifying model mis-specification. Here, for both models, the p-values are above the typical significance levels (0.1, 0.05, 0.01), and therefore the null hypothesis cannot be rejected. This suggests that there is no evidence of model mis-specification.}}
\end{spacing}
\end{table}


\begin{table}[h!]
\centering
\def\sym#1{\ifmmode^{#1}\else\(^{#1}\)\fi}
\caption{Placebo test for pre-treatment trends}
\label{tab:placebo}
\begin{spacing}{1} 
\begin{tabular}{>{\def\arraystretch{1.5}}l>{\def\arraystretch{1.5}}c} 
\hline
                    &\multicolumn{1}{c}{Placebo Test} \\
\hline
\textit{Treated} & $0.0285$ \\
                 & ($0.0493$) \\
\textit{Treated $\times$ Relative years pre-fake-treatment} & $0.0001$ \\
                                                              & ($0.0010$) \\
\hline
Observations & $22,804$ \\
\hline
\multicolumn{2}{l}{\footnotesize Standard errors in parentheses}\\
\multicolumn{2}{l}{\footnotesize \sym{*} \(p<0.10\), \sym{**} \(p<0.05\), \sym{***} \(p<0.01\)}\\
\end{tabular}
\end{spacing}
\begin{spacing}{1} 
\footnotesize
\begin{tabularx}{\textwidth}{@{}X@{}}
The table presents the key results from the placebo test conducted to check for pre-treatment trends. We assigned a fictitious treatment period before the actual implementation of the treatment to observe any effects that should not exist if the parallel trends assumption holds. The coefficients for the placebo treatment (\textit{Treated}) and its interaction with relative years pre-treatment (\textit{Treated $\times$ Relative years pre-fake-treatment}) are statistically insignificant, indicating no pre-treatment differences between treated and control groups. These results reinforce the validity of our main findings.\\
\end{tabularx}
\end{spacing}
\end{table}
\clearpage

\subsection{\textbf{Spillover Checks}}
Cross-state telehealth practice in the 2010--2019 sample period was institutionally restricted, not merely economically unlikely. Before the COVID-19 pandemic, physicians were required by each state's medical practice act to hold a license in the state where the patient is located \citep{Mehrotra2021NEJM}. According to the Federation of State Medical Boards Telemedicine Overview (2015), 80\% of states required out-of-state clinicians offering telehealth to be licensed in the patient's residing state. The only legal channel through which cross-state telehealth practice could occur within the sample period is the Interstate Medical Licensure Compact, which began admitting states in 2015 and 2016; the specification controls for this channel through the post-IMLC indicator. Even after the pandemic, when federal emergency waivers temporarily relaxed these licensure restrictions, only 5.0\% of Medicare telemedicine visits were out-of-state, and 57.2\% of those were concentrated among patients living within 15 miles of a state border \citep{Mehrotra2022JAMA}. Under the stricter pre-pandemic rules in force throughout the sample period, cross-state telehealth service spillovers were negligible and are controlled for by the post-IMLC indicator in all specifications.

Physician spatial spillovers---whereby physicians relocate across state borders in response to neighboring states' parity law adoptions---require a separate empirical check. The U.S. county adjacency file was used to identify counties bordering other states and to compute interstate spillover variables, including the count of neighboring states adopting telehealth parity laws (TPL, denoted as \texttt{neighbor\_tpl\_count}, where TPL refers to the main policy indicator \texttt{provpar} encompassing all framing types, including ``No Framing''). These variables were merged with the main dataset to flag border counties (those with at least one interstate neighbor) and add spillover terms. Three PPML regressions are estimated using the \texttt{ppmlhdfe} estimator, incorporating county and year fixed effects and state-clustered standard errors: (1) the ``$\texttt{No Border}$'' model, excluding counties bordering other states to test sensitivity to potential interstate spillovers that could bias estimates due to physician mobility or telehealth practice diffusion; (2) the ``$\texttt{With Spill}$'' model, the spillover interaction term (\texttt{c.neighbor\_tpl\_count\#c.bbdT1std}) incorporation, where \texttt{neighbor\_tpl\_count} represents the number of neighboring states with TPL and \texttt{bbdT1std} is standardized broadband access) alongside the original regressors to capture how neighbors' TPL adoption, moderated by local broadband, affects physician counts; and (3) the ``$\texttt{placebo}$'' model, applied to untreated border counties (those with \texttt{provpar=0} and bordering other states) to check for SUTVA violations through spillover contamination, where significant effects would indicate that neighbors' TPL influences untreated outcomes, violating the no-interference assumption. The three spillover checks in \textit{Table~\ref{tab:spillover}}---excluding border counties, adding a neighboring-state TPL adoption interaction, and a placebo test on untreated border counties---confirm that physician spatial spillovers are minimal and that the ATT estimates are not driven by interstate physician mobility: the spillover term is insignificant in both the $\texttt{With Spill}$ (0.0094, SE 0.0238) and $\texttt{placebo}$ ($-$0.0214, SE 0.0382) models, while main policy effects remain consistent in sign and significance across all models (e.g., Post Price Floor $\times$ Broadband remains positive and significant, with magnitudes ranging from 0.0223 to 0.1658).

\vspace{1em}
\begin{table}[h!]
\begin{small}
\def\sym#1{\ifmmode^{#1}\else\(^{#1}\)\fi}
\caption{Spillover Robustness Checks}
\label{tab:spillover}
\setlength{\tabcolsep}{1.2pt}
\begin{spacing}{1}
\begin{tabular}{@{}l*{3}{c}@{}}
\hline\hline
& \multicolumn{1}{c}{(1)} & \multicolumn{1}{c}{(2)} & \multicolumn{1}{c}{(3)} \\
&\multicolumn{1}{c}{No Border} &\multicolumn{1}{c}{With Spill} &\multicolumn{1}{c}{Placebo} \\
\hline
Neighborhood TPL $\times$ Broadband &  & $0.0094$ & $-0.0214$ \\
&  & ($0.0238$) & ($0.0382$) \\
Neighborhood Price Control $\times$ Broadband &  & $-0.0134$ & $-0.0191\sym{*}$ \\
&  & ($0.0267$) & ($0.0109$) \\
Neighborhood Cost Control $\times$ Broadband &  & $0.0146$ & $0.0318\sym{**}$ \\
&  & ($0.0189$) & ($0.0159$) \\
Post Payment Parity $\times$ Broadband & $0.0002$ & $0.0046$ &  \\
& ($0.0039$) & ($0.0037$) &  \\
Post Provider Parity $\times$ Broadband & $0.0014$ & $-0.0040$ &  \\
& ($0.0039$) & ($0.0035$) &  \\
Post Price Floor $\times$ Broadband & $0.0223\sym{***}$ & $0.1658\sym{***}$ &  \\
& ($0.0060$) & ($0.0235$) &  \\
Post Price Ceiling $\times$ Broadband & $0.0355\sym{***}$ & $0.0303\sym{***}$ &  \\
& ($0.0040$) & ($0.0053$) &  \\
Post Cost Parity $\times$ Broadband & $-0.0812\sym{***}$ & $-0.0260\sym{***}$ &  \\
& ($0.0019$) & ($0.0059$) &  \\
Post Cost Ceiling $\times$ Broadband & $0.0016$ & $-0.0029$ &  \\
& ($0.0037$) & ($0.0042$) &  \\
\hline
Observations & 13058 & 7620 & 3760 \\
\hline
Time Fixed Effects  & Yes & Yes & Yes \\
County Fixed Effects & Yes & Yes & Yes \\
\hline\hline
\end{tabular}
\bigskip
\footnotesize
\begin{tabularx}{\textwidth}{@{}X@{}}
\smallskip
\footnotesize \sym{*} \(p<0.10\), \sym{**} \(p<0.05\), \sym{***} \(p<0.01\) \\
\smallskip
\textit{Note:} Standard errors in parentheses are clustered at the state level. The outcome variable is physician count. All the controls are included but not shown. \\
\end{tabularx}
\end{spacing}
\end{small}
\end{table}


\newpage

\subsection{Leave-One-State-Out Jackknife: Non-Metro Sample}
Table~\ref{tab:jackknife} reports the distribution of Post $\times$ Regulation Type $\times$ Broadband coefficients from a leave-one-state-out jackknife restricted to non-metro counties. Each of the 37 replications drops all counties belonging to one state. All coefficients are stable across replications, confirming that no single state drives the non-metro results.

\begin{table}[htbp]
\centering
\caption{Leave-One-State-Out Jackknife: Post $\times$ Regulation Type $\times$ Broadband Coefficients, Non-Metro Sample}
\label{tab:jackknife}
\begin{threeparttable}
\begin{tabular}{lcccc}
\toprule
\textbf{Coefficient} & \textbf{Mean} & \textbf{Std. Dev.} & \textbf{Min} & \textbf{Max} \\
\midrule
Post TPL $\times$ Broadband                & $-$0.0005 & 0.0218 & $-$0.0959 &  0.0702 \\
Post Price Floor $\times$ Broadband        &  0.3856 & 0.0234 &  0.2832 &  0.4403 \\
Post Price Parity $\times$ Broadband       & $-$0.1342 & 0.0405 & $-$0.2365 &  0.0232 \\
Post Price Ceiling $\times$ Broadband      &  0.2270 & 0.0222 &  0.1565 &  0.3205 \\
Post Cost Ceiling $\times$ Broadband       & $-$0.1072 & 0.0292 & $-$0.1665 & $-$0.0168 \\
\bottomrule
\end{tabular}
\begin{tablenotes}
\small
\item \textit{Notes:} Each row reports the distribution of coefficients from 37 jackknife replications, each dropping all counties belonging to one state. Dependent variable is CHSPI. Sample restricted to non-metro counties. Post Cost Parity $\times$ Broadband is omitted: states adopting Cost Parity framing have no non-metro counties in the sample. Post Price Ceiling $\times$ Broadband reports 36 observations as one state drop renders the coefficient undefined.
\end{tablenotes}
\end{threeparttable}
\end{table}


\vspace{2em}
\section{Data}\label{sec:data}

\vspace{1em}
\begin{figure}[h!]
    \centering
        \caption{County-wise Degree of Urbanization and Bi-variate Distribution of Broadband Penetration and Physician Counts}
        \vspace{1em}
    \label{fig:bivariate}
    \begin{minipage}[t]{0.48\textwidth}
        \centering
        \includegraphics[width=\textwidth,height=0.22\textheight]{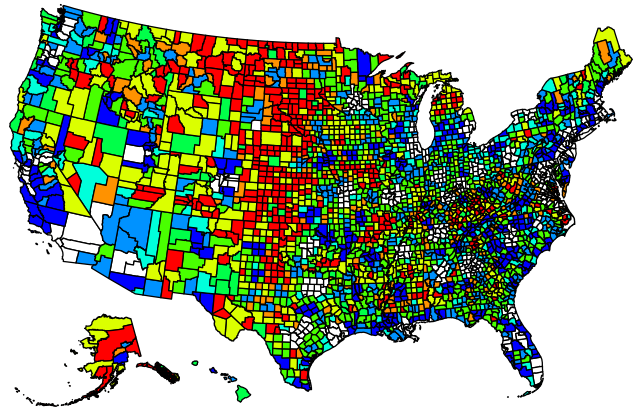}
     \includegraphics[width=0.9\textwidth,height=0.10\textheight]{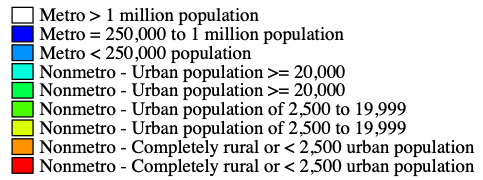}
        \subcaption[]{}
    \end{minipage}
    \hspace{0.01\textwidth} 
    \begin{minipage}[t]{0.48\textwidth}
        \centering
        \includegraphics[width=\textwidth,height=0.20\textheight]{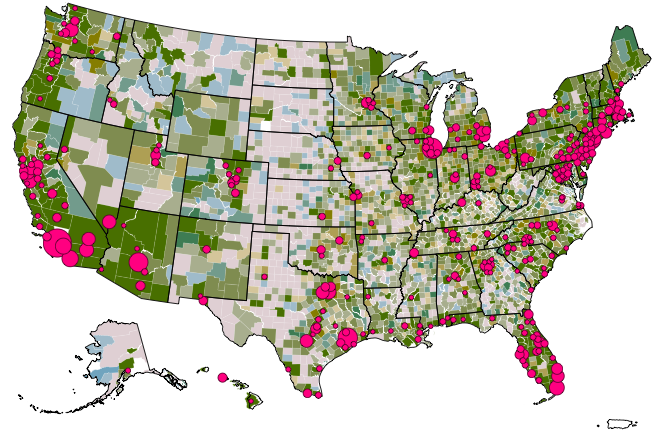}
        \includegraphics[width=0.52\textwidth,height=0.12\textheight]{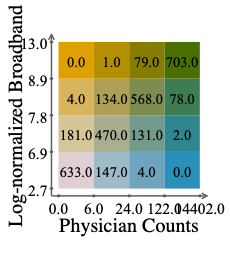}
        \subcaption[]{}
    \end{minipage}
    \noindent
    \begin{minipage}{\textwidth}
        \begin{spacing}{1.5}
        \footnotesize
        \justifying
        \textit{Note}: Panel (a) shows the metro and non-metro, rural, and urban areas. The bluish shades correspond to the metro counties. The reddish-orange counties are non-metro rural and greenish-yellowish ones are non-metro urban. Panel (b) shows the bivariate distribution of broadband penetration and physician counts for 2019. The Y-axis of the legend indicates log-normalized broadband score, while the X-axis of the legend indicates physician count. The numbers inside the legend box are the number of counties corresponding to each row-column intersection. The pink bubbles represent the densely populated areas with a population of more than 200,000, with the size of the bubbles proportional to the population of the county. The maps are generated using the Stata package from \cite{Naqvi_Stata_package_bimap_2023} using shapefiles from the \cite{tiger2016}.
        \end{spacing}
    \end{minipage}
\end{figure}

\subsection{Metro and Non-Metro County Classification} The study uses the 2013 Rural-Urban Continuum Codes \citep{rucc2013}. It categorizes the U.S. counties based on their degree of urbanization, population size, and proximity to metro areas. This data enables evaluation of spatial distinctions such as metro and non-metro in the U.S. \textit{Panel (a), Figure \ref{fig:bivariate}}, displays categories of urbanization in further detail. \textit{Table \ref{tab:sumstats}} shows the summary statistics.

\vspace{1em}
\subsection{Telehealth Parity Laws (TPL) Data Description}\label{sec:paritylawsdata}
The insights into the TPL come from the survey report \cite{foley2021}, from where the TPL language has been put together in table form by \cite{dills2021}. 

\begin{table}[h!]
\caption{Statewide Adoption Timeline and Framing of TPL}
\label{tab:framingtimeline}
\small
\begin{tabular}{|>{\raggedright}p{0.9in}|>{\centering\arraybackslash}p{0.6in}|>{\centering\arraybackslash}p{0.7in}|>{\centering\arraybackslash}p{0.35in}|>{\centering\arraybackslash}p{0.35in}|>{\centering\arraybackslash}p{0.35in}|>{\centering\arraybackslash}p{0.35in}|>{\centering\arraybackslash}p{0.5in}|}
\hline
\multirow{2}{*}{\textbf{State}} & \multirow{2}{*}{\textbf{Year Adopted}} & \multirow{2}{*}{\textbf{Payment Parity}} & \multicolumn{3}{c|}{\textbf{Provider Reimbursement (Price Controls)}} & \multicolumn{2}{c|}{\textbf{Deductibles, Copays, Coinsurance (Cost Controls)}} \\
\cline{4-8}
& & & \textbf{Same Rate as} & \textbf{Does Not Exceed} & \textbf{At Least the Same} & \textbf{Same Rate as} & \textbf{Does Not Exceed} \\
\hline
Alaska & 2016 & & & & & & \\
Arizona & 2013 & x & & & & & x \\
Arkansas & 2016 & x & & & x & & x \\
California & 1997 & x & x & & & x & \\
Colorado & 2016 & x & x & & & & x \\
Connecticut & 2016 & x\textsuperscript{a} & & & x & & \\
Delaware & 2016 & x & x & & & x & \\
District of Columbia & 2013 & & & & & & x \\
Georgia & 2005 & x & & & x & & \\
Hawaii & 1999 & x & x & & & & \\
Illinois & 2021 & x\textsuperscript{c} & x & & & & x \\
Indiana & 2015 & & & & & & x \\
Iowa & 2021 & x & x & & & x & \\
Kansas & 2019 & & & & & & \\
Kentucky & 2001 & x & & x & & & x \\
Louisiana & 1995 & x & & x\textsuperscript{d} & & & \\
Maine & 2010 & x\textsuperscript{e} & & & & & x \\
Maryland & 2012 & x\textsuperscript{f} & x & & & & \\
Massachusetts & 2021 & x\textsuperscript{g} & & & x & & x \\
Michigan & 2012 & & & & & & \\
Minnesota & 2016 & x & x & & & & x \\
Mississippi & 2013 & limited & x & & & & x \\
Missouri & 2014 & x\textsuperscript{e} & & & & & x \\
Montana & 2014 & & & & & & \\
Nebraska & 2017 & x\textsuperscript{h} & & & & & \\
Nevada & 2015 & PHE only & x & & & & \\
New Hampshire & 2009 & x & & x &  & x & \\
New Jersey & 2017 & x & & x & & x & \\
New Mexico & 2013 & x & & & x & & x \\
New York & 2016 & x\textsuperscript{e} & & & & & x \\
North Dakota & 2017 & & & & & & \\
Ohio & 2021 & x\textsuperscript{e} & & & & & x \\
Oregon & 2010 & & & & & & \\
Rhode Island & 2018 & x & & & x & & \\
South Dakota & 2020 & x\textsuperscript{e} & & & & & x \\
Tennessee & 2015 & x & & x & & & \\
Texas & 1997 & x\textsuperscript{e} & & & & & x \\
Vermont & 2012 & x\textsuperscript{i} & x & & & & x \\
Virginia & 2011 & x\textsuperscript{e} & & & & & x \\
Washington & 2017 & x & x\textsuperscript{b} & & & & \\
West Virginia & 2021 & x\textsuperscript{e} & & & & x & \\
\hline
\end{tabular}
\footnotesize
\begin{tabularx}{\textwidth}{@{}X@{}}
\textit{Note:} Source: Lacktman et al., ``50-State Survey". \\
Utah's payment parity mandate applies only to mental health. Oklahoma's payment parity mandate becomes effective January 1, 2022. PHE = public health emergency. \\
\textsuperscript{a} Effective through June 30, 2023. \textsuperscript{b} Large healthcare providers permitted to negotiate rates. \textsuperscript{c} Effective through January 1, 2028. \textsuperscript{d} Not less than 75 percent. \textsuperscript{f} Payment parity only for insured. \textsuperscript{g} Applies only to behavioral health. \textsuperscript{h} Applies only to mental health. 
\end{tabularx}
\end{table}

\vspace{1em}
\subsection{Broadband Data Description}\label{sec:bbddata}

The dataset originates from the Federal Communications Commission's (FCC) Form 477 County Data on Internet Access Services \citep{fcc2024}. This dataset is combined with the Staff Block Estimates \citep{staffblockestimates} from where the county level control variables are obtained.

\begin{figure}[htbp!]
    \centering
    \caption{Health Related Usage of the Internet}
    \includegraphics[width=\textwidth, height=0.31\textheight]{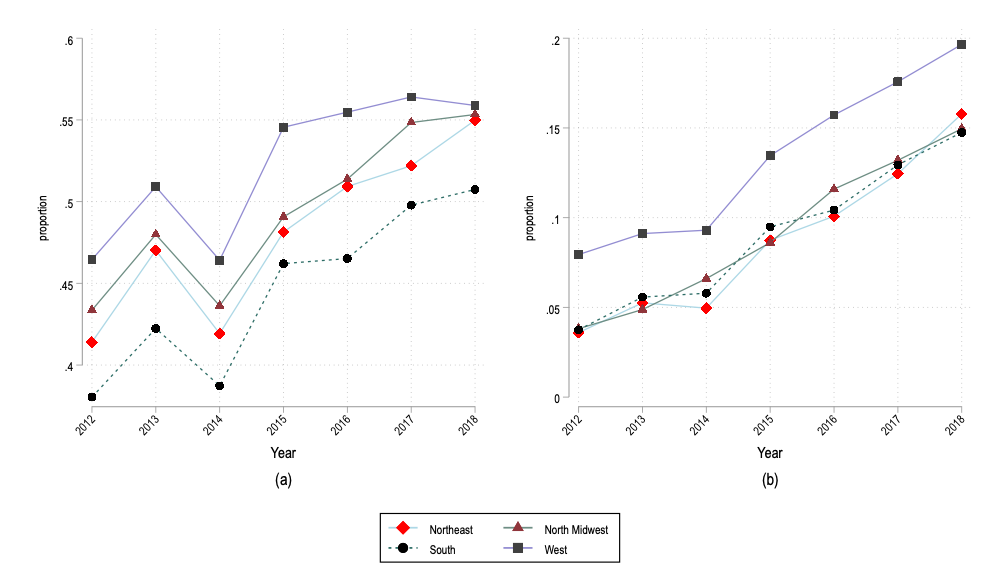} 
    \label{fig:intuse}
    \parbox{\textwidth}{
        \begin{spacing}{1.5}
        \justifying
        \footnotesize
        \textit{Note}: Panel (a) shows the region-wise proportion of people in the sample who “looked up health information on the internet in the past 12 months,” and Panel (b) shows the region-wise proportion of people who “scheduled an appointment with a health care provider on the internet in the past 12 months.” The data comes from the American Time Use Survey \citep{Blewett2023}.
        \end{spacing}}
\end{figure}

The dataset delineates four categorical variables, \textit{Tier 1} through \textit{Tier 4}, each accounting for the number of residential fixed broadband connections per 1000 households at different downstream speeds. Specifically, \textit{Tier 1} covers connections with at least 200 Kbps, \textit{Tier 2} includes those with a speed of 10 Mbps or more, \textit{Tier 3} represents at least 25 Mbps, and \textit{Tier 4} captures 100 Mbps and above.
These Tiers are further subdivided into categories ranging from `0' to `5', each signifying a specific range of connections per 1000 households in each county for the respective speed tier. Category `0' denotes no connections, Category 1 signifies up to 200 household connections per 1000 households, Categories 2 to 4 represent 201-400, 401-600, and 601-800 household connections per 1000 households respectively, while Category 5 encompasses all situations where the connection exceeds 800 per 1000 households. This meticulous classification affords a detailed overview of U.S. household broadband connection distribution patterns. It is pertinent to note that Tier 1 inherently includes Tier 2.\footnote{Despite the lack of explicit firm identifiers within these data files, the potentiality exists for the data to be combined with other data sources, thereby providing insights into a provider's operations in a specific region. In light of such considerations and as per email correspondence from FCC, to assure firm confidentiality, the data regarding the 100 Mbps speed tier for the period from June 2014 to June 2016 has been redacted. The FCC also made a decision not to disclose data related to the 100 Mbps speed tier prior to December, 2016. It is worth noting that the compilation of historical file (2008–2013) and the current file (2014 – present) employs different speed tiers. For the purpose of our study, the sample is restricted to years 2009 to 2019. For instance, Tier 4 in the historical file includes connections with a downstream speed of at least 10 Mbps, while the same tier in the current file represents connections with a downstream speed of at least 100 Mbps. Therefore, to construct a consistent time-series analysis from 2009 to 2019 without data imputation, reliable usage of two speeds can be made: connections of at least 200 Kbps (Historical Tier 1 and Current Tier 1) and those of at least 10 Mbps (Historical Tier 4 and Current Tier 2).}

\begin{figure}[h!]
\centering
{\includegraphics[width=\textwidth, height=0.38\textheight]{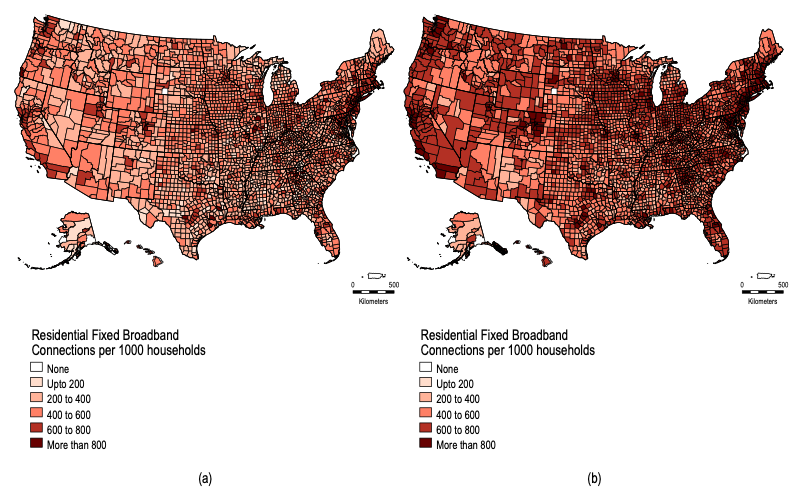}}
\caption{Broadband Penetration}
\label{fig:bbdpen}
\footnotesize
\begin{spacing}{1.5}
\justifying
\textit{Note}: Panel (a) and Panel (b) represent year 2010 and 2019 respectively. A visual comparison shows the increased broadband penetration owing to Federal and State level policy efforts. The Stata code used to generate all graphs and maps is available in the online supplementary file \href{https://livealbany-my.sharepoint.com/:u:/g/personal/pgade_albany_edu/Eabo0W2R5jpGgzm1H9vvTDkB5la43yJ7mATe83taOW1kgQ?e=4CJmCr}\texttt{Figures\_Script\_SupplementaryData.do.}
\end{spacing}
\end{figure}

It is important to recognize that the four speed tiers are not mutually exclusive. All the connections accounted for in Tiers 2, 3, and 4 are inherently contained within Tier 1. In light of the minimum downstream speed associated with each tier, any connection with a downstream speed equal to or greater than the stipulated cutoff is included within that tier's dataset.
For example, a connection offering a downstream speed of 75 Mbps would be incorporated within the Tier 1, 2, and 3 of the current dataset, since 75 Mbps exceeds the thresholds of 200 kbps, 10 Mbps, and 25 Mbps. However, this particular connection would be excluded from Tier 4 of the current dataset because the 75 Mbps speed does not meet the required 100 Mbps.

As defined by the FCC, broadband connections are lines (or wireless channels) that terminate at an end-user location and enable the end user to receive information from and/or send information to the Internet at information transfer rates exceeding 200 kilobits per second (kbps) in at least one direction.\footnote{For further details, please refer to \url{https://transition.fcc.gov/form477/477glossary.pdf}} Tier 1, which encompasses all other tiers and qualifies as ``broadband" according to the FCC definition, provides a comprehensive measure of broadband residential connections and is available for both historical and current periods. Therefore, this study adopts Tier 1 as the measure of broadband penetration. A crucial step in the data transformation process involved accounting for the number of households in each observation when measuring broadband connectivity. The broadband tier data, sourced from the Federal Communications Commission, is expressed in terms of residential fixed broadband connections per 1,000 households.

To reconcile these measurements with the number of households represented in each dataset observation, a unique weighting variable, \textit{hhweight}, was created. This variable was calculated by dividing the total number of households in each county by 1,000. The weighting factor was then applied to convert broadband connections into units compatible with the household counts. Specifically, the original \textit{Tier 1} variable was multiplied by \textit{hhweight}, yielding new weighted variables for each tier. By this method, the transformed variables represent the number of broadband connections proportionally adjusted to the number of households in each spatial unit of analysis, facilitating accurate cross-county comparisons.

Subsequently, the weighted variables were standardized, using the mean and standard deviation of the full 2010--2019 construction panel prior to the sample restrictions that produce the estimation sample, so that one unit of the standardized variable is one standard deviation of that panel. The z-score transformation adds interpretability by indicating whether an observation's value is above or below the mean and by how many standard units (standard deviations). This approach is particularly beneficial for datasets involving different spatial units with potentially vastly differing numbers of households. $B$ measures the size of a county's broadband-connected market rather than a coverage rate. This is the object the mechanism requires, since the broadband-enhanced telehealth supply elasticity (\hyperref[ass:betsea]{\textit{BETSEA}}) depends on the pool of connected patients and providers that a telehealth practice can draw on, which grows with both the connection rate and the number of households in the county. Because the variable counts actual residential connections rather than provider-reported availability, it is not subject to the concern that Form 477 deployment data overstate coverage.

Broadband is also predetermined with respect to parity law adoption and is not a ``bad control'' in the sense of \citet{AngristPischke2009}. County broadband depends on Federal Communications Commission and state infrastructure programs and on pre-existing telephone and cable network topology, not on the adoption of telehealth parity laws, so its inclusion does not introduce post-treatment bias. The literature identifies the effects of broadband penetration from such predetermined infrastructure, through natural experiments in network rollout and instruments built from pre-existing networks \citep{HjortPoulsen2019, AkermanGaarderMogstad2015, Czernich2011, FalckGoldHeblich2014}. Figure \ref{fig:floorceiling} illustrates the spatial differences in broadband penetration in 2010 and 2019, respectively. Although the county-level spatial disparity is evident in both panels, the increase in broadband penetration across the country from 2010 to 2019 is quite pronounced.

\vspace{1em}
\subsection{Area Health Resource file (AHRF) Data Description}\label{sec:ahrfdata}

The data containing aggregate and specialty-wise county level count of physicians, outpatient visits, and the population demographics used as control variables come from the Area Health Resource file (AHRF) \citep{AHRF2021_2022}. Each county is uniquely classified by a \textquote{County Code}, in accordance with Federal Information Processing Standards (FIPS). The sample used in the study spans from 2010 to 2019.

\vspace{1em}
\begin{figure}[h!]
    \caption{Density Plot of Physician Counts}
    \centering
    \includegraphics[width=0.8\textwidth, height=0.35\textheight]{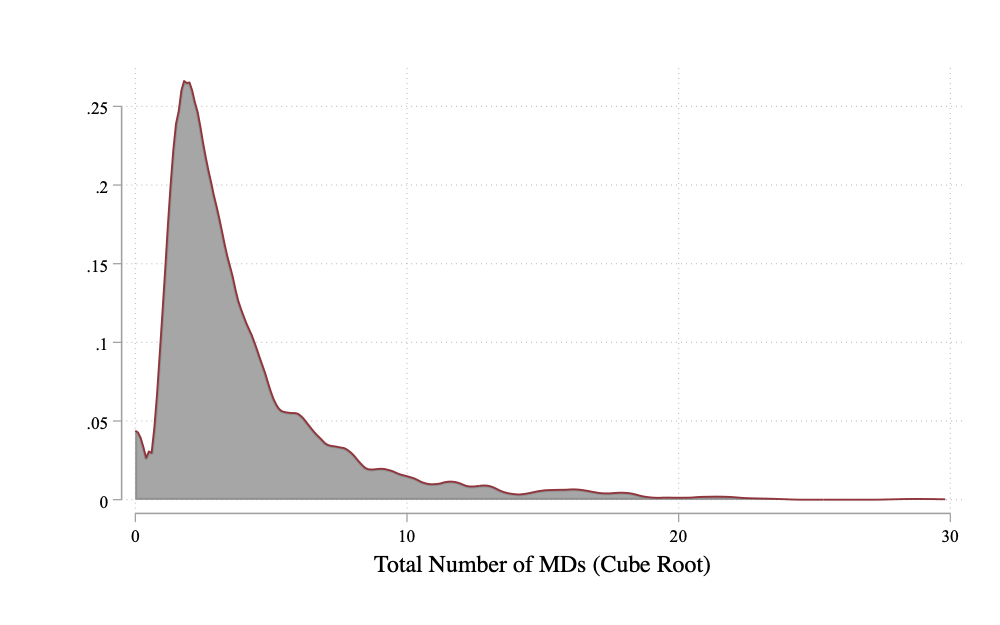}
    \label{fig:densityplots}
\noindent
\begin{minipage}{\textwidth}
    \begin{spacing}{1.5}
    \footnotesize
    \justifying
    \textit{Note}: The figure shows the total non-federal and federal MDs for specialties. The cube root transformation moderates the large values, so that the non-zero values are spread out to allow for a more distinctive visualization of the distribution, enabling better identification of patterns in the data.
    \end{spacing}
\end{minipage}
\end{figure}

The sample of  contains numerous health-related facets, including characteristics of the labor force such as the total count of individuals employed and unemployed who are aged 16 or older, and the rate of unemployment. Moreover, it provides poverty statistics, represented as the percentage of persons living in poverty, as well as important economic indicators like per capita personal income and median household income.
Details of health insurance coverage segmented by different age groups and information pertaining to Medicare beneficiaries and costs are also included.

The dataset gives insights into hospital utilization rates across different ranges, data about inpatient days in various types of hospitals and nursing homes, and the total number of hospitals along with characteristics about each type of hospital. One of the most significant features of the data is the count of medical practitioners at a county level which is classified according to their type - whether they are Federal or Non-Federal, their field of specialty, their age group, their gender, and so on. Including both outpatient and hospital-based physicians, this measure captures comprehensive physician service capacity. The data also has 4 to 5 digit county FIPS codes consisting of 3 digit county code preceded by 1 or 2 digit State FIPS codes. 
To address missing values for ``the percentage of people aged 65 and older without health insurance", for the years 2010–2012, a fixed-effects regression model was employed using Stata. The dataset was structured as panel data with countyfips as the panel identifier and year as the time variable. The variable was regressed on year for the period 2013 and onwards, controlling for unobserved heterogeneity and using robust standard errors. Predicted values were generated and used to impute the missing observations for 2010–2012, ensuring continuity and leveraging data from subsequent years for accurate interpolation.

\vspace{1em}
\begin{figure}[h!]
    \centering
    \caption{Time Trends in Physician Count}
    \includegraphics[scale=.3]{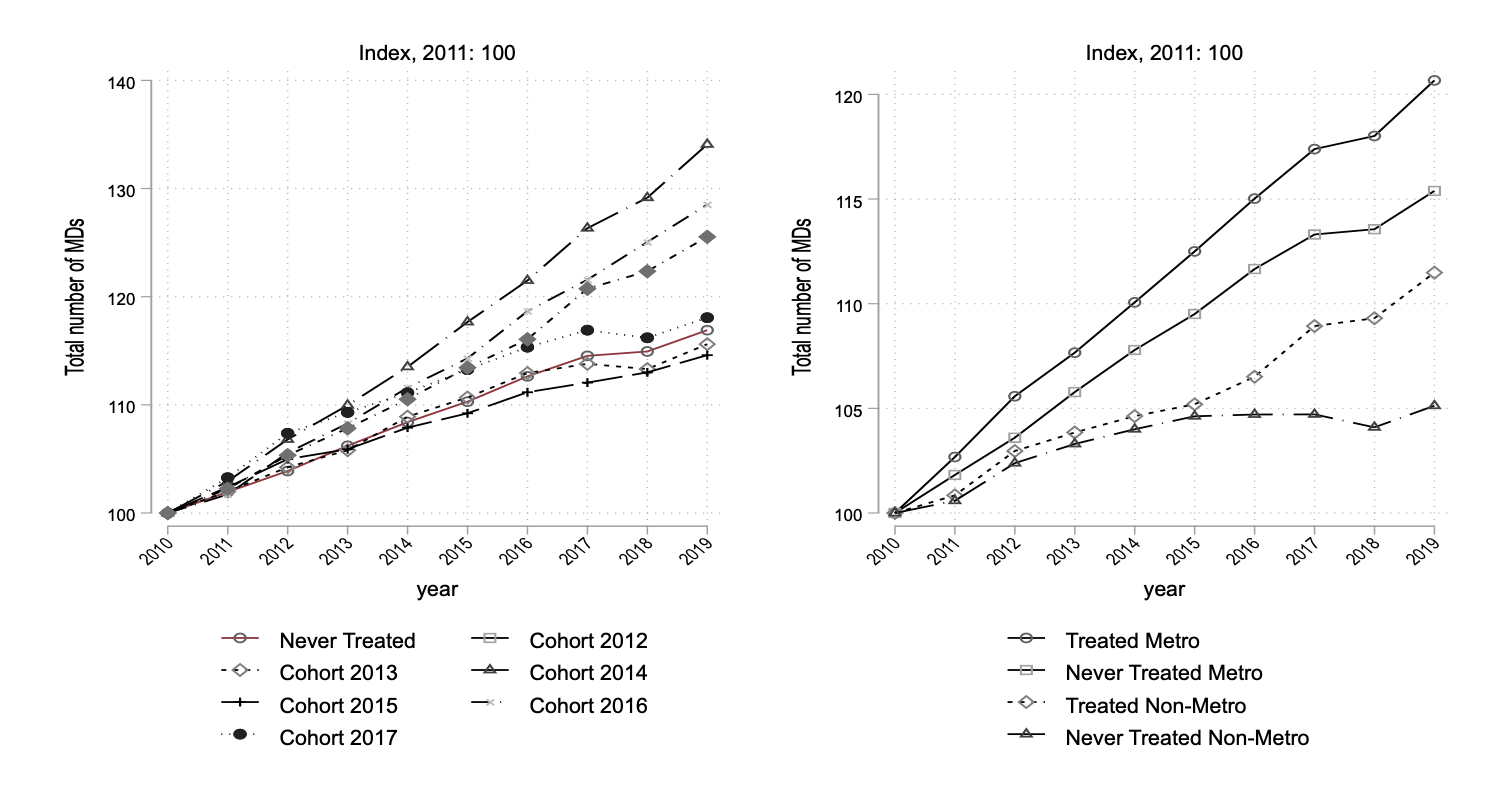}
    \label{fig:physiciantrends}
\noindent
\begin{minipage}{\textwidth}
    \begin{spacing}{1.5}
    \footnotesize
    \justifying
    \textit{Note}: The first panel shows the indexed trends in physician count according to the cohorts (formed according to the year in which a unit was first treated) including the never treated units. The second panel shows the count of physicians for treated and control (never treated) groups for metro and non-metro areas.
    \end{spacing}
\end{minipage}
\end{figure}

Note that there are non-physician-provided healthcare services such as nursing, therapy, pharmacy, or other alternative modes of healthcare that are not relevant to this study. Although DO (Doctor of Osteopathic Medicine) is also considered a physician, the study focuses only on MD (Doctor of Medicine) only, since all specialty-wise variables are not present for DO. The inclusion of DO does not affect our results.

\vspace{1em}
\subsection{IMLC Data Description}\label{sec:imlcdata}

\begin{figure}[htbp!]
\centering
\includegraphics[scale=.3]{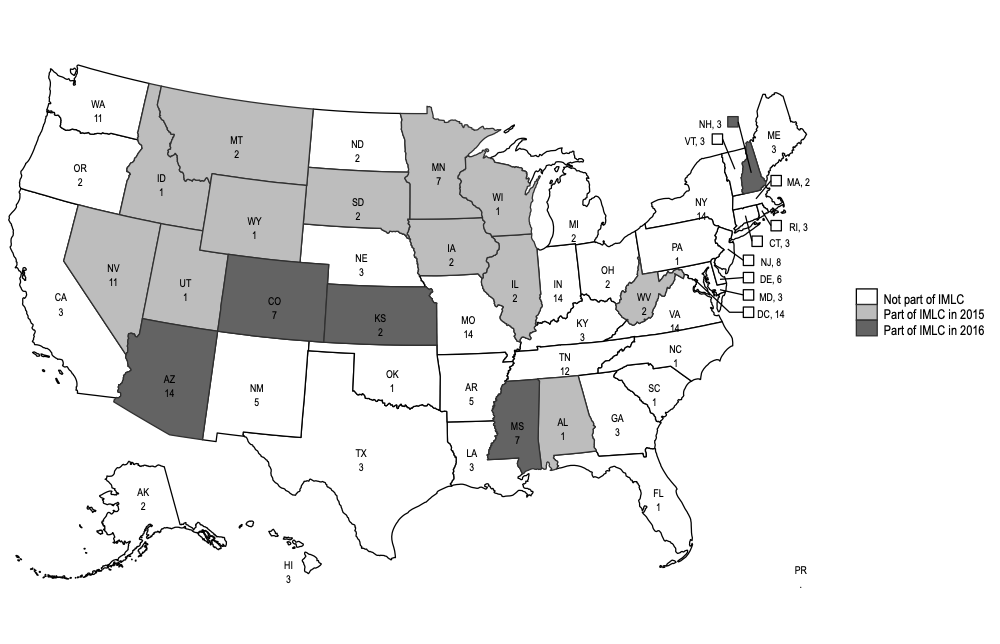}
\caption{Inter State Licensure Compact States}
\label{fig:IMLCC}
\parbox{\textwidth}{
\begin{spacing}{1.5}
\justifying
\footnotesize
\textit{Note}: The map indicates the states that joined the compact in 2015 and 2016. To accommodate the constraints of the map representation, Alaska and Hawaii are positioned below the contiguous United States rather than at their actual locations. A more detailed exploration can be obtained from: \url{https://www.imlcc.org/news/press-releases-and-publications/}.
\end{spacing}}
\end{figure}

\vspace{1em}
\textit{Licensure, credentialing and privileging}: The Interstate Medical Licensure Compact is a cooperative tool among participating U.S. states and territories that simplifies the licensing process for physicians practicing in multiple states. The data for the states who joined the compact in 2015 and 2016 comes from \cite{IMLCC2015_2016}, official website.

State licensure requirements could potentially hinder the broader usage of telehealth and affect the physician's ability to practice telehealth. The administrative burden of licensure laws could deter physicians from utilizing telehealth.\footnote{According to the Federation of State Medical Boards Telemedicine Overview (2015), 80\% of states require out-of-state clinicians offering telehealth to be licensed in the patient's residing state.} To address this, a model \enquote{Interstate Licensure Compact} was drafted, which aims to streamline interstate licensing and expand telehealth usage where the hospital where the patient is located to have the ultimate authority in decision-making for \enquote{privileging}.\footnote{Credentialing involves the verification of a provider's qualifications, while privileging decides what specific procedures or services the provider can offer based on those credentials.} The license agreements or `compacts', offer a more efficient route to practicing telehealth across multiple states. For instance, Wisconsin became part of IMLC and enacted the medical licensing legislation to successfully address doctor shortage in the area. By controlling for the effects of the Interstate Medical Licensure Compact, the impact of TPL can be isolated by ensuring that the results are not confounded by simultaneous policy changes.

\label{tab:IMLCC}
\begin{longtable}{|l|l|l|l|l|l|l|}
    \caption{List of State Enactments}\\
    \hline
    \textbf{State} & \textbf{Join Date} & \textbf{Type} & & \textbf{State} & \textbf{Join Date} & \textbf{Type} \\
    \hline\hline
    \endfirsthead

    \hline
    \textbf{State} & \textbf{Join Date} & \textbf{Type} & & \textbf{State} & \textbf{Join Date} & \textbf{Type} \\
    \hline\hline
    \endhead

    \hline
    \endfoot

    \hline
    \endlastfoot

    WY & 2/27/2015 & Composite & & NE & 1/5/2017 & Composite \\
    \hline
    SD & 3/12/2015 & Composite & & WA & 1/17/2017 & MD \& DO \\
    \hline
    UT & 3/20/2015 & MD \& DO & & TN & 2/9/2017 & MD \& DO \\
    \hline
    ID & 3/25/2015 & Composite & & DC & 3/7/2017 & Composite \\
    \hline
    WV & 3/31/2015 & MD \& DO & & ME & 4/6/2017 & MD \& DO \\
    \hline
    MT & 4/8/2015 & Composite & & GU & 6/26/2017 & Composite \\
    \hline
    AL & 5/19/2015 & Composite & & VT & 12/21/2017 & MD \& DO \\
    \hline
    MN & 5/19/2015 & Composite & & MD & 1/19/2018 & Composite \\
    \hline
    NV & 5/27/2015 & MD \& DO & & KY & 12/13/2018 & Composite \\
    \hline
    IA & 7/2/2015 & Composite & & ND & 1/15/2019 & Composite \\
    \hline
    IL & 7/21/2015 & Composite & & GA & 1/17/2019 & Composite \\
    \hline
    WI & 12/14/2015 & Composite & & OK & 1/18/2019 & MD \& DO \\
    \hline
    NH & 5/5/2016 & Composite & & MI & 6/1/2019 & MD \& DO \\
    \hline
    AZ & 5/11/2016 & MD \& DO & & LA & 10/28/2020 & Composite \\
    \hline
    KS & 5/13/2016 & Composite & & TX & 6/7/2021 & Composite \\
    \hline
    MS & 5/17/2016 & Composite & & DE & 6/23/2021 & Composite \\
    \hline
    CO & 6/8/2016 & Composite & & OH & 7/1/2021 & Composite \\
    \hline
    PA & 10/26/2016 & MD \& DO & & NJ & 1/10/2022 & Composite \\
    \hline
    FL & 3/21/2024 & MD \& DO & & IN & 3/10/2022 & Composite \\
    \hline
    HI & 6/22/2023 & Composite & & CT & 5/13/2022 & Composite \\
    \hline
    MO & 7/6/2023 & Composite & & RI & 6/29/2022 & Composite \\
    \hline
\end{longtable}

\footnotesize
\noindent\textit{Note}: This table shows the state-level enactments for the Interstate Medical Licensure Compact (IMLC). States listed joined the IMLC in 2015 and 2016, and the data was provided by IMLC Executive Director Marschall S. Smith.
\normalsize

\end{document}